\documentclass[fleqn,usenatbib]{mnras}

\usepackage{newtxtext,newtxmath}

\usepackage[T1]{fontenc}
\usepackage{ae,aecompl}

\usepackage{float}
\usepackage{hyperref}
\usepackage{color}
\usepackage{mathrsfs}
\usepackage{epsfig}
\usepackage{epstopdf}
\usepackage{multirow}
\usepackage{longtable}
\usepackage{booktabs}
\usepackage{lscape}
\usepackage{threeparttablex}
\usepackage{caption}
\usepackage{subcaption}

\usepackage{graphicx}	
\usepackage{amsmath}	

\newcommand{\mum}{${\rm \mu m}$}
\newcommand{\Lx}{$L_{\rm X}$}
\newcommand{\Ledd}{$\lambda_{\rm Edd}$}
\newcommand{\z}{$z$}
\newcommand{\Msun}{$M_\odot$}
\newcommand{\ergps}{erg s$^{-1}$}
\newcommand{\Lsun}{$L_\odot$}
\newcommand{\Tdust}{$T_{\rm dust}$}
\newcommand{\Lir}{$L_{\rm IR}$}

\newcommand{\Lbol}{$L_{\rm bol}$}
\newcommand{\eqwspd}{${\rm EW}^{\rm PAH}_{\rm 6.2 \mu m}$}
\newcommand{\LxChandra}{$L_{\rm X}^{2\--10}$}
\newcommand{\LxSwift}{$L_{\rm X}^{14\--195}$}
\newcommand{\Lpah}{$L_{\rm PAH}$}
\newcommand{\Ldust}{$L_{\rm dust}$}

\title[The post-{\it Herschel} view of intrinsic AGN emission]{The post-{\it Herschel} view of intrinsic AGN emission: constructing templates for galaxy and AGN emission at IR wavelengths}

\author[E.~Bernhard et al.]{
E. Bernhard,$^{1}$\thanks{E-mail: e.p.bernhard@sheffield.ac.uk \newline The templates associated with this work are available in the supplementary material, or directly at \url{https://tinyurl.com/yawp96qc}. The SED fitting code {\sc iragnsep} is distributed via \url{https://pypi.org/project/iragnsep/}. We used the version 7.2.0 for this work.}
C. Tadhunter,$^{1}$
J. R. Mullaney,$^{1}$
L. P. Grimmett,$^{1,2}$
D. J. Rosario$^{3}$
\newauthor
and D. M. Alexander$^{3}$.
\\
$^{1}$Department of Physics $\&$ Astronomy, University of Sheffield, Sheffield S3 7RH, UK\\
$^{2}$Department of Statistics, School of Mathematics, University of Leeds, Leeds LS2 9JT UK\\
$^{3}$Centre for Extragalactic Astronomy, Department of Physics, Durham University, South Road, Durham DH1 3LE, UK\\
}

\date{Accepted XXX. Received YYY; in original form ZZZ}

\pubyear{2020}

\begin{document}
\label{firstpage}
\pagerange{\pageref{firstpage}--\pageref{lastpage}}
\maketitle

\begin{abstract}
Measuring the star-forming properties of AGN hosts is key to our understanding of galaxy formation and evolution. However, this topic remains debated, partly due to the difficulties in separating the infrared (i.e. 1--1000~\mum) emission into AGN and star-forming components. Taking advantage of archival far-infrared data from {\it Herschel}, we present a new set of AGN and galaxy infrared templates, and introduce the spectral energy distribution fitting code {\sc iragnsep}. Both can be used to measure infrared host galaxy properties, free of AGN contamination. To build these, we used a sample of 100 local (\z~$<$~0.3), low-to-high luminosity AGNs (i.e. \Lbol$~\sim~10^{42\--46}$~\ergps), selected from the 105-month {\it Swift}--BAT X-ray survey, which have archival {\it Spitzer}--IRS spectra and {\it Herschel} photometry. We first built a set of seven galaxy templates using a sample of 55 star-forming galaxies selected via infrared diagnostics. Using these templates, combined with a flexible model for the AGN contribution, we extracted the intrinsic infrared emission of our AGN sample. We further demonstrate that we can reduce the diversity in the intrinsic shapes of AGN spectral energy distributions down to a set of three AGN templates, of which two represent AGN continuum, and one represents silicate emission. Our results indicate that, on average, the contribution of AGNs to the far-infrared ($\lambda~\gtrsim$~50~\mum) is not as high as suggested by some recent work. We further show that the need for two infrared AGN continuum templates could be related to nuclear obscuration, where one of our templates appears dominated by the emission of the extended polar dust.
\end{abstract}




\section{Introduction}

Multiple lines of empirical evidence now suggest that the growth of supermassive black holes (SMBHs) at the centres of galaxies is connected to the growth of galaxies themselves \citep[see][for a review]{Harrison2017}. For instance, a number of studies have shown that the redshift evolution of the star formation rate density of the Universe is closely related to that of the black hole accretion rate density, with both peaking at \z~$\sim$~2 and tracing each other, yet shifted by a factor of roughly 1500 \citep[e.g.][]{Aird2015}. However, the precise connection between SMBH growth (observed as active galactic nuclei, or AGN) and galaxy properties continues to be debated, in part due to contradictory results.

To better understand the connection between galaxies and AGNs, major efforts have focused on using the infrared (IR) part of the electromagnetic spectrum to measure the star-forming properties of AGN host galaxies. The reason for this is twofold. Firstly, the IR luminosity of a galaxy (i.e. integrated between 8~\mum\ and 1000~\mum; \Lir) is believed to be a well-calibrated proxy of star formation rate (SFR) for star-forming galaxies \citep[e.g.][]{Kennicutt1998}. This is due to the ultraviolet (UV) light produced by young massive stars being re-processed by dust grains surrounding the star-forming regions within galaxies. Secondly, the thermal emission from the close proximity of AGNs (e.g. circumnuclear\footnote{Here nuclear is used to refer to the central engine of the galaxy.} torus) is, in theory, too hot to contaminate large portions of the IR spectrum, especially in the far-infrared (i.e. $\gtrsim~30$~\mum; FIR; e.g. \citealt{Rosario2012}). Nevertheless, the dust in the regions surrounding the AGN central engine can generate sufficient thermal emission at near-to-mid IR wavelengths to dominate over the host galaxy \citep[e.g.][]{Smith2007, Sales2010, Lambrides2019}. In addition, some recent studies suggest that, in at least some cases, an AGN is able to heat dust on much larger scales (i.e. galactic scales), thereby generating thermal emission at longer wavelengths (i.e. $\lambda~\gtrsim~$80~\mum; e.g. \citealt{Tadhunter2007, Schweitzer2008, Dicken2009, Mor2009, Symeonidis2016}). Therefore, even in the FIR, AGN contamination potentially introduces a number of limitations and uncertainties when determining the star-forming properties of the host galaxies, which ultimately hampers the search for any connection between AGN and galaxy properties.

Recent studies have gone to some lengths in attempting to measure the properties of AGN host galaxies while accounting for AGN contamination. Some, for example, model the full (i.e. from UV to radio) spectral energy distribution (SED) of AGN host galaxies using several components that can each be attributed to either a galactic or a nuclear origin \citep[e.g.][]{Noll2009, CalistroRivera2016, Robotham2020, Yang2020}. Often, however, these depend on modelled grids of templates to represent galaxy and AGN emission, which are idealised cases containing many free parameters, and hence subject to uncertainties and degeneracies.

An alternative approach involves focusing on the IR part of the spectrum and using empirical templates to account for any AGN contamination at these wavelengths. For example, \cite{Mullaney2011} constructed a set of empirical templates that represent galaxy and AGN emission at mid-to-far-IR wavelengths (i.e. 6~\mum\ to 100~\mum), enabling the decomposition of the IR SED of AGN-hosting galaxies, and the subsequent measurement of the host galaxy SFR free of AGN contamination. To do this, they used a sample of typical X-ray AGNs (i.e. \LxChandra~$\sim~10^{42\--44}$~\ergps, where \LxChandra\ is the X-ray 2--10~keV luminosity) for which they had {\it Spitzer}--IRS spectra and {\it IRAS} photometry (i.e. photometry up to 100~\mum). The mean average AGN template of \cite{Mullaney2011} was found by extracting any excess in the flux observed in the IR SEDs of AGN-dominated sources once the host galaxy contribution had been removed. However, \cite{Mullaney2011} did not have access to wavelengths long-ward of 100~\mum, since at the time {\it Herschel}\footnote{{\it Herschel} is an ESA space observatory with science instruments provided by European-led Principal Investigator consortia and with important participation from NASA.} data were not available. This is problematic, as these longer wavelengths are crucial to accurately probe the dust heated by star formation \citep[e.g.][]{Ciesla2015}. In addition, the spatial resolution of {\it IRAS} data used by \cite{Mullaney2011} is roughly 20 times worse than that of {\it Herschel} at similar wavelengths, potentially leading to contamination from neighbouring sources.

More recently, using {\it Herschel} data, \cite{Symeonidis2016} suggested that AGN emission is cooler (and thus contributes more strongly to the FIR) than previously found. These findings could alter significantly the SFRs measured from FIR observations of AGN-hosting galaxies, and therefore greatly impact our understanding of AGN--galaxy co-evolution. However, we stress that the method used in \cite{Symeonidis2016} to derive their template is controversial, and that inconsistencies with other observations have been pointed out in several studies \citep[e.g.][]{Lani2017, Lyu2017a, Stanley2018, Schulze2019}. Instead, some studies that use {\it Herschel} \citep[e.g.][]{Lani2017, Lyu2017a, Xu2020} report an AGN contribution to the FIR that is similar to that obtained by \cite{Mullaney2011}. Most of these studies, however, focus on optically bright quasars (QSOs), which are extreme sources and do not represent the majority of the AGN population.

Here, we built upon the work of \cite{Mullaney2011} by constructing a new library of templates for galaxy and AGN emission at IR wavelengths that can be applied to more typical AGNs. With the availability of {\it Herschel} data, we can now decompose the IR SEDs into AGN and galaxy components without having to extrapolate the FIR from shorter wavelengths, as in \cite{Mullaney2011}. In \S\ref{sec:sample} we describe how we selected our sample and constructed SEDs of X-ray-selected AGNs for which we have mid-IR (MIR) spectra and FIR photometry. We then present in \S\ref{sec:templBuild} how we defined a set of templates for the IR emission of non-AGN galaxies that we then fit to our sample of AGN IR SEDs, in doing so extracting the AGN components which we then used to define a set of templates for the IR emission of AGNs. In \S\,\ref{sec:results} we compare our AGN templates to those reported in previous studies and discuss some aspects of our new library of AGN templates. Finally, we present our conclusions in \S\,\ref{sec:conclusion}. Throughout, we adopted a WMAP--9 year cosmology ($H_0$~=~69.33~km~s$^{-1}$~Mpc$^{-1}$, $\Omega_{\Lambda}$~=~0.712, $\Omega_b$~=~0.0472, $\Omega_c$~=~0.2408; \citealt{Hinshaw2013}) and a \cite{Chabrier2003} initial mass function when calculating galaxy properties.

\section{Sample selection and data description}
\label{sec:sample}

\subsection{X-ray data}
To investigate the typical IR emission of AGNs and that of their host galaxies, we used a sample of X-ray AGNs selected from the 105-month {\it Swift}--BAT all-sky hard X-ray survey of \cite{Oh2018}. The Burst Alert Telescope (BAT) on board the {\it Swift} gamma-ray burst space observatory \citep{Gehrels2004} is conducting a uniform all-sky survey at energies 14--195~keV to unprecedented sensitivities (e.g. 90~per~cent of the sky to a depth of 8.40$\times 10^{-12}\,{\rm erg\, s^{-1} cm^{-2}}$). Although the mission was not specifically designed for AGNs (see \citealt{Barthelmy2005} for the mission primary goal), the energies covered by {\it Swift}--BAT correspond to those emitted by AGNs at wavelengths almost free from absorption. This, combined with the all-sky survey strategy, provides one of the best opportunities to define a clean sample of mostly low redshift AGNs. By cross-matching the hard X-ray detections with archival data at various wavelengths (i.e. from softer X-ray to optical), \cite{Oh2018} have been able to robustly classify the {\it Swift}--BAT sources into different types \cite[see Table 1 in][]{Oh2018}.

Out of the 1632 sources reported in \cite{Oh2018}, we retained Seyfert galaxies (819 sources), LINERs (6 sources), and unknown AGNs (114 sources)\footnote{Unknown AGNs are X-ray sources associated with galaxies whose optical spectra and type classifications are not known (see \citealt{Oh2018}).}. All other classes, including beamed AGNs, such as Blazars and flat-spectrum radio quasars, were discarded to avoid complications arising from the potential non-thermal contamination to the FIR\footnote{Discarded classes were: Unknown class I, Unknown class II, Unknown class III, Multiple, Galactic centre, Galaxy Cluster, Beamed AGN (Blazar/FSRQ), Cataclysmic Variable Star (CV), Symbiotic Star, Other Star, Open Star Cluster, Starburst Galaxy, Compact Group of Galaxies, Pulsar, Supernova Remnant (SNR), Nova, High Mass X-ray Binary (HMXB), Low Mass X-ray Binary (LMXB), Other X-ray Binary (XRB), Globular Cluster (GC), Molecular Cloud, and Gamma-ray Source.}. In addition, after visual inspection, we removed Cygnus A which was the only source that showed obvious non-thermal contamination in the FIR. We therefore retained 938 hard-X-ray selected AGNs which constitute our parent sample, the redshifts of which span 0~$<~$\z$~<$~4.7, as reported by the NASA/IPAC Extragalactic Database\footnote{The NASA/IPAC Extragalactic Database (NED) is operated by the Jet Propulsion Laboratory, California Institute of Technology, under contract with the National Aeronautics and Space Administration.} (NED) or the SIMBAD \citep{Wenger2000} database \citep{Oh2018}.

\subsection{IR data}

\subsubsection{Far-IR photometry}
\label{subsubsec:dataFIRphoto}

We used FIR photometry (i.e. 70~\mum~$<~\lambda_{\rm obs}~<~$500~\mum) to characterise the cold dust emission of our sample of AGNs. As we will perform SED fits to separate out the AGN contribution (see \S\,\ref{subsubsec:AGNfit}), we aimed to obtain at least three FIR photometric measurements (considering 3$\sigma$ detections), in addition to the {\it Spitzer}-IRS spectra characterising the MIR emission of our AGN sample (see \S\,\ref{subsubsec:dataIRSspec} for the MIR spectra). The Highly Processed Data Products\footnote{The explanatory supplements for the PACS point source catalogue can be found at \url{http://archives.esac.esa.int/hsa/legacy/HPDP/PACS/PACS-P/PPSC/HPPSC_Explanatory_Supplement_v2.2.pdf}}$^{,}$\footnote{The explanatory supplements for the SPIRE point source catalogue can be found at \url{http://archives.esac.esa.int/hsa/legacy/HPDP/SPIRE/SPIRE-P/SPSC/SPIREPointSourceCatalogExplanatorySupplementFull20170203.pdf}} (HPDPs) provide the most complete and uniform database of reduced fluxes for both the PACS \citep[covering the 70~\mum, 100~\mum, and 160~\mum\ bands;][]{Poglitsch2010, Marton2017} and SPIRE \citep[covering the 250~\mum, 350~\mum, and 500~\mum\ bands;][]{Griffin2010, Schulz2017} instruments on-board the {\it Herschel} Space Observatory \citep{Pilbratt2010}.

We identified FIR counterparts to the X-ray selected AGNs by cross-matching their positions using a 5\arcsec\ matching radius. We found 112 sources out of the 938 of our parent sample that followed our criteria for SED fits. However, we also found that the HPDP point source catalogues largely failed to recognise the extended nature of sources, in particular at \z~$\lesssim$~0.3, where our final AGN sample lies (see \S\,\ref{subsec:obsSED} for the properties of our final AGN sample). As a consequence, the vast majority of our sources suffered from under-estimated fluxes in the HPDPs, since they had been regarded as point sources in their reduction.

To obtain accurate fluxes (i.e. accounting for extended emission, where necessary), we used {\it Herschel}--PACS and {\it Herschel}--SPIRE images from the ``{\it Herschel} high level images'' products of the NASA/IPAC Infrared Science Archive (IRSA) database (see Appendix\,\ref{app:IRflux} for more details on the images). The {\it Herschel} fluxes for extended sources were calculated following the method presented in Appendix\,\ref{app:IRflux}, and are reported in Table\,\ref{table:AGNFlux}. Overall, we found that 30~per~cent, 20~per~cent, 55~per~cent, 62~per~cent, 58~per~cent, and 45~per~cent of our sources have been misclassified as point sources in the HPDPs at 70~\mum, 100~\mum, 160~\mum, 250~\mum, 350~\mum, and 500~\mum, respectively. Some fluxes showed up to a factor of 10 increase after accounting for the spatially extended emission. We also found very good agreements between our fluxes and those calculated for extended sources reported in \cite{Melendez2014} and \cite{Shimizu2016} for {\it Herschel}--PACS and {\it Herschel}--SPIRE, respectively, for the 93~per~cent of our sample which overlap with theirs (see Fig.\,\ref{fig:fluxCompAGN}).

\subsubsection{mid-IR spectra}
\label{subsubsec:dataIRSspec}
As AGN-heated dust is expected to radiate significantly at MIR wavelengths, we completed the IR coverage using {\it Spitzer}-IRS low resolution spectra \citep{Houck2004}. The {\it Spitzer}--IRS is made of two separate low resolution spectrograph modules, Short--Low (SL) and Long--Low (LL), covering wavelengths from 5.3~\mum\ to 38~\mum, and for which the resolution ranges between $R~=~\lambda/\Delta\lambda~\sim$~90--600 depending on wavelength (with shorter wavelengths having higher spectral resolution). Each of the SL and LL modules is also divided into two sub-slits SL1/SL2 and LL1/LL2 covering different wavelength ranges within the aforementioned full wavelength range (see Table~1 in \citealt{Houck2004} for the properties of the {\it Spitzer}--IRS). When a source is observed within the sub-slits of the second order, a short section of the first order is also illuminated. Therefore, the {\it Spitzer}--IRS offers bonus orders at wavelengths 7.3~\mum\ to 8.7~\mum\ and 19.4~\mum\ to 21.7~\mum\ for the SL and the LL modules, respectively. These were used to improve the overlap between the first and the second orders.

The {\it Spitzer}--IRS spectra of all 112 of the sources in our sample that also have at least three {\it Herschel} FIR measurements (see \S\,\ref{subsubsec:dataFIRphoto} for the {\it Herschel} photometry) were extracted from the Combined Atlas of Sources with {\it Spitzer}--IRS Spectra \citep[CASSIS;][]{Lebouteiller2011}. The CASSIS pipeline offers post-basic calibrated data that can handle a large variety of sources, from barely detected to bright sources, and from point sources to extended sources (see \citealt{Lebouteiller2010} for details on the tools used by CASSIS). From the CASSIS database, we downloaded the optimal spectrum for each of our AGNs that have been observed with {\it Spitzer}-IRS. This included optimal point source extraction for unresolved sources, and tapered column source extraction for extended sources (see \citealt{Lebouteiller2010} for more details on the source extraction method).

\subsection{Observed IR SEDs of AGNs}
\label{subsec:obsSED}

To produce observed IR SEDs we combined the {\it Spitzer}--IRS spectra with the {\it Herschel} photometry. To do this, we first corrected for flux mismatch between the SL and LL arms of the {\it Spitzer}-IRS spectra due to the different slit widths\footnote{The strategy adopted by the CASSIS team was to reduce independently the two arms of the spectrograph. Therefore, the connection between the SL and LL spectra is not guaranteed since slit losses can arise within the narrower slit (i.e. the SL arm).} (the slit widths of SL1 and SL2 are 3.7\arcsec\ and 3.6\arcsec, respectively, and those of LL1 and LL2 are 10.7\arcsec\ and 10.5\arcsec, respectively). As the LL slit width is larger, and hence likely to contain more light from the host galaxy, we renormalised the fluxes in the SL arm to ensure continuity between the two arms. This method is referred to as stitching. To do this we fit independently smoothed splines to the two arms, and interpolated at 14~\mum, which corresponds to the wavelength at which the two arms connect. We then calculated the correction needed for the SL arm to join the LL arm. As the presence of large polycyclic aromatic hydrocarbon (PAH) features can challenge the accuracy of the spline fitting, we visually inspected each of the stitched spectra, and corrected roughly half of the sample by a few per~cent to increase the stitching accuracy. Overall, we found a median correction factor of 1.10, with a maximum of a factor of five.

The full observed IR SED of each source was obtained by combining the {\it Spitzer}--IRS spectra with the {\it Herschel} photometry. To assess the overall normalisation of the spectra, we collected archival {\it Spitzer}-MIPS 24~\mum\ fluxes, only considering those derived accounting for the spatially extended flux ($\sim$70~per~cent of our sample), and compared to the synthetic 24~\mum\ {\it Spitzer}--MIPS flux (calculated by convolving the synthetic {\it Spitzer}--MIPS 24~\mum\ filter with the {\it Spitzer}--IRS spectra). We found that, on average, the observed 24~\mum\ photometry was $\sim$4~per~cent lower than the synthetic photometry. Therefore, we corrected the spectra of those sources for which we had a corresponding 24~\mum\, photometry measurement, but did not apply any corrections to the rest of the sample since the typical correction factor is negligible.

\begin{figure}
 \centering
 \includegraphics[width=0.5\textwidth]{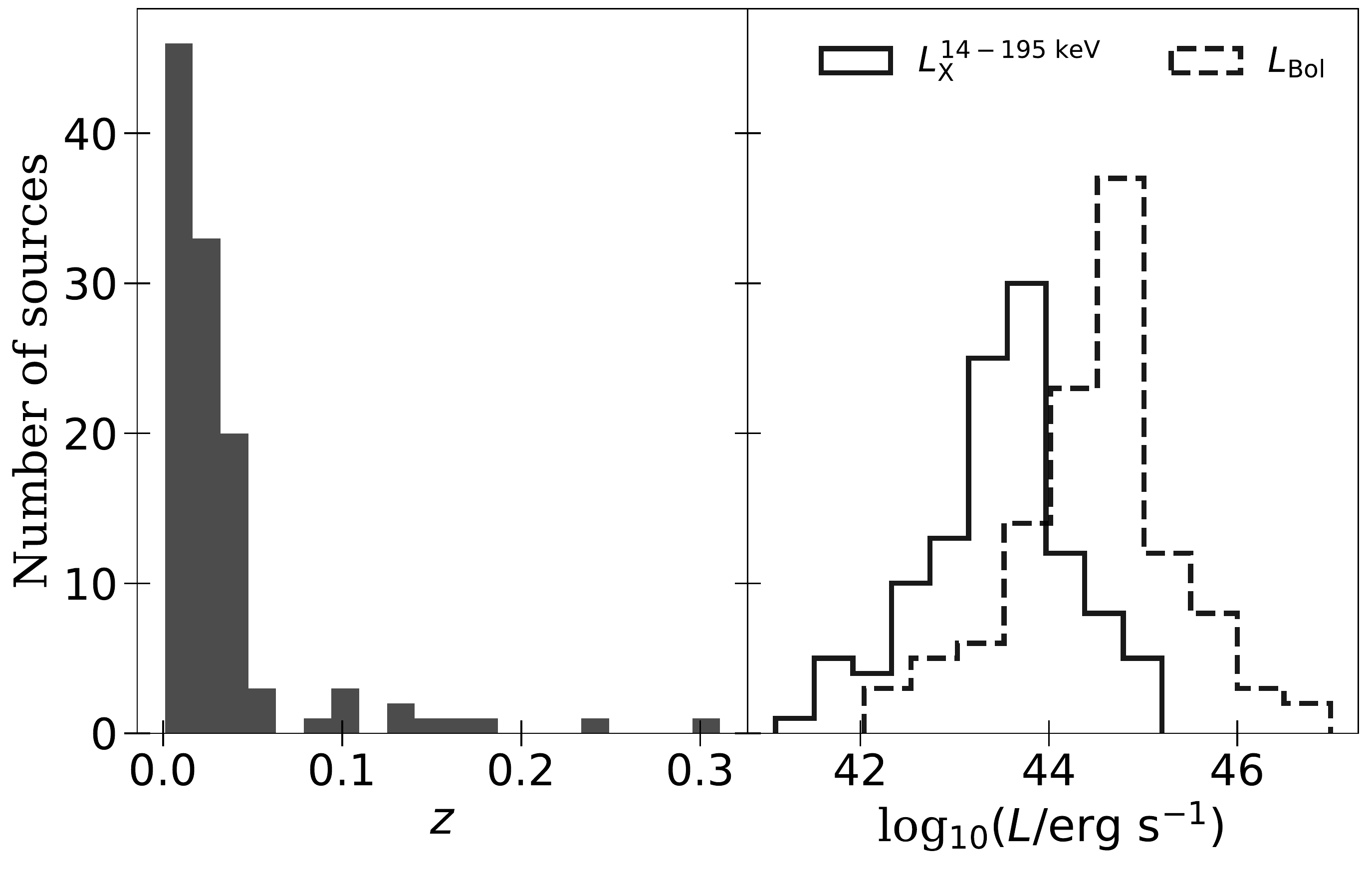}
 \caption{The redshift (left-hand panel), the {\it Swift}--BAT X-ray 14--195~keV luminosity (right-hand panel; continuous line distribution), and the bolometric luminosity (right-hand panel; dashed line distribution) distributions of our final sample of 112 AGNs. The latter is typical of low-to-high luminosity AGNs in the local Universe. \label{fig:sample}}
\end{figure}

\begin{figure}
 \centering
 \includegraphics[width=0.5\textwidth]{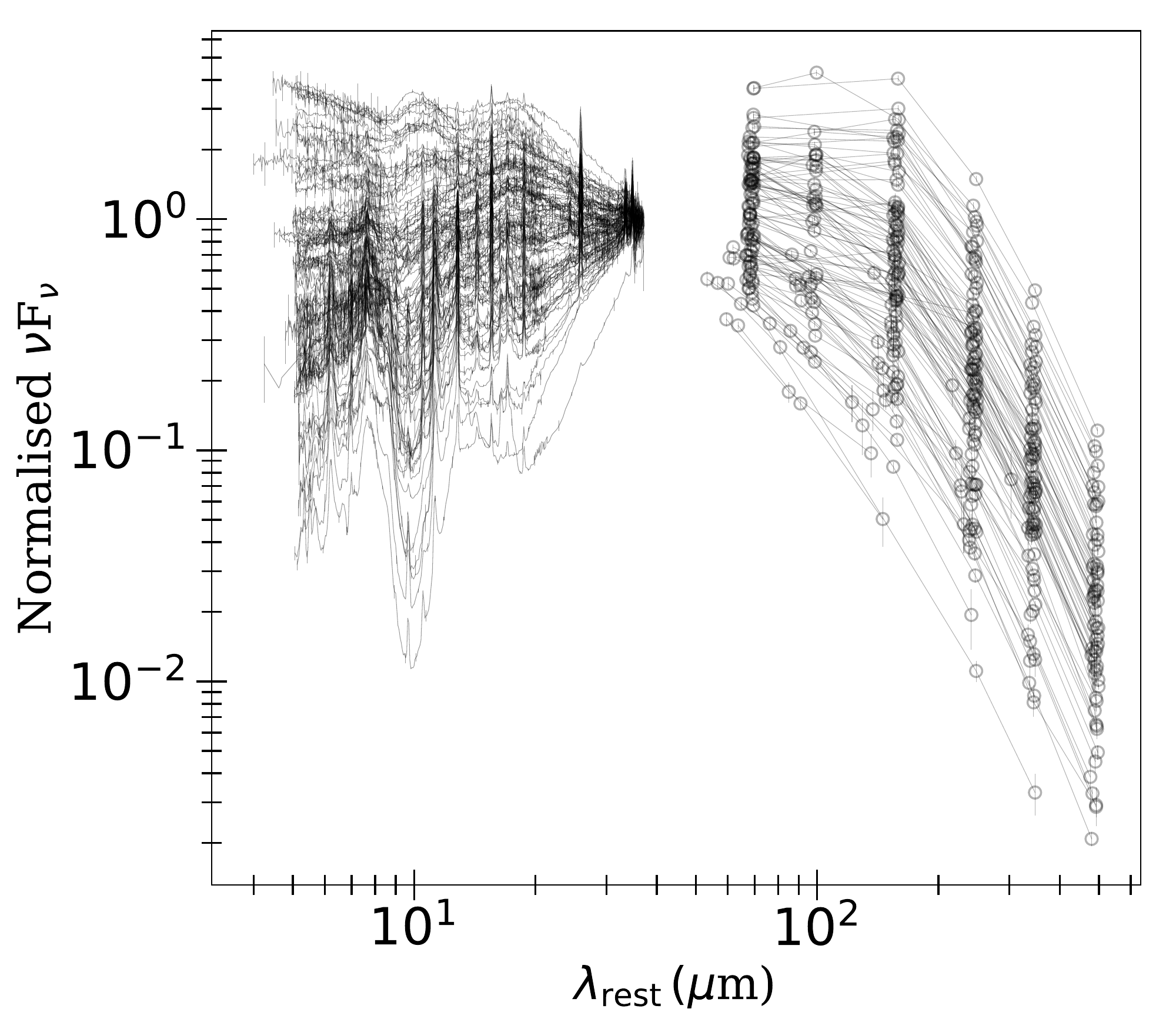}
 \caption{The IR SEDs of our full AGN sample normalised at 40~\mum. These were obtained by combining the low resolution {\it Spitzer}--IRS spectra with corresponding {\it Herschel} photometry. The latter has been corrected for spatially extended emission that was largely missed in the HPDPs. The thin lines show a linear interpolation between points and serve only to guide the eyes. We note that there is a large diversity in the IR SEDs of AGNs, once normalised at 40~\mum. \label{fig:allSEDs}}
\end{figure}

The main properties of the 112 sources in our AGN sample are reported in Table\,\ref{table:AGNprop}. We note that the sample is typical of local (i.e. \z~$<$~0.3, but the vast majority of the sample lies at \z~$<$~0.1) low-to-high luminosity AGNs with \LxSwift~$\sim~10^{41\--45}$~\ergps\ (where \LxSwift\ is the {\it Swift}--BAT X-ray 14--195~keV luminosity). This roughly corresponds to bolometric luminosities of \Lbol~$\sim~10^{42\--46}$~\ergps, after using the photon indices reported in \cite{Oh2018} to convert \LxSwift\ to \Lbol. We show in Fig\,\ref{fig:sample} the redshift, {\it Swift}--BAT luminosity, and bolometric luminosity distributions of our AGN sample. We note that our sample could be biased toward bright IR sources since we required {\it Herschel} detections. In Fig.\,\ref{fig:allSEDs} we show all\footnote{An extra 10 AGN SEDs are not shown as they will be removed from the sample after their fits revealed some discrepancies between the {\it Spitzer}--IRS spectra and the {\it Herschel} photometry (see \S\,\ref{subsubsec:AGNfit}).} our SEDs normalised at rest-frame 40~\mum. We note in particular the large diversity in the IR SEDs of our sample of AGNs, and that some are clearly flatter in the MIR, suggesting AGN-dominated MIR sources.

\section{Building libraries of templates for galaxy and AGN emission}
\label{sec:templBuild}

In this section we describe how we built sets of templates which are representative of galaxy and AGN emission at IR wavelengths. To do this, we first selected a clean sample of 55 star-forming galaxies with {\it Spitzer}--IRS spectra and {\it Herschel} photometry (see \S\,\ref{subsubsec:galSample}), from which we extracted seven different galaxy templates that capture the full diversity of this sample of star-forming galaxies (see \S\,\ref{subsubsec:Templ}). By then fitting our sample of 112 AGN SEDs (see \S\,\ref{subsubsec:AGNfit}) with each of these galaxy templates and a flexible model for AGN emission (see \S\,\ref{subsubsec:model}), we extracted three AGN IR templates (see \S\,\ref{subsubsec:AGNtempl}).

\subsection{IR templates for galaxy emission}
\label{subsec:hostTempl}

\subsubsection{Sample of IR SEDs for star-forming galaxies}
\label{subsubsec:galSample}

As previously mentioned, we carefully built a set of templates to represent galaxy emission at IR wavelengths. To do this we required a set of galaxy-dominated SEDs. Our starting point was the work from \cite{Lambrides2019}, who have recently classified 2\,015 {\it Spitzer}--IRS spectra according to whether they are star-forming or AGN-dominated by using the equivalent width of the PAH emission feature at 6.2~\mum\, (\eqwspd). The strength of the 6.2~\mum\, emission feature is thought to be related to the level of star formation \citep[e.g.][]{Peeters2004}. In the presence of an AGN, the dust heated in the central region of the galaxy enhances the underlying continuum and therefore reduces the equivalent width of the 6.2~\mum\, feature \citep[e.g.][]{Nenkova2008}. As a consequence, the \eqwspd\ is thought to be a good indicator of the level of AGN contribution, {\it at least at wavelengths close to 6~\mum} \citep[e.g.][]{Armus2007, Spoon2007, Lambrides2019}.

\begin{figure}
 \centering
 \includegraphics[width=0.48\textwidth]{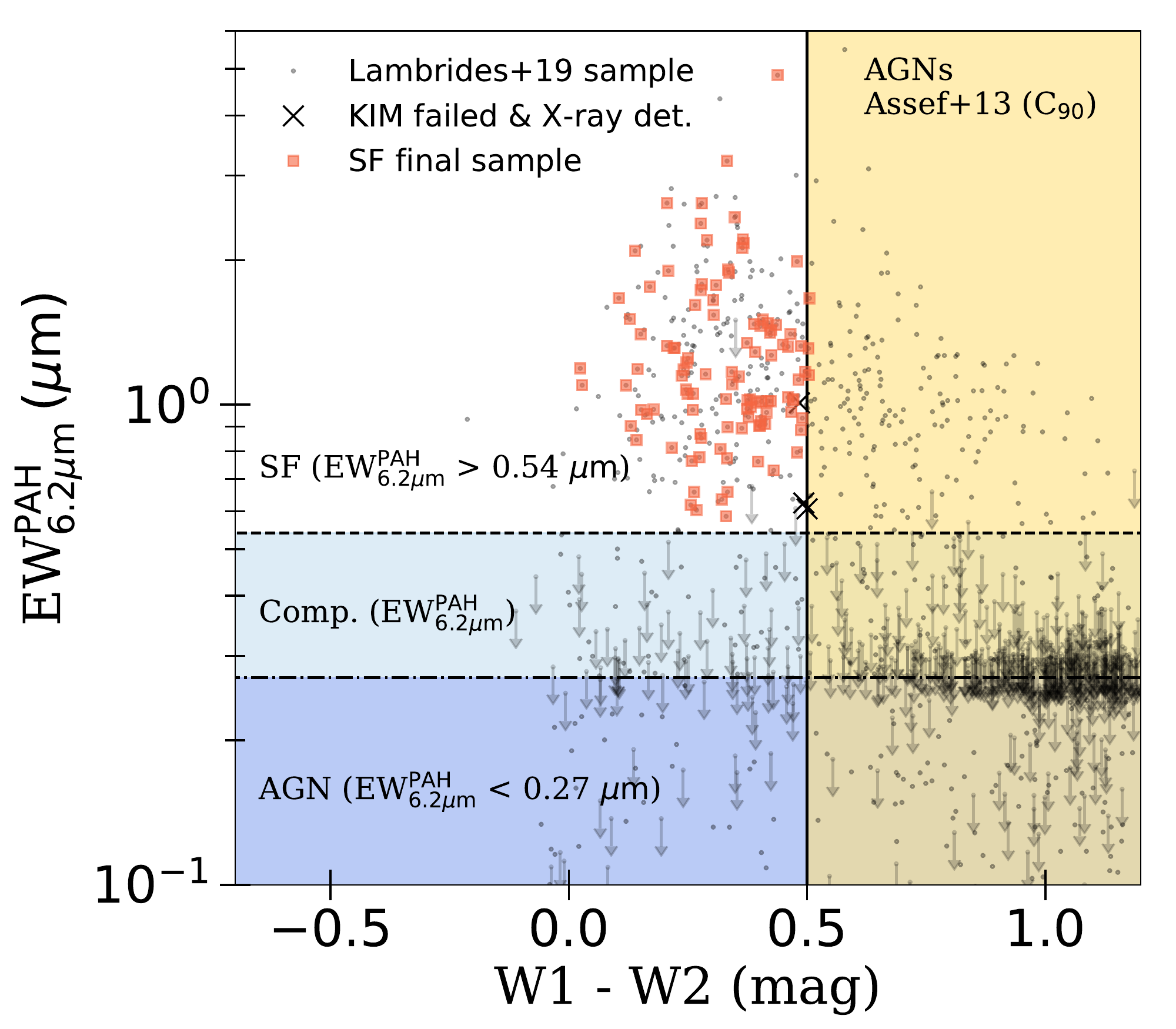}
 \caption{The \eqwspd\ versus the WISE colours W1~$-$~W2 of the sample from L19 (grey dots). The arrows indicate upper limits on \eqwspd. For clarity, we only show the top-left part of the full parameter space covered by the 2,015 sources from L19, where star-forming galaxies lie. Our sample of 110 pre-selected star-forming galaxies was first selected based on \eqwspd, where sources below the dot-dashed line are AGNs, sources above the dashed line are star-forming galaxies, and sources between these lines are composite galaxies. The values employed to define these limits are from \protect\citeauthor{Armus2007} (\protect\citeyear{Armus2007}; see text). We then employed the WISE colour selection criterion of \protect\cite{Assef2013}, shown with a vertical black line, where star-forming galaxies lie on the left-hand side of this line. In addition, one source was discarded as it failed the MIR KIM test of \protect\cite{Messias2012}, and two sources were discarded since detected by the {\it Swift}-BAT (black crosses). The red squares show our final sample of 110 pre-selected star-forming galaxies with {\it Spitzer}-IRS spectra and at least three {\it Herschel} photometric measurements. \label{fig:SFsel}}
\end{figure}

We retained all the sources in \citeauthor{Lambrides2019} (\citeyear{Lambrides2019}; hereafter L19) that had \eqwspd~$>$~0.54~\mum, where \eqwspd~=~0.54 corresponds to the average value found for pure starburst galaxies \citep{Armus2007}. Above this threshold, the MIR emission of galaxies is thought to be fully dominated by star-forming processes. Amongst the 2\,015 sources in L19, we found 464 that satisfy this criterion (see Fig\,\ref{fig:SFsel}). However, as measuring \eqwspd\ is challenging (e.g. the continuum must be estimated), we double-checked the values of \eqwspd\ by cross-matching our pre-selected sample of 464 star-forming galaxies with the sample of \cite{Samsonyan2016} -- who have compiled \eqwspd\ measurements from \cite{Sargsyan2011}, \cite{Stierwalt2014}, and the CASSIS database -- finding, in total, 107 overlapping sources. In doing so, we found 53 ambiguous sources that were classified as star forming in L19, and as composite or AGN-dominated in \cite{Samsonyan2016}. These 53 were discarded from our pre-selected sample of pure star-forming galaxies, leaving a sample of 411 star-forming galaxies selected purely on \eqwspd.

As recent evidence suggests that some AGNs can show \eqwspd\ typical of star-forming galaxies (e.g. L19), we further used IR colour selection to discard other potential AGNs from our pure star-forming sample. Following \cite{Assef2013}, AGNs can be separated from star-forming galaxies using the WISE W1 (3.4~\mum) and W2 (4.6~\mum) colours. We used the criterion from \cite{Assef2013} which guarantees 90~per~cent completeness and $\sim$80~per~cent reliability at the W2 magnitudes of our sample (typically W2~$<$~15~mag; L19), and only retained sources with W1~$-$~W2~$\leq$~0.5 (see Fig\,\ref{fig:SFsel}). The main caveat of such a selection is that we potentially retain 10~per~cent of AGNs in the sample of pre-selected star-forming galaxies, as a corollary of the 90~per~cent completeness reported in \cite{Assef2013}. However, we stress that after fitting the SEDs, we will carefully discard the excess of AGNs found in the star-forming sample selected purely based on the \eqwspd\ and the WISE colours (see \S\,\ref{subsubsec:SFfit}).

By using WISE colours we discarded 139 galaxies out of the 411 galaxies that were selected based purely on their \eqwspd. In addition, by considering archival K-band magnitudes, {\it Spitzer}-IRAC, and {\it Spitzer}-MIPS fluxes from L19, we also discarded one extra source which does not pass the IR colour ``KIM'' criterion of \cite{Messias2012}. Finally, we discarded two more sources that were detected by {\it Swift}--BAT. After making the aforementioned cuts, we were left with a total 269 galaxies in our pre-selected sample of star-forming galaxies.

Amongst the 269 pre-selected star-forming galaxies, we only retained 110 that have {\it Spitzer}--IRS spectra and at least three {\it Herschel} photometric measurements from the HPDPs. As for our AGN sample, we subsequently updated the {\it Herschel} fluxes to account for spatially extended emission (see \S\,\ref{subsubsec:dataFIRphoto} and Appendix\,\ref{app:IRflux} for the extended emission). The Table \,\ref{table:preSF} presents some basic properties of this sample of 110 galaxies, and the Table\,\ref{table:galFlux} reports the {\it Herschel} fluxes. We show in Fig.\,\ref{fig:SFsel} the \eqwspd\ of our full sample of 110 pre-selected star-forming galaxies plotted against the WISE colours. The galaxy sample extends to \z~$\sim$~1, although most sources lie at \z~$<$~0.3, which is similar to that of our AGN sample (see \S\,\ref{subsec:obsSED}). To combine the {\it Spitzer}--IRS spectra with the {\it Herschel} photometry we repeated the process outlined in \S\,\ref{subsec:obsSED}.

\subsubsection{Prescriptions for IR obscuration}
\label{subsubsec:obsCorr}

\begin{figure}
 \centering
 \includegraphics[width=0.48\textwidth]{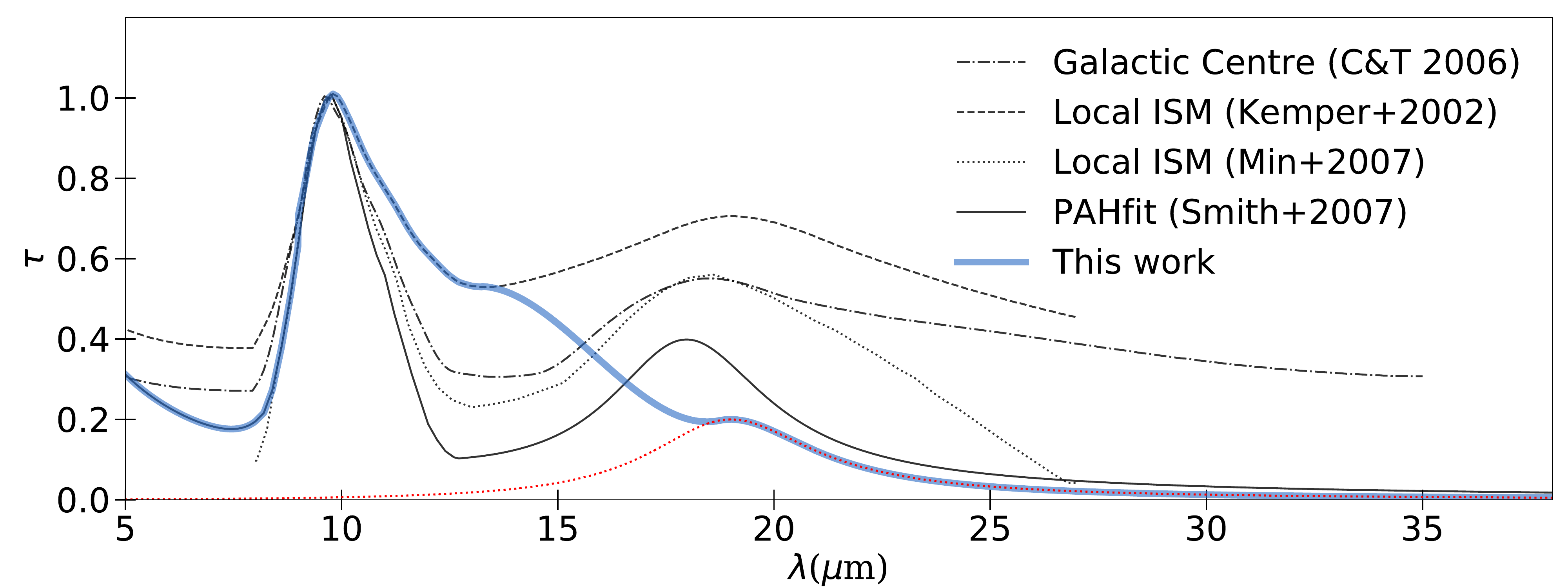}
 \caption{Dust opacity per unit wavelength (extinction curves) normalised at 9.7~\mum, and derived for the Galactic centre \protect\citep[][C\&T~2006 in keys]{Chiar2006}, for the local ISM \protect\citep{Kemper2002, Min2007}, and from PAHfit \protect\citep{Smith2007}, as indicated in the top-right keys. The pseudo--extinction curve adopted in this work is shown with a thick blue line, and is a combination of several of these curves (see text). The dotted red line shows the re-scaled and shifted extinction curve from PAHfit reflecting the relative extinction at 18~\mum\ observed in our sample of star-forming galaxies. \label{fig:dustOpacity}}
\end{figure}

In order to perform SED fits, we estimated the level of IR obscuration affecting the observed SEDs of our galaxies. First, we must assume a dust opacity curve which provides the relative level of obscuration per unit wavelength. We note that, in addition to significant disagreements between curves derived for the local ISM or for the Galactic centre \citep[e.g.][]{Chiar2006}, disagreements are found amongst curves for the local ISM \citep[e.g.][]{Kemper2002, Min2007}. We show in Fig.\,\ref{fig:dustOpacity} a number of these curves, in addition to that used in PAHfit \citep{Smith2007}, which is a fitting routine for the {\it Spitzer}-IRS spectra, and for which the attenuation curve has been derived from a combination of several curves (see \citealt{Smith2007} for details).

We found that none of these curves provided a satisfactory model for obscuration amongst our star-forming galaxies while fitting their SEDs (see \S\,\ref{subsubsec:SFfit} for the fits). In particular, we noted an excess of model flux around $\sim$13~\mum\ while using the attenuation curve of PAHfit. Instead, the extinction curve of \cite{Chiar2006}, which had a higher level of obscuration around $\sim$13~\mum, provided a better fit at these wavelengths, but a deficit of model flux at longer wavelengths, when compared to observations. To reconcile these, we derived our own fully empirically-motivated extinction curve. While we assumed that the discrepancies between the observed and the model fluxes, corrected for extinction, were due to the lack of a universal extinction law, we stress that the models of galaxies used to fit our SEDs have not been calibrated against MIR spectral observations. The latter could, to some extent, also explain the discrepancies between the observed and the model fluxes. However, we note that the galaxy templates used to fit our SEDs have been successfully tested against the stacked {\it Spitzer}--IRS spectra of luminous and ultra-luminous IR galaxies at \z~=~1 and \z~=~2, respectively \citep[][see \S\,\ref{subsubsec:SFfit} for the galaxy templates]{Schreiber2018}. Our extinction curve uses a combination of that from PAHfit and \cite{Kemper2002}. At wavelengths longer than 15~\mum, we renormalised the extinction curve of PAHfit to match the apparent obscuration measured in our sample of galaxies, and we shifted the silicate absorption peak from 18~\mum\ to 19~\mum\ (see Fig.\,\ref{fig:dustOpacity}). We refer to this extinction curve as ``pseudo--extinction curve'', as it is fully empirically motivated, and can potentially be related as much to the model used to fit our SEDs (see \S\,\ref{subsubsec:SFfit} for the fits) as to the properties of the obscuring dust.

With the extinction curve defined, we must estimate the total amount of obscuration within the galaxy. We have that,

\begin{equation}
\label{Eq:obs}
{\rm F}_\nu^{\rm~Obs}~\equiv~{\rm F}_\nu^{\rm~Int}~\times~\Theta(\tau_\lambda),
\end{equation}

\noindent where ${\rm F}_\nu^{\rm~Obs}$ and ${\rm F}_\nu^{\rm~Int}$ are the observed (i.e. attenuated) and intrinsic (i.e. unattenuated) flux densities, and $\Theta(\tau_\lambda)$ is the total attenuation at wavelength $\lambda$ given the optical depth $\tau_\lambda$. For the galaxy, we assumed a slab of uniformly mixed stars and dust \citep[e.g.][]{Li2001, Smith2007}. Therefore, for an observer looking perpendicular onto the screen, and in the case of pure absorption, the solution to the transfer equation leads to \citep[e.g.][]{Krugel2009},

\begin{equation}
\label{eq:solTransEqGal}
\Theta(\tau_\lambda)~=~\frac{(1~-~e^{-\tau_\lambda})}{\tau_\lambda}.
\end{equation}

\noindent We refer the reader to \S\,\ref{subsubsec:AGNfit} for the treatment of the AGN obscuration, as the assumption of a uniformly mixed medium breaks.

To estimate $\Theta(\tau_\lambda)$ and correct the full SED, we first calculated the total attenuation at 9.7\mum, $\Theta(\tau_{9.7})$, using the silicate absorption feature observed at that wavelength. To determine the unabsorbed flux at 9.7\mum, we fit a first-order spline to the observed continuum at rest wavelengths 6.8~\mum\ and 30~\mum, which shows minimum obscuration (see Fig.\,\ref{fig:dustOpacity}). The total attenuation at 9.7\mum, $\Theta(\tau_{9.7})$, was therefore given by the ratio between the observed to the intrinsic flux at that wavelength (${\rm F}_{9.7\mu {\rm m}}^{\rm~Obs}~/~{\rm F}_{9.7\mu {\rm m}}^{\rm~Int}$ in Eq.\,\ref{Eq:obs}). The latter was inferred from the fit to the estimated continuum. We stress that this does not provide the {\it exact amount} of full obscuration, as {the continuum} is estimated using wavelengths that are themselves slightly affected by obscuration. Despite this, we find that it provides an overall good estimate, based on the goodness of the SED fits (see \S\,\ref{subsubsec:SFfit}).

Since we have that,
\begin{equation}
\Theta(\tau_{9.7})~=~(1~-~e^{-\tau_{9.7}})/\tau_{9.7} \equiv~{\rm F}_{9.7\mu {\rm m}}^{\rm~Obs}~/~{\rm F}_{9.7\mu {\rm m}}^{\rm~Int},
\end{equation}
then the optical depth at 9.7~\mum, $\tau_{9.7}$, was calculated using interpolation on a log-grid in steps of 0.01. Once $\tau_{9.7}$ was measured, we used the aforementioned pseudo--extinction curve to calculate the obscuration at all IR wavelengths.

\subsubsection{Fit to the IR SEDs for star-forming galaxies}
\label{subsubsec:SFfit}

Our approach to building a set of typical IR templates for non-AGN-hosting galaxies is based on the work of \citeauthor{Schreiber2018} (\citeyear{Schreiber2018}, hereafter S18; and references therein). S18 provide two independent grids of IR templates, each varying with dust temperature, \Tdust, and corresponding to different emission mechanisms present in star-forming galaxies. The first grid represents large and small grain emission (hereafter dust continuum), where the amorphous carbon dust models of \cite{Galliano2011} were adopted, with the mass-fraction of thermalised carbonated to silicate grains fixed to the Milky Way value, and the mass fraction of the non-thermalised silicate grains fixed to zero (see~\S\,3.1.1 in S18).

The second grid is used to model the PAH features, and was also adopted from \cite{Galliano2011}, with the fraction of neutral-to-ionised molecules fixed to the Milky Way value (see \S\,3.1.2 in S18). The free parameter in each of these models is the radiation intensity, $U$, and composite SEDs have been built by assuming a distribution for $U$, as in \citeauthor{Dale2001} (\citeyear{Dale2001}; see \S\,3.1.3 in S18). However, for simplicity, S18 fixed the relative contributions of intensities $U$ (i.e. $\alpha_{\rm SF}~=~2.6$, where $\alpha_{\rm SF}$ is the power law index of the distribution of $U$), and was left with a single parameter (\Tdust) to control the shape of their SEDs, split in terms of PAH and dust continuum emission\footnote{The templates can be found at: \url{https://github.com/cschreib/s17-irlib}}.

Using this two-component model, S18 found that it is possible to reliably fit the stacked IR broadband SEDs (i.e. not including {\it Spitzer}--IRS spectra) of Main Sequence galaxies (i.e. typical star-forming galaxies observing a relationship between stellar mass and SFR, e.g. \citealt{Schreiber2015}), split in terms of redshifts and stellar masses, out to \z~$\sim$~5. In addition, S18 showed that this two-component model was able to fit starburst galaxy SEDs, which are galaxies experiencing a burst of star formation, and located above the MS at a given stellar mass and redshift. However, we could not use the S18 galaxy templates directly on our AGN sample because of the large number of templates, making it difficult to achieve convergence and avoid degeneracies while fitting the SEDs of AGN-hosting galaxies. To overcome this, we first used the S18 galaxy templates to fit the 110 SEDs of our pre-selected star-forming galaxy sample, and extracted a smaller, more representative set of galaxy IR templates. As we describe later in \S\,\ref{subsec:agntemp}, these templates were then used, together with a flexible model for AGN emission, to fit our sample of AGN SEDs.

The fits of S18 templates to our star-forming SEDs were performed using maximum likelihood estimation\footnote{Because our SEDs include {\it Spitzer}--IRS spectra and {\it Herschel} photometry, we weighted the spectral data so that the MIR and the FIR part of the SEDs were equally fit. The weights corresponded to the inverse of the number of points in the spectrum.} (MLE). We incorporated the possibility of including flux upper limits in the fit so that we were able to replace the {\it Herschel} flux at 160~\mum\ for galaxies at \z~$<$~0.3 with an upper-limit. We did this as at \z~$<$~0.3 the {\it Herschel}-PACS band at 160~\mum\ could be contaminated by the [C~II] emission line at 158~\mum\ which can be strong in star-forming galaxies \citep[e.g.][]{Smail2011}. Because our likelihood function includes upper limits it could not be maximised analytically. Instead, we maximised it by randomly sampling the posterior distribution of each parameter, using the affine invariant ensemble sampler of \cite{Goodman2010} fully implemented into {\sc emcee}\footnote{{\sc emcee} is publicly available at \url{https://emcee.readthedocs.io/en/v3.0.2/}} \citep{Foreman-Mackey2013}. The median values of the posterior distributions were taken as the best-fitting parameters, and the standard deviations as their 1$\sigma$ uncertainties. 

We fit our sample of 110 IR SEDs of star-forming galaxies, including {\it Spitzer}--IRS spectra, using the combination of the dust grain continuum and the PAH emission templates of S18, independently adjusting the normalisation and the dust temperature of each of the two components. Prior to the fits, we affected the dust continuum templates by obscuration following the prescriptions presented in \S\,\ref{subsubsec:obsCorr}. In fact, better fit were obtained when the PAH templates were left free of obscuration. This is in agreement with recent work providing evidence that PAHs are from more diffuse and less dense regions \citep[e.g.][]{Hirashita2020}.

\begin{figure}
 \centering
 \includegraphics[width=0.5\textwidth]{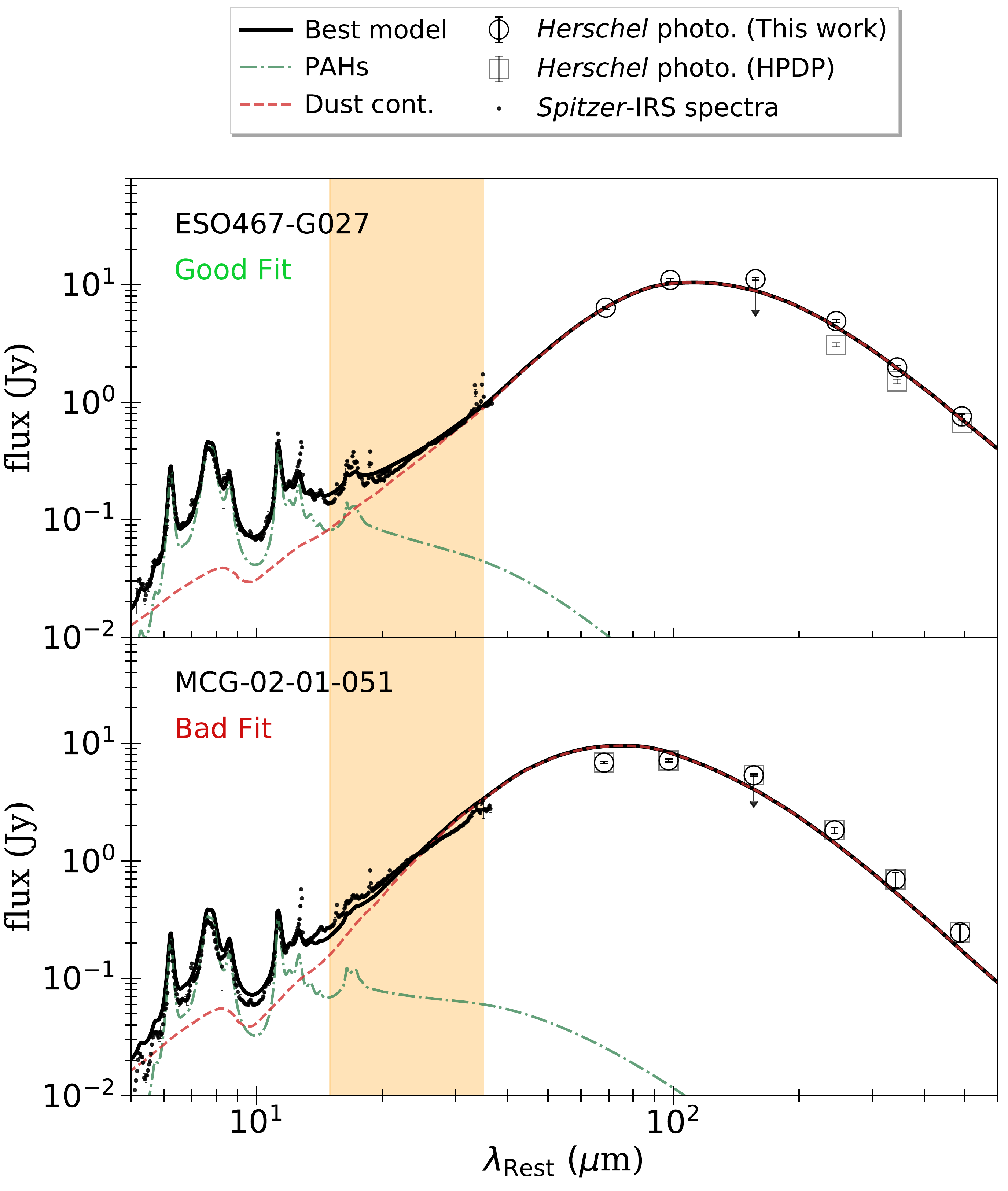}
 \caption{An example of two SED fits from our sample of star-forming galaxies. The observed {\it Spitzer}--IRS data are shown with small black dots, and the {\it Herschel} photometry, corrected for spatially extended emission, with large open circles. We also show with large open squares the {\it Herschel} photometry found in the HPDPs, which failed to identify the extended nature of the galaxy presented in the top panel (see \S\,\ref{subsubsec:dataFIRphoto}). The arrow on the flux at 160~\mum\ indicates that the flux has been included in the fit as an upper limit (see \S\,\ref{subsubsec:SFfit}). In each panel, the black line shows the best fit, decomposed into dust continuum (dashed red line) and PAH emission (dot-dashed green line). {The model is affected by obscuration as explained in the text.} The top panel shows an example of a good fit to the data using the model of S18, as found for 50~per~cent of our sample, and the bottom panel shows an example of a poorer fit to the data, as found for the rest of the sample. The name of the two sources are indicated in the top-left hand corner of each panel. The orange area highlights the differences in the observed SEDs between the two sources, where the slope of the MIR continuum and the relative level of MIR emission differ leading to a poorer fit for MCG-02-01-051. For the latter, it was found that the MIR emission is contaminated by AGN. \label{fig:SFfit}}
\end{figure}

In the top panel of Fig.\,\ref{fig:SFfit} we show an example of a best SED fit, separated between the dust continuum (dashed red line) and the PAH emissions (dot-dashed green line). However, we found that the model of S18 was able to satisfactorily fit 50~per~cent of our sample of pre-selected star-forming galaxies. For the rest of the sample, significant discrepancies between the observed SEDs and the models were noted, as shown for MCG-02-01-051 in the bottom panel of Fig.\,\ref{fig:SFfit}. In particular, the majority of the failed fits were showing a MIR excess and a shallower MIR slope, which could be due to mild AGN contamination (see the filled orange area in Fig\,\ref{fig:SFfit}).

For the 50~per~cent of sources that could not be fit using the model of S18, we inspected their SEDs and searched the {\sc NED} database to find potential evidence for misclassification. The full analysis is presented in Appendix \ref{app:AGNinter} (see also Table\,\ref{table:preSF}). Briefly, in addition to finding sources that had intrinsic issues with their SEDs (e.g. FIR fluxes contaminated by companion galaxies), the vast majority (i.e. $\sim$60~per~cent of the badly fit sources) showed AGN contribution to the MIR. The AGN contributions to the total IR luminosities (i.e. integrated from 8~\mum\ to 1000~\mum) were found to be typically below $\sim$10--20~per~cent, as revealed via SED fitting \citep{Magdis2013, Vika2017}, or by combining multiple MIR diagnostics (\citealt{Diaz2017}; see Appendix \ref{app:AGNinter} for more details). We note that these were identified mostly via SED fits, which in turns depend on the models used to fit the SEDs. However, by looking at the individual SEDs, we noted the presence of multiple faint high-ionisation emission lines, such as [NeV] at 24~\mum\ and [OIV] at 26~\mum, in half of the galaxies that were poorly fit by the model of S18. This strongly suggests the presence of an AGN. Nevertheless, we cannot exclude the possibility that, at least for some of these SEDs, the reported AGN contributions result from fits based on photometric measurements affected by systematics.

We further stress that we preferred not to directly pre-select our star-forming sources in \S\,\ref{subsubsec:galSample} using these results as (1) the AGN fractions were often low and consistent with zero once the uncertainties accounted for, (2) we found some large discrepancies between the AGN fractions reported for the same source across studies (i.e. sometimes up to $\sim$60~per~cent differences), and (3) we found some SEDs with ambiguous AGN fractions reported in previous work that could be fit with the model of S18, and that would have been otherwise discarded. Instead, we preferred to keep our pre-selection based on the \eqwspd\ and the WISE colours, and discard the sources that could not be fit using the model of S18 after finding that some potential AGN contamination was reported in previous work.

The large number of AGN interlopers, mainly revealed by attempting to fit models of pure galaxy emission, highlights the challenges in detecting faint AGN IR emission. This further supports the necessity to carefully decompose SEDs into AGN and host contribution. These sources were discarded from our sample of pure star-forming galaxies, so the full sample now contains 55 sources for which we have detailed fits of their SEDs\footnote{The full set of 55 SEDs can be found in the supplementary material, or directly at \url{https://tinyurl.com/ybu9jegl}.}. The results of the fits using the model of S18 are presented in Table\,\ref{table:fitgalParam}. We stress that we demonstrate in \S\,\ref{subsubsec:galTemplComp} that the galaxy templates extracted from this sample covers a wide range of observed SEDs for typical star-forming galaxies in the local Universe.

\subsubsection{Templates for star-forming galaxies}
\label{subsubsec:Templ}

\begin{figure*}
 \centering
 \includegraphics[width=\textwidth]{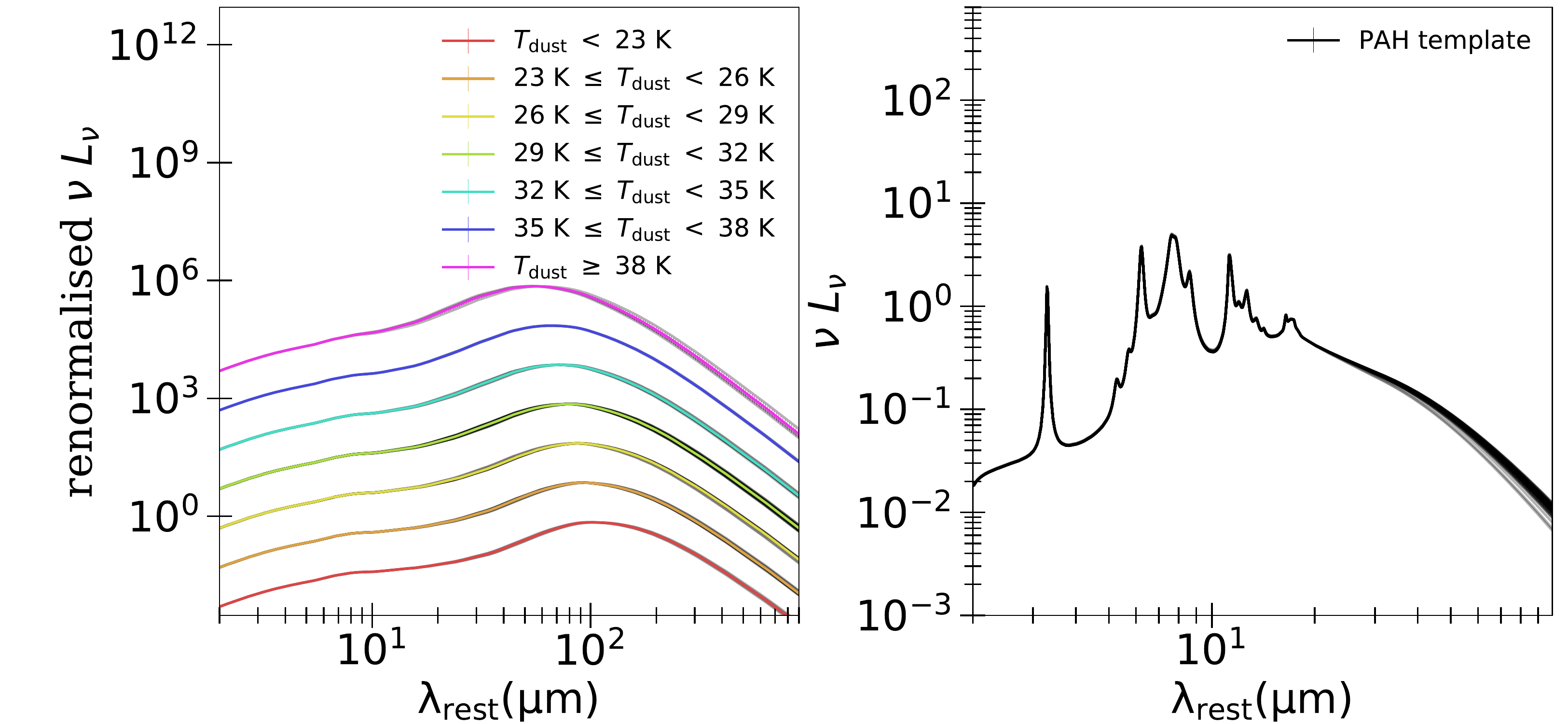}
 \caption{The eight IR SED templates for galaxies split in terms of dust continuum (left-hand side panel) and PAH emission (right-hand side panel). For the dust continuum, we split in terms of \Tdust\ (see keys). For the PAHs we adopted a generic PAH template by combining all the best fits together. The thin black lines show the individual contributions of each of our galaxies. \label{fig:B18templ_split}}
\end{figure*}

We used the aforementioned 55 SED fits to build a set of templates for star-forming galaxies, based on the model of S18. To do this, we considered the PAH and the dust emission independently. For the PAH emission, we found that, once normalised to the IR luminosity (i.e. integrated from 8~\mum\ to 1000~\mum; \Lir), a universal PAH template worked for all galaxies, consistent with some previous work \citep[e.g.][]{Xie2018}. We constructed our typical PAH emission template by averaging over the \Lir\ normalised PAH component of our best fits. We defined 1$\sigma$ uncertainties on the PAH template by measuring the standard deviation arising from combining multiple best fits (see the right-hand side panel in Fig.\,\ref{fig:B18templ_split} for the PAH template).

To construct the dust continuum templates we combined together SED models in bins of \Tdust, as \Tdust\ defines the shapes of the dust continuum components in S18. We found for our best fits that 20~K~$<$~\Tdust~$<$~45~K, with a mean at \Tdust~$\sim$~33~K. This range of \Tdust\ is consistent with that observed for general populations of galaxies, and the average \Tdust\ is consistent with that measured in starburst populations \citep[e.g.][]{Orellana2017}. We defined seven bins of \Tdust\ which constitute our seven dust continuum templates, and are representative of our sample of star-forming galaxies (see the left-hand side panel in Fig.\,\ref{fig:B18templ_split} for the bins of \Tdust). As for our PAH template, we constructed our dust continuum templates by averaging over the normalised dust components of our best fits, once split in terms of \Tdust, and calculated 1$\sigma$ uncertainties. Therefore, we reduced the full set of templates from S18 to a set of eight templates, of which one represents the PAH emission and seven represent the dust continuum at various \Tdust\ (see Fig.\,\ref{fig:B18templ_split}).

\begin{figure}
 \centering
 \includegraphics[width=0.5\textwidth]{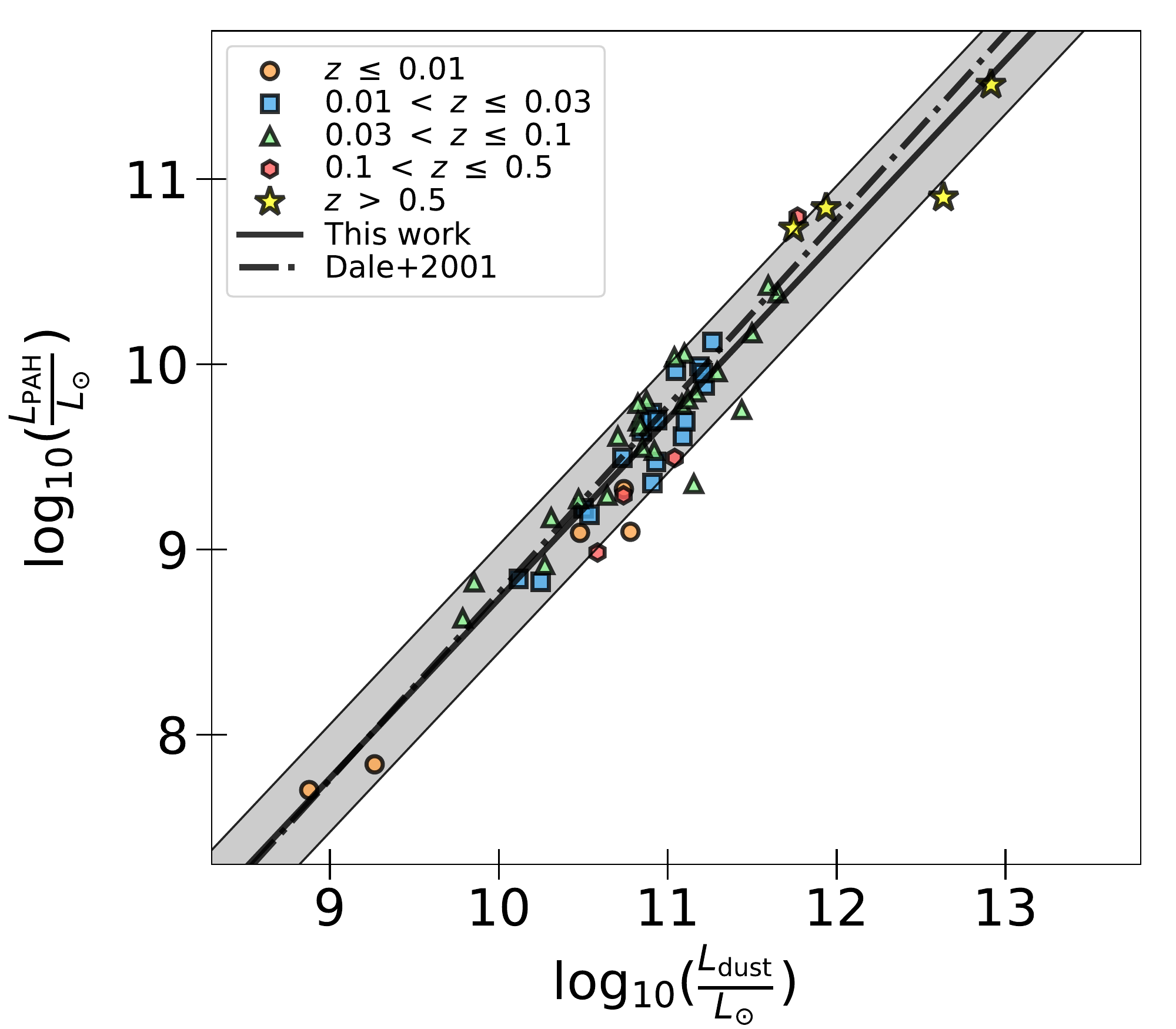}
 \caption{The relationship between the luminosity of the dust continuum and of the PAH observed while – independently – fit to our sample of star-forming galaxies. We split in terms of redshift bins (see keys) to verify that this relationship was not a consequence of the Malmquist bias. The black line shows the average best fit represented by Eq.\,\ref{eq:PAHcorrDust}, and the greyed area shows the scatter of 0.3~dex. The dot-dashed black line shows the relationship between the PAH and the dust expected in \protect\cite{Dale2001}. \label{fig:PAHcorrDust}}
\end{figure}

A major challenge in interpreting the emission from an AGN-hosting galaxy is being able to determine the balance between the dust heating mechanisms (i.e. AGN or stars). This often becomes heavily degenerate, in particular at wavelengths $\lambda~>~70$~\mum\ where this balance is determined using SEDs that are sampled by only few photometric points. The fact that, so far, our templates are built using fully independent dust continuum and PAH emission templates can increase the degeneracies while fitting AGN SEDs. We show in Fig.\,\ref{fig:PAHcorrDust} that there is a correlation between the IR luminosity of the dust continuum (\Ldust) and that of the PAHs (\Lpah) when considering both these components independently. We tried splitting in terms of redshifts to test whether this is a consequence of the Malmquist bias, and found that this relationship holds, at least at \z~$\lesssim$~1 where our sample lies. In addition, we note that there is no evidence of a dependence with the total luminosity (i.e. \Lir~$\equiv$~\Lpah~+~\Ldust), and our relationship appears to be valid from \Lir~$\sim~10^{9}$~\Lsun\ to \Lir~$\sim~10^{13}$~\Lsun. 

In light of the this, we defined the relationship between \Lpah\ and \Ldust\ expressed as:

\begin{equation}
\log_{10}(\frac{L_{\rm PAH}}{L_\odot}) \equiv~0.97~\times\ \log_{10}(\frac{L_{\rm dust}}{L_\odot})~-~0.95.
\label{eq:PAHcorrDust}
\end{equation}

\begin{figure}
 \centering
 \includegraphics[width=0.5\textwidth]{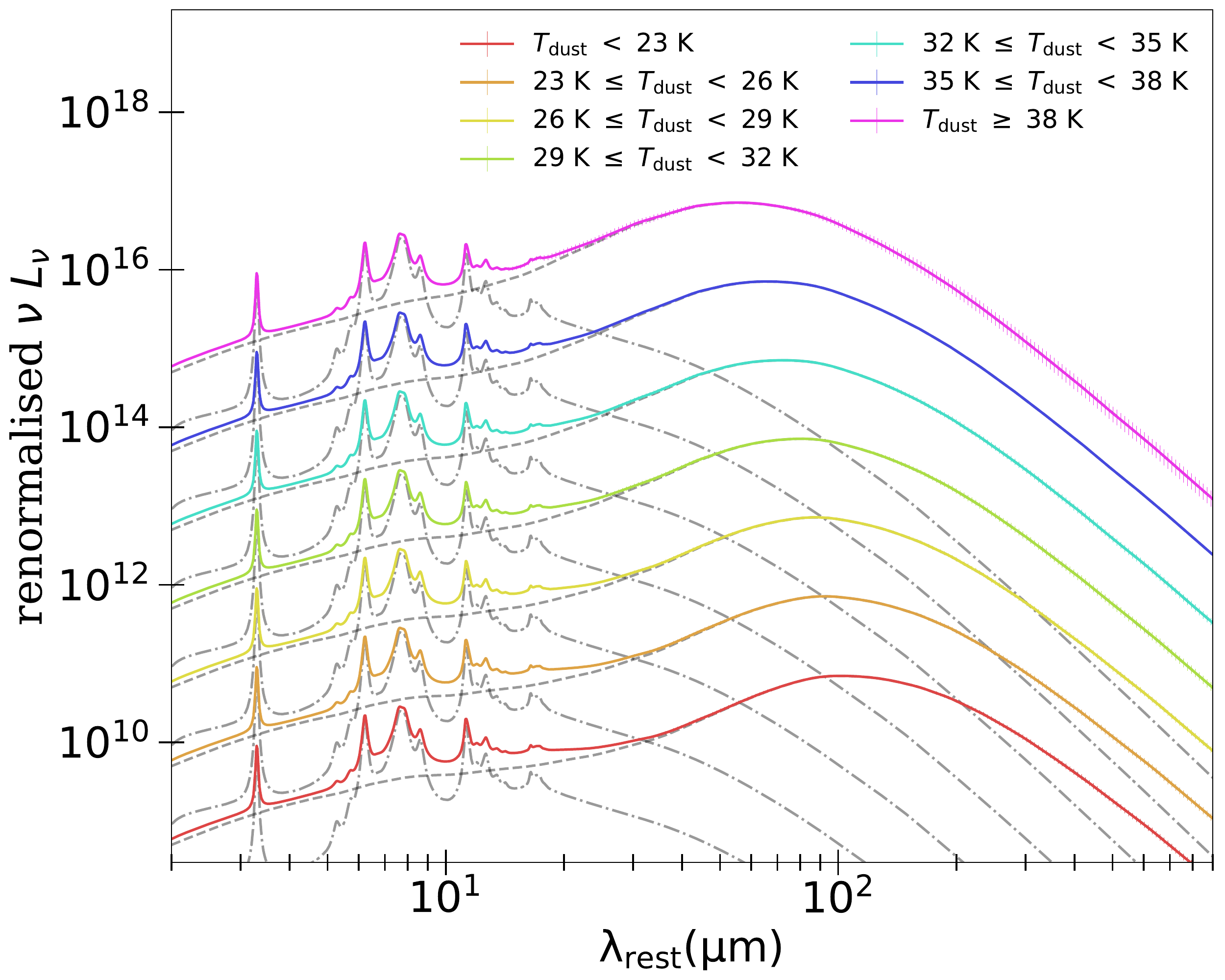}
 \caption{Our set of seven templates for galaxies where Eq.\,\ref{eq:PAHcorrDust} was used to calculate the normalisation of the PAHs assuming \Ldust$~=~10^{11}$ \Lsun. The different colours correspond to different dust continuum templates (see keys). The dashed and dot-dashed lines correspond to the dust continuum and PAH emission components, respectively. \label{fig:B18templ}}
\end{figure}

\noindent This relationship is shown in Fig.\,\ref{fig:PAHcorrDust}, where we also show that it is fully consistent with that reported in \cite{Dale2001}. In addition, to allow our PAH and dust continuum templates to combine with more extreme relative luminosities, we assumed a normally distributed scatter of $\sigma~=~$0.3~dex on Eq.\,\ref{eq:PAHcorrDust}, as observed in our sample of star-forming galaxies, and shown in Fig.\,\ref{fig:PAHcorrDust}. The relationship defined in Eq.\,\ref{eq:PAHcorrDust} and its scatter will be used while fitting the SEDs of our AGN sample (see \S\,\ref{subsec:agntemp}). We show in Fig.\,\ref{fig:B18templ} the seven templates for star-forming galaxies after using Eq.\,\ref{eq:PAHcorrDust}, assuming \Ldust$~=~10^{11}~$\Lsun, to connect the dust continuum to the PAH emission.

To summarise, our full set of galaxy templates contains eight individual templates spanning the wavelength range 1~\mum$~<~\lambda~<$~1000~\mum, of which seven represent the dust continuum at different \Tdust, and one represents a generic PAH emission. Eq.\,\ref{eq:PAHcorrDust} can be used to constrain the dust-to-PAH luminosity ratio, allowing for a scatter of 0.3~dex. We re-fit our sample of 55 star-forming galaxies with this set of templates and find good fits to all of these galaxies’ SEDs.

\subsubsection{Comparison between our non-AGN galaxy templates to previous work}
\label{subsubsec:galTemplComp}

\begin{figure}
 \centering
 \includegraphics[width=0.5\textwidth]{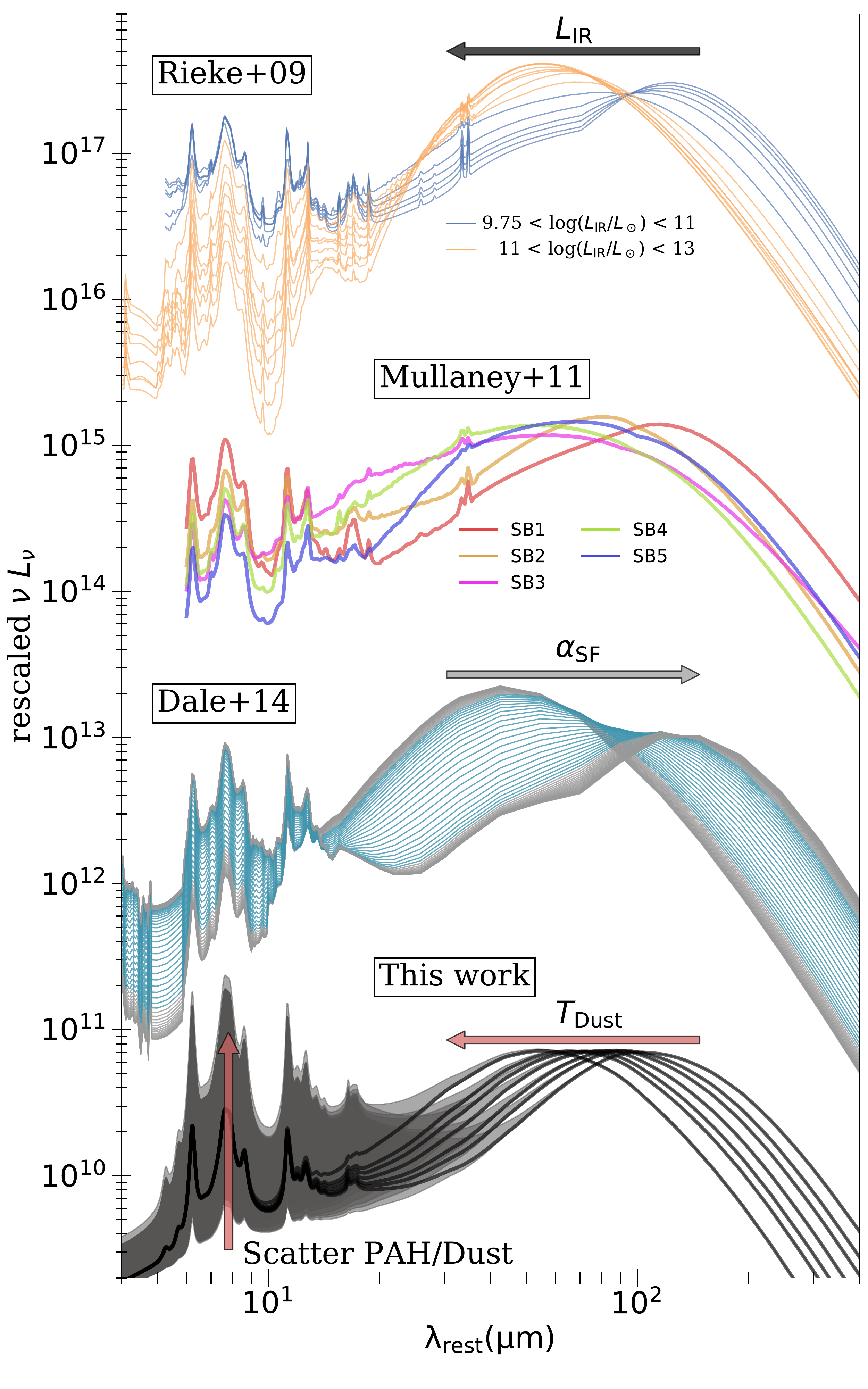}
 \caption{Comparison between our newly derived set of seven galaxy templates and several widely used libraries of templates for star-forming galaxies. {\it From top to bottom}: the set of templates of \protect\cite{Rieke2009} for LIRGs and ULIRGs shown with blue and orange lines, respectively, and where \Lir\ controls the shape of the SEDs in the range \Lir\,=~$10^{9.75\--13}~$\Lsun\ (see the black arrow at the top of the figure); the set of templates of \protect\cite{Mullaney2011}, colour coded by their individual names (see corresponding keys); the set of templates of \protect\cite{Dale2014}, where $\alpha_{\rm SF}$ controls the shape of the SEDs between 0.0625~$<~\alpha_{\rm SF}~<$~4 (see the grey arrow), and where the templates representing the majority of the star-forming galaxies are highlighted in blue; and our set of templates controlled by \Tdust\ (see the horizontal red arrow at the bottom of the figure), assuming \Ldust$~=~10^{11}$~\Lsun, and where Eq.\,\ref{eq:PAHcorrDust} was used to calculate the relative amount of PAHs, and the grey area shows the 3$\sigma$ scatter (see the vertical red arrow at the bottom left-hand side of the figure). \label{fig:SFtemplComp}}
\end{figure}

\begin{figure}
 \centering
 \includegraphics[width=0.5\textwidth]{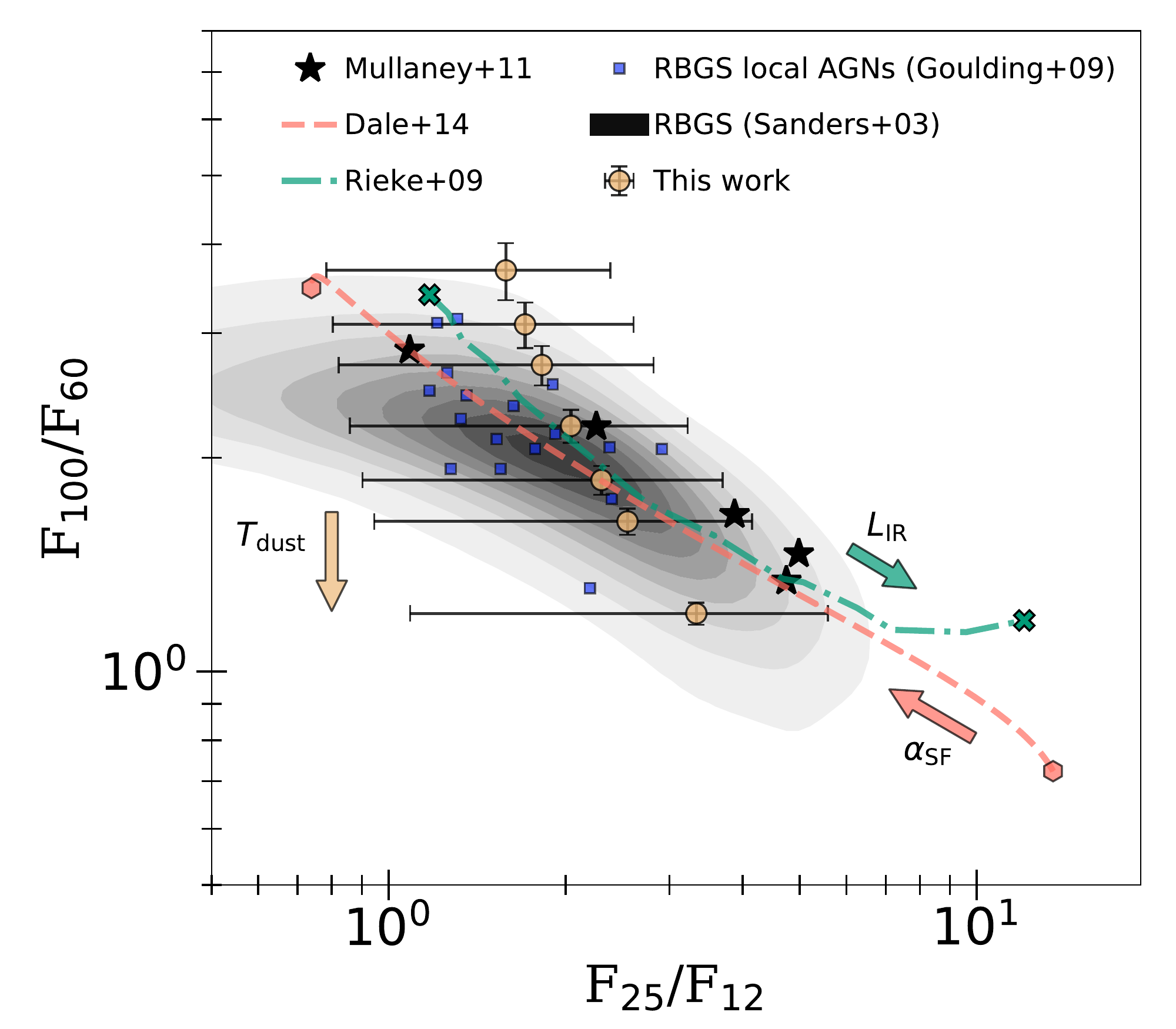}
 \caption{Plot showing the {\it IRAS} colour-colour parameter space, where the RBGS sample of \protect\cite{Sanders2003b} is shown as a kernel density estimation. The colour-colour parameter space covered by our set of average templates are shown with filled orange circles. The error bars illustrate departures from these average values once assuming the scatter on the PAH-to-dust continuum relationship. For illustrative purposes, we show the 3$\sigma$ scatter of 0.9~dex. The black filled stars show that of \protect\cite{Mullaney2011}, the dot-dashed green line, terminated by green filled crosses, shows that of \protect\cite{Rieke2009}, for \Lir\,=~$10^{9.75\--13}~$\Lsun, and the dashed pink line, terminated by pink filled hexagons, shows that of \protect\cite{Dale2014}, with 0.0625~$<~\alpha_{\rm SF}~<$~4. The arrows indicate the direction of change in the {\it IRAS} colours when the parameter controlling the shape of each of the corresponding set of templates is changed. The small filled blue squares show the {\it IRAS} colours of galaxies hosting AGNs in the complete volume limited sub-sample of \protect\cite{Goulding2009}.\label{fig:iras}}
\end{figure}

In this section we compare our new galaxy SED templates to some previous widely used libraries. To do this, we first compare in Fig.\,\ref{fig:SFtemplComp} the $\nu L_\nu$ shapes of our templates to those from \cite{Rieke2009}, \cite{Mullaney2011} and \citeauthor{Dale2014} (\citeyear{Dale2014}; hereafter, R09, M11, and DH14, respectively) for which different methods were used to extract the galaxy templates. We then plot in Fig.\,\ref{fig:iras} the {\it IRAS} F$_{100}$/F$_{60}$, F$_{25}$/F$_{12}$ colour-colour space for each set of templates, where F$_{12}$, F$_{25}$, F$_{60}$, and F$_{100}$ are the {\it IRAS} fluxes at 12~\mum, 25~\mum, 60~\mum, and 100~\mum, respectively, and compare against the Revised Bright Galaxy Sample (RBGS) of \cite{Sanders2003b}. We do this as the {\it IRAS} colours provide a simple way to quantify the mid-to-far-IR differences observed in Fig.\,\ref{fig:SFtemplComp}.

In R09, 11 luminous and ultra-luminous IR galaxies (LIRGs and ULIRGs, respectively) with MIR spectra (i.e. from {\it Spitzer} or {\it ISO}), mid-to-far-IR photometry (i.e. from {\it Spitzer}-IRAC and {\it Spitzer}-MIPS out to 70~\mum), and radio 1.4~Ghz were used to construct 11 individual templates (see R09 for details). We note that R09 had to rely on a modified black-body curve to extrapolate the FIR, as {\it Herschel} was not available at the time. Using this set of 11 templates, and aided by previous IR SED templates (see R09 and references therein), R09 built a new set of average templates for (U)LIRGs, defined as having \Lir~=~10$^{9.75\--13}$~\Lsun, and shown at the top of Fig.\,\ref{fig:SFtemplComp}, where LIRGs and ULIRGs are separated by different colours as derived using different prescriptions (see R09 for more details). 

To derive a set of five templates for the IR emission of galaxies, M11 used a sub-sample of 10 star-forming galaxies taken from \cite{Brandl2006} which, in turn, were extracted from the RBGS, plus four other ``cold'' galaxies, aiming to sample the full {\it IRAS} colour-colour space, and with \Lir\ ranging from $\sim$2~$\times$~10$^{10}$~\Lsun\ to $\sim$2~$\times$~10$^{11}$~\Lsun. {Similarly to R09, M11 used a modified black-body curve to extrapolate the FIR beyond the 100~\mum\ {\it IRAS} photometry}. These are shown as second from the top in Fig.\,\ref{fig:SFtemplComp}.

The DH14 IR templates are originally from \cite{Dale2001} and \cite{Dale2002}. In DH14 they updated the MIR to the star-forming spectra of \cite{Spoon2007}. The templates have been built using a series of local SEDs from dust exposed to a wide range of heating intensities, and calibrated on a sample of 69 star-forming galaxies. The parameter which controls the shape of the SED is the power-law index, $\alpha_{\rm SF}$, defining the relative contribution of each of the heating components (i.e. the hardness of the field), which is also proportional to ${\rm F_{60}/F_{100}}$. In essence, the method used to derive this set of templates is the closest to that of used here, since models of galaxy emission have been used. The full set of templates from DH14 is shown as third from the top in Fig.\,\ref{fig:SFtemplComp}.

At the bottom of Fig.\,\ref{fig:SFtemplComp}, we show our set of seven average IR SED galaxy templates, for which we used Eq.\,\ref{eq:PAHcorrDust} to determine the relative amount of PAHs assuming \Ldust~=~10$^{11}$~\Lsun. In addition, we added the grey areas which show the effect of the scatter on Eq.\,\ref{eq:PAHcorrDust}. To show the extended effect that the scatter has on the shape of our templates, Fig.\,\ref{fig:SFtemplComp} shows a 3$\sigma$ scatter of 0.9~dex. By design, our average templates correspond to different \Tdust, from \Tdust~$\sim$~20~K to \Tdust~$\sim$~40~K (see \S\,\ref{subsubsec:Templ}).

We find some differences between the various sets of templates, particularly when R09 and M11 are compared to DH14 and the templates defined in this work. Perhaps one of the most striking differences is at mid-to-far-IR wavelengths (i.e. from 20~\mum\ to 70~\mum), where the templates from R09 at lower \Lir, and some of the M11 templates (e.g. ``SB3''), have significantly flatter $\nu L_{\nu}$. In fact, some peculiar galaxies (e.g. strongly starbursting galaxies or Wolf-Rayet galaxies) are known to show such flatter MIR $\nu L_{\nu}$. By considering the scatter on the PAH-to-dust continuum ratio, some of our templates show a flatter $\nu L_\nu$ (see Fig.\,\ref{fig:SFtemplComp}). One feature of our set of templates is that, at a given \Tdust, they offer a continuous range of MIR slopes (to within the PAH-to-dust continuum scatter), in contrast with previous templates that offered a unique MIR slope once the free parameter fixed.

By comparing our set of templates to the higher luminosity ones of R09, we note that the latter are strongly affected by dust obscuration (i.e. large absorption feature are 9.7~\mum; see Fig.\,\ref{fig:SFtemplComp}). This is a direct consequence of R09 using observed spectra, including highly obscured galaxies such as Arp220, which have not been corrected for obscuration. The effect of obscuration is also seen in some of the templates from M11 (e.g. ``SB5''). In contrast, we have considered obscuration separately, such that, SEDs that are strongly affected by obscuration can also be reproduced by our set of templates after applying our prescriptions (see \S\,\ref{subsubsec:obsCorr}). Again, this has the advantage of offering a much larger range of SED shapes compared to R09 or M11.

We find that our set of average templates is broadly consistent with that of DH14, although the latter appears to offer a wider range of FIR dust emission. However, we stress that the majority of normal star-forming galaxies have 1~$<~\alpha_{\rm SF}~<$~2.5 \citep{Dale2002}, corresponding to the blue coloured DH14 templates in Fig.\,\ref{fig:SFtemplComp}, and in better agreement with our set of average templates. Another difference is that in DH14 the mid-to-far-IR balance changes with $\alpha_{\rm SF}$, as some suppression of PAHs in harsher environments has been implemented. However, our set of templates offer a wider, more continuous range of mid-to-far-IR balance via the scatter on the PAH-to-dust continuum relationships.

In Fig.\,\ref{fig:iras} we show the locations of the different sets of galaxy templates on the {\it IRAS} colour-colour parameter space, overlaid on the full RBGS sample of \cite{Sanders2003b}, the latter shown as a kernel density estimation (grey contours in Fig.\,\ref{fig:iras}). For each set of templates, we varied the parameter controlling the SED shapes, when necessary, and calculated the corresponding synthetic flux in each of the {\it IRAS} filters. For our set of templates, we assumed \Ldust~=~$10^{11}$~\Lsun, and calculated the corresponding amount of PAHs using Eq.\,\ref{eq:PAHcorrDust}, allowing for a 3$\sigma$ scatter of 0.9~dex, as shown by the error bars in Fig.\,\ref{fig:iras}.

We note that, by design, the M11 templates are representative of the RBGS sample since they were selected to cover most of this parameter space. However, since M11 only considered a small set of five fixed galaxy templates, their SEDs can only accommodate for a small number of shapes. In contrast, while our average templates cover a smaller fraction of the {\it IRAS} colour-colour parameter space when compared to M11, the scatter in the PAH-to-dust continuum ratio allows our templates to take a large variety of shapes, and to represent most of the RBGS sample.

When compared to the galaxy templates of R09, we also find that our templates cover a larger fraction of the {\it IRAS} colour-colour space, due to the flexibility in terms of SED shapes. We note that the highest \Lir\ templates of R09 cover a wider range of F$_{25}$/F$_{12}$ colours, which are not covered by any of our galaxy templates, even after accounting for the scatter in the relationship between the PAH and the dust continuum (see Fig.\,\ref{fig:iras}). However, we stress that the highest \Lir\ templates in R09 are generally affected by obscuration (see Fig.\,\ref{fig:SFtemplComp}). This results in an artificially depleted flux at 12~\mum, relative to the other {\it IRAS} wavelengths that are less affected by obscuration. Our templates are able to reach such extreme MIR {\it IRAS} flux ratios once our prescriptions for obscuration is applied (see \S\,\ref{subsubsec:obsCorr} for the obscuration prescriptions).

Our set of templates also covers a larger part of the RBGS colour-colour space than the DH14 templates. We note in Fig.\,\ref{fig:iras} that for $\alpha_{\rm SF}\sim0.1$, the colour-colour ``sequence'' of the DH14 templates extend to the bottom right-hand corner, which is not covered by our templates. However, we stress that model SEDs with $\alpha_{\rm SF}<1$ in DH14 are atypical, since the majority of normal star-forming galaxies have 1$~<~\alpha_{\rm SF}<$~2.5 \citep{Dale2002}, which is consistent with the coverage of the {\it IRAS} colour-colour space probed by our galaxy templates.

Finally, we note that the {\it IRAS} colours of our average galaxy templates, taken as an ensemble, seem to deviate from the bulk of the RBGS sample of galaxies, while the other sets of templates closely follow the central density (see Fig\,\ref{fig:iras}). In fact, the RBGS sample also contains AGNs, while our sample of star-forming galaxies have been carefully built by discarding potential IR AGNs via a series of stringent selections (see \S\,\ref{subsubsec:galSample} and Appendix\,\ref{app:AGNinter} for the selection of our sample of star-forming galaxies). In Fig.\,\ref{fig:iras} we also show results from \cite{Goulding2009}, where a volume-limited (D~$<$~15~Mpc) complete sub-sample of galaxies (regardless of whether they host an AGN or not) was selected from the RBGS, and used to investigate the ubiquity of AGNs in star-forming galaxies. To find AGNs, \cite{Goulding2009} used the detection of the high-ionisation emission line [NeV]. The location of these AGNs is shown with filled blue squares in Fig.\,\ref{fig:iras}. It appears that the selections applied to define our sample of star-forming galaxies have naturally avoided {\it IRAS} colours consistent with the bulk of these local AGNs found in the RBGS sample by \cite{Goulding2009}. However, we stress that local AGN-free galaxies in \cite{Goulding2009} partly occupy the same locus as the local AGNs on the {\it IRAS} colour plot (not shown in Fig.\,\ref{fig:iras} for clarity). We further emphasise that our templates are able to represent these galaxies via the inclusion of the scatter on the PAH-to-dust continuum ratio.

\subsection{IR templates for AGN emission}
\label{subsec:agntemp}

In \S\,\ref{subsec:hostTempl} we defined a set of templates for (non-AGN) star-forming galaxy IR SEDs based on the results of L19 and S18. In this section, we explain how we used this set of templates to fit the IR SEDs of our sample of local AGNs defined in \S\,\ref{subsec:obsSED} (see also Table\,\ref{table:AGNprop}). The AGN emission was accounted-for with a flexible IR model (see \S\,\ref{subsubsec:model}), which we combined with our galaxy templates to construct an SED representing the whole system (i.e. galaxy $+$ AGN; see \S\,\ref{subsubsec:AGNfit}). We then used the results from the fits to build three generic templates which are representative of AGN emission at IR wavelengths (see \S\,\ref{subsubsec:AGNtempl}).

\subsubsection{A flexible model for AGN IR emission}
\label{subsubsec:model}

To account for the possibility of AGN emission in our observed IR SEDs, we defined a model which is flexible enough to fit a diverse range of SED shapes. In fact, previous studies have found a wide variety of SED shapes to represent the IR emission of AGNs, depending on whether their focus was on bright QSOs \citep[e.g.][]{Richards2006, Netzer2007, Schweitzer2008, Lani2017, Symeonidis2016, Lyu2017b, Lyu2018}, or more typical Seyfert galaxies \citep[e.g.][]{Mullaney2011, Mor2012}.

\paragraph{AGN IR continuum}
The AGN IR continuum can have a complex shape as a result of the various dusty structures on different scales that might be illuminated by the AGN. In particular, dust close to the centre (e.g. the inner part of the dusty torus) will emit at shorter IR wavelengths, while dust located farther away from the centre (e.g. in the narrow line region, or even at galactic scales) will emit at longer IR wavelengths \citep[e.g.][]{Schweitzer2008, Honig2013, Lyu2018}. In fact, the connection between these emitting regions is far from clear. Therefore, we defined a model to represent the AGN continuum which can easily account for such complexity. We adopted three smoothly-connected broken power-laws. {\it Each} of these is expressed as,

\begin{equation}
B_{\nu} = \lambda^{\alpha_{\rm i}}~\times~\left(~1~+~\left(\frac{\lambda}{\lambda_{\rm break}}\right)^{|\alpha_{\rm j}~-~\alpha_{\rm i}|~\times~s} \right)^{\frac{\rm sgn(\alpha_{\rm j}~-~\alpha_{\rm i})}{s}},
\label{eq:PL}
\end{equation}

\noindent where $B_{\nu}$ is the flux density, $\alpha_{\rm i}$ and $\alpha_{\rm j}$ are the slopes before and after the break, respectively, $\lambda_{\rm break}$ is the position of the break, $s$ is a smoothing factor, and sgn($\alpha_{\rm j}~-~\alpha_{\rm i}$) is the sign of the difference between the slopes after and before the break, respectively.

The positions of the breaks were fixed to 11~\mum\ and 18~\mum\ for the first and the second broken power-laws, respectively, and free to change at $\lambda~>~30~$\mum\ for the third broken power-law. To ensure continuity, the slopes were defined such that, if we call the slopes of the first broken power-law $\alpha_{\rm i}~=~\alpha_1$ and $\alpha_{\rm j}~=~\alpha_2$ in Eq.\,\ref{eq:PL}, then $\alpha_{\rm i}~=~\alpha_2$ and $\alpha_{\rm j}~=~\alpha_3$, and $\alpha_{\rm i}~=~\alpha_3$ and $\alpha_{\rm j}~=~\alpha_4$ in Eq.\,\ref{eq:PL}, for the second and the third broken power-laws, respectively.

By defining the positions of the breaks at 11~\mum, 18~\mum, and $\lambda~>~30$~\mum, for the first, the second and the third broken power-laws, respectively, we allow our model for the AGN continuum to adopt different relative contributions of the hot, warm and cold dust. The choice of fixing the positions of the breaks to wavelengths corresponding to the peak of the silicate emission (see below for the model of the silicate emission) was purely empirical, yet allowed our model for the continuum to have extra flexibility at these particular wavelengths, and avoid over-estimating the strength of the silicate emission. More importantly, the break at 18~\mum\ for the second broken power-law was to account for the ``IR bump'' observed in some AGNs \cite[e.g.][]{Deo2009}, and now believed to be related to some extended AGN-heated dust emission \cite[e.g.][]{Honig2013, Lyu2018}. The latter remains poorly explored, and should not be confused with the silicate emission peak at 18~\mum\ (e.g. \citealt{Deo2009, Prieto2010}). We note that shifting the positions of the fixed breaks by a few microns do not have any impacts on our results. Finally, $\lambda_{\rm break}$ for the third broken power-law was free to change at $\lambda~>~30~$\mum\ as there is no consensus on the universality of this value \citep[e.g.][]{Richards2006, Mullaney2011, Lani2017, Symeonidis2016}.

The slopes $\alpha_1$, $\alpha_2$ and $\alpha_3$ were free to change, and we fixed the last slope to $\alpha_4$~=~-3.5. The latter was to mimic the Rayleigh-Jeans tail of a modified black body with $\beta~=~1.5$. We do this to reduce degeneracies, as the flexibility of our AGN continuum model is such that over-fitting could arise at longer wavelengths if no PAH emission were available to better constrain the dust continuum. The value of $\alpha_4~=~-3.5$ was adopted as this is the value that is often empirically assumed in studies characterising the IR emission of AGNs \citep[e.g.][]{Mullaney2011, Symeonidis2016}. This is also consistent with the slope at longer wavelengths found by radiative transfer applied to dusty torus in AGNs \citep[e.g.][]{Siebenmorgen2005}.

The full model for AGN IR continuum was defined by connecting these three broken power-laws, at 14.5~\mum\ for the first and the second broken power-law, and at 24~\mum\ for the second and the third broken power-laws. These were chosen to fall between the breaks (i.e. in the linear regime of each of the broken power-laws) to ensure a smooth connection. The full model for AGN continuum was then normalised at 14.5~\mum. The overall normalisation, the slopes $\alpha_1$, $\alpha_2$, $\alpha_3$, and the position of the break for the last broken power-law are the free parameters. We show in Fig.\,\ref{fig:AGNfitEx} with dot-dashed blue lines a few examples of the various shapes that our model for the AGN IR continuum can take.

\paragraph{AGN silicate emission}
In the case of some AGNs, two main MIR silicate emission features have been observed, one around $\sim$~10--11~\mum, and another around $\sim$~17--18~\mum, and there is no consensus to the exact wavelength at which each of these silicate emission features peak \citep[e.g.][]{Hao2005, Sturm2005, Hao2007, Schweitzer2008, Hatziminaoglou2015, MartinezParedes2020}. Hereafter, we refer to the silicate features at shorter and longer wavelengths as the 11~\mum\ and 18~\mum\ features, respectively. Although these silicate emission features have been extensively observed, their exact shapes are largely unknown, and vary depending on the models \citep[see e.g.][and references therein]{MartinezParedes2020}. In addition, the determination of the exact shape of the 18~\mum\ silicate emission feature is made more difficult by the presence of some continuum features observed at similar wavelengths \citep[e.g.][]{Deo2009, Hatziminaoglou2015}.

To account for the silicate emission features, we adopted a profile that was inspired by previous work \citep{Schweitzer2008, Hatziminaoglou2015}. Because the silicate emission at 11~\mum\ is easily noticeable, we defined the shape of the silicate emission based on that at 11~\mum, and assumed that the 18~\mum\ feature had the same shape. Based on previous work, and using the SEDs in our sample for which the {\it Spitzer}-IRS spectrum was clearly dominated by silicate emission (e.g. Ark120, ESO548-81, Mrk50, Fairall9, and some PG QSOs), we found that the silicate emission at 11~\mum\ was best reproduced when we used a broken power-law (i.e. Eq.\,\ref{eq:PL}) peaking at 11~\mum, and with $\alpha_{\rm i}~=~18$, $\alpha_{\rm j}~=~-5$, and $s$~=~0.6. The silicate emission at 18~\mum\ was reproduced by using a broken power-law with the same parameters as that at 11~\mum, but shifting the peak to 18~\mum.

In addition, in the fits, we will also allow for the peaks of the silicate emission to shift from their native values of 11~\mum\ and 18~\mum, by a maximum of $\pm$1~\mum, which is consistent with observations, though currently unexplained \citep[e.g.][]{Hatziminaoglou2015}. The same shift was applied to both the 11~\mum\ and the 18~\mum\ silicate features, as there is no reason to why they should be different. We show in Fig.\,\ref{fig:AGNfitEx} with dot-dashed green lines a few examples of shapes for the silicate emission, where the only free parameters are the normalisations of the silicate emission features, and the shift of the peaks.

\paragraph{Full AGN model}
Our full model for AGN emission was therefore defined by the sum of our AGN continuum model and our silicate emission features. So far, this model contains in total eight free parameters, which are: the overall normalisation for the continuum (defined at 14.5~\mum), the three slopes of the continuum ($\alpha_1$, $\alpha_2$, and $\alpha_3$), the position of the break at longer wavelengths ($\lambda_{\rm break}$), the two normalisations for the silicate emission, and the shift of the peaks of the silicate emission. This AGN model is referred to as our ``full AGN model'', to distinguish it from our AGN templates defined in \S\,\ref{subsubsec:AGNtempl}.

\subsubsection{Fits to the AGN SEDs}
\label{subsubsec:AGNfit}

\begin{figure}
 \centering
 \includegraphics[width=0.48\textwidth]{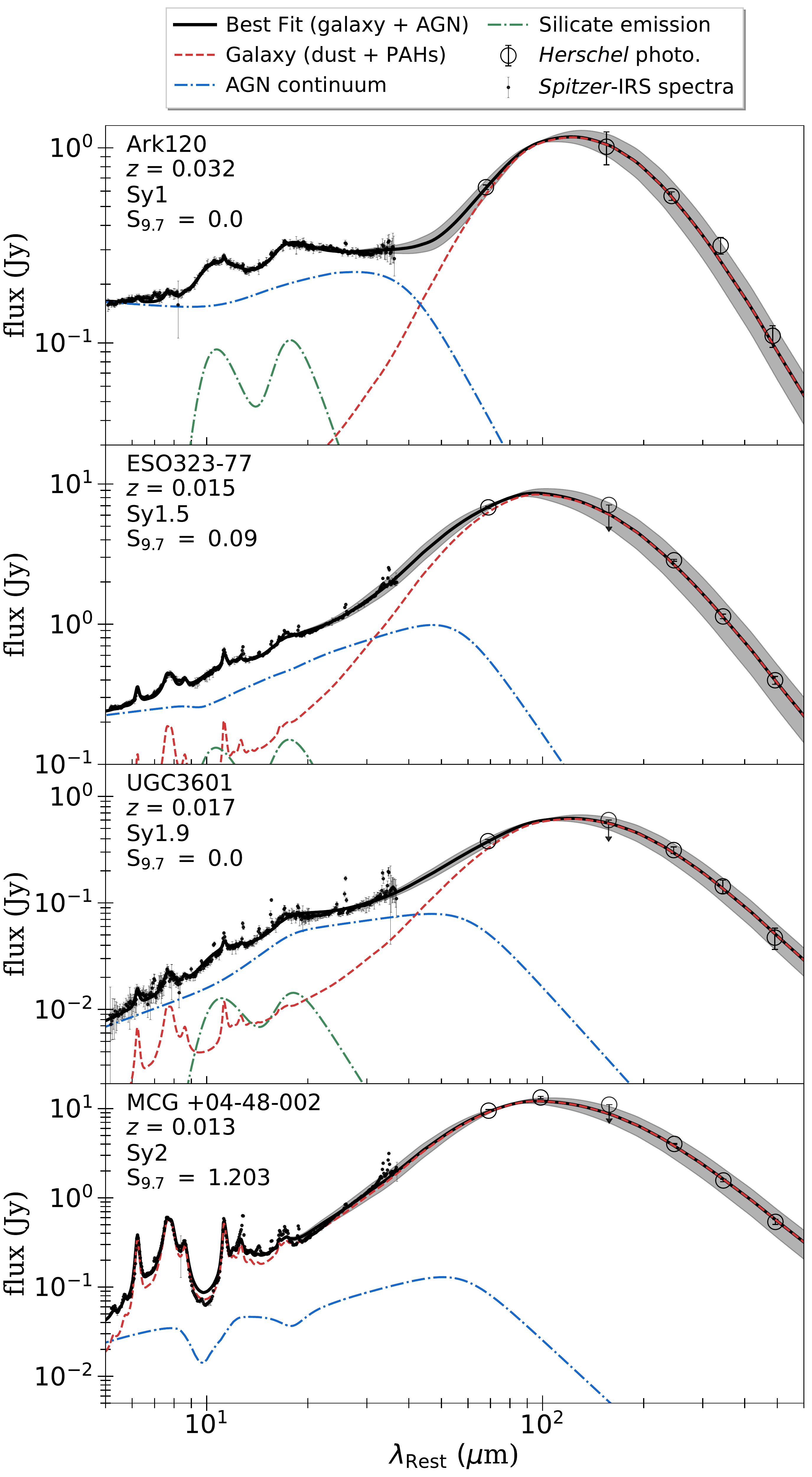}
 \caption{Examples of four best fits of AGN IR SEDs using a combination of our galaxy templates (see \S\,\ref{subsec:hostTempl}) and our full AGN model (see \S\,\ref{subsubsec:model}). The observed {\it Spitzer}--IRS data are shown with small black dots and the {\it Herschel} photometry with large open circles. The arrow on the flux at 160~\mum\ indicates that it has been set as an upper limit in the fits (not due to non-detection, but instead due to the possible boosting by [CII]; see \S\,\ref{subsubsec:SFfit}). In each panel, the black line shows the attenuated full weighted model (i.e. galaxy + AGN; see \S\,\ref{subsubsec:AGNfit}), the red dashed line shows the attenuated galaxy contribution, and the dot-dashed blue and green lines show the attenuated AGN continuum and silicate emission, respectively. The grey area shows the weighted uncertainties carried forward from the galaxy templates. The names of the sources, the redshifts, the optical types, and the absorption at 9.7~\mum\ (see Eq.\,\ref{eq:s9p7} for S$_{9.7}$) are indicated in the top-left corner of each panel. \label{fig:AGNfitEx}}
\end{figure}

Our approach to extracting the intrinsic AGN emission was to fit our AGN SEDs with each of our galaxy templates defined in \S\,\ref{subsec:hostTempl} combined with our full AGN model (see \S\,\ref{subsubsec:model}). We stress that we used a fitting approach (as opposed to simply subtracting the host contribution to extract the AGN emission), as the IR can be a complicated mixture of each of these contributions, and these can ``play off’’ against each other. So far, our whole model (i.e. galaxy $+$ AGN) contains ten free parameters. In particular, two parameters correspond to the galaxy emission (i.e. one normalisation for the galaxy templates and one for the normalisation of the PAH template, see \S\,\ref{subsec:hostTempl}), and eight correspond to our full AGN model (see \S\,\ref{subsubsec:model}).

To find the best fit parameters we employed an MLE approach of a similar design to that presented in \S\,\ref{subsec:hostTempl}. As for our galaxy sample, we replaced fluxes at 160~\mum\ with upper limits in the fits for sources at \z~$<$~0.3 (see \S\,\ref{subsubsec:SFfit}). Prior to the fits, we measured the level of obscuration present in the AGN SEDs by using the observed 9.7~\mum\ silicate absorption feature, as described in \S\,\ref{subsubsec:obsCorr}. However, the assumption of a uniformly mixed absorbing and emitting material, as used for the galaxy (see \S\,\ref{subsubsec:obsCorr}), does not hold for the AGN obscuration. Instead, for the AGN system, we used a more appropriate assumption which is a source behind a screen of dust, and for which the solution to the transfer equation is given by \citep[e.g.][]{Calzetti2001, Krugel2009},

\begin{equation}
\label{eq:solTransEqAGN}
\Theta(\tau_\lambda)~=~e^{-\tau_\lambda},
\end{equation}

\noindent instead of eq.\,\ref{eq:solTransEqGal}. Therefore, while we assume that the system AGN and galaxy are affected by the same amount of obscuration at 9.7~\mum\ (defined by the ratio between the observed to the intrinsic fluxes at 9.7~\mum), the optical depth of the two systems, $\tau_{9.7}$, are different. In addition, we found significantly better fits to the AGN SEDs, when we used the extinction curve of PAHfit \citep{Smith2007}, instead of that defined for the galaxy in \S\,\ref{subsubsec:obsCorr}.

To summarise, the level of obscuration in the whole system (AGN + galaxy) was calculated using the ratio of the observed to the intrinsic fluxes at 9.7~\mum. Subsequently, the optical depth at 9.7~\mum, $\tau_{9.7}$, was calculated by using Eq.\,\ref{Eq:obs}, where $\Theta(\tau_\lambda)$ was substituted by the expression in Eq.\,\ref{eq:solTransEqGal} (i.e. assuming uniformly mixed absorbing and emitting materials) and that in Eq.\,\ref{eq:solTransEqAGN} (i.e. assuming a screen of dust), for the galaxy and the AGN, respectively. Then, the pseudo--extinction curve presented in \S\,\ref{subsubsec:obsCorr} was used on the galaxy dust continuum templates, within the context of a uniformly mixed obscuring and emitting material, and using the corresponding value of $\tau_{9.7}$ (we recall that the PAHs are assumed to be free of obscuration; see \S\,\ref{subsubsec:SFfit}). Similarly, the extinction curve of PAHfit was used on the AGN model (continuum and silicate emission), within the context of a screen of dust, and using the corresponding $\tau_{9.7}$. This allows our model to fit SEDs displaying a large amount of obscuration, as is the case for ``MCG+04-48-002'', which is shown in the bottom panel of Fig.\,\ref{fig:AGNfitEx}. In addition, we note that our model was able to fit the SED of the peculiar Mrk~231, which is classified as an optically unobscured source, yet shows a large silicate absorption feature at 9.7~\mum. The latter was not in our original sample of AGNs.

Due to the large number of free parameters in the AGN model, we must define some boundaries (i.e. priors) to avoid degeneracies while fitting the AGN SEDs. In particular, we found that if the position of the break ($\lambda_{\rm break}$) was unconstrained at longer wavelengths (i.e. at $\lambda~>~30$~\mum), then a number of fits became fully degenerate. This was particularly true for SEDs whose MIR emission is dominated by silicate emission, and which lacked PAH features. By using fits in which we had no significant degeneracies, we found, on average, that $\lambda_{\rm break}~=~40$~\mum. In light of this, we constrained $\lambda_{\rm break}$ by using a normally-distributed prior centred at $\lambda_{\rm break}^{\rm prior}~=~40$~\mum\ and with a standard deviation of $\sigma_{\rm break}^{\rm prior}~=~10$~\mum. We stress that this prior is weak enough to accept fits with $\lambda_{\rm break}~>~40$~\mum\ when required by the data. In doing so, we find that the value of the break can extend out to $\lambda_{\rm break}~\sim~100$~\mum\ in the case of some of our AGNs.

We also placed boundaries and prior information on the slopes $\alpha_1$, $\alpha_2$, and $\alpha_3$ to remove non-physical regions of parameter space and avoid over-fitting. We defined $-3.5~<~\alpha_1~<~3.5$, $-3.5~<~\alpha_2~<3.5$, and $-3.5~<~\alpha_3~<1$. These are in agreement with the range of values observed in previously measured AGN SEDs \citep[e.g.][]{Mullaney2011, Lyu2017b, Xu2020}, or in models of dusty torus emission \citep[e.g.][]{Siebenmorgen2005, Honig2010}.

We assumed that the strengths of the silicate features at 11~\mum\ and 18~\mum\ are identical by using a single normalisation for both. This choice was made to avoid over-fitting that at 18~\mum, since found to be blended with some continuum features. In fact, relatively little attention has been paid to the silicate emission at 18~\mum\ \citep[e.g.][]{Hatziminaoglou2015}. However, some recent studies that carefully decomposed the {\it Spitzer}-IRS spectra of Type-I QSOs showed that the strengths of the two features are consistent to within the uncertainties \citep{MartinezParedes2020}, supporting our assumption. In addition, this reduces the number of free parameters of our full AGN model (see \S\,\ref{subsubsec:model} for the AGN model) from ten to nine.

The relationship between the relative level of PAHs to dust continuum (see \S\,\ref{subsec:hostTempl}) was incorporated by using Eq.\,\ref{eq:PAHcorrDust}. To do this, we set a normally distributed prior, centred on the value returned by Eq.\,\ref{eq:PAHcorrDust} for a given \Ldust, and with a standard deviation of 0.3~dex. As such, values of \Ldust\ and \Lpah\ which deviate from Eq.\,\ref{eq:PAHcorrDust} are less likely. The inclusion of such relationship between the PAH and the dust continuum reduced degeneracies, and helped the determination of the level of FIR emission of dust heated by regions of star formation when PAH features were observed in the {\it Spitzer}-IRS spectra.

We fit each of our AGN SEDs separately, with each of our galaxy templates plus our full AGN model. To test whether the AGN component was necessary, we also independently fit our AGN SEDs with each of our galaxy templates alone (i.e. without including the AGN model). We therefore performed a total of 14 possible fits for each source, seven of which included an AGN component (i.e. one for each galaxy component). To select the best fit amongst these various attempts, we used the Akaike Information Criterion \citep[e.g.][AIC]{Akaike1973, Akaike1994}. The AIC is a function of the log-likelihood returned by the fit, and enables the comparison of models that have a different number of free parameters. The best model has the minimum AIC, and each model can be weighted accordingly using the Akaike weight \citep[e.g.][]{Akaike1973, Akaike1994}. Therefore, instead of using the best fits (i.e. minimum AIC), and if not otherwise stated, all our results are based on a weighted sum of the 14 models.

We show in Fig.\,\ref{fig:AGNfitEx} four examples of best weighted fits, split in terms of galaxy and AGN contributions. We note that our full AGN model is able to accommodate a large variety of AGN SEDs, whether the AGN contribution is minimal or dominates, and allows for different levels of AGN continuum to silicate emission \footnote{The individual fits for each of our AGN SEDs are available at \url{https://tinyurl.com/1f7lrd7t}}.

Out of our full sample of 112 AGNs, we found two sources for which the {\it Herschel}-SPIRE fluxes were contaminated by nearby companions (i.e. MCG-1-24-12 and NGC2992), potentially affecting their SED fits. We discarded these two sources from further analysis to avoid biasing the construction of our IR AGN templates (see \S\,\ref{subsubsec:AGNtempl} for the AGN templates). In addition, we found ten SEDs that were poorly fit by our model\footnote{These are: ESO511-30, Mrk1018, NGC3147, NGC788, NGC2655, NGC4258, NGC4939, NGC4941, NGC7682, and Z121-75}. Amongst those, six showed potential inconsistencies in the {\it Herschel} photometry, perhaps arising from the images used to calculate their fluxes (these are flagged with a dag in Table\,\ref{table:AGNFlux}). Furthermore, we found that the ten sources that were poorly fit are the most spatially extended, and we suspect that their {\it Spitzer}-IRS observations only captured the nuclear emission dominated by the AGN. In such cases, the simple re-scaling of the {\it Spitzer}-IRS spectra to match the photometry is not sufficient, as substantial PAH emission from the galaxy could have been missed. Therefore, we also discarded these ten sources from further analysis, and only considered the remaining 100 AGNs for which we have detailed SED fits.

\subsubsection{Constructing IR templates for AGN emission}
\label{subsubsec:AGNtempl}

\begin{figure}
 \centering
 \includegraphics[width=0.5\textwidth]{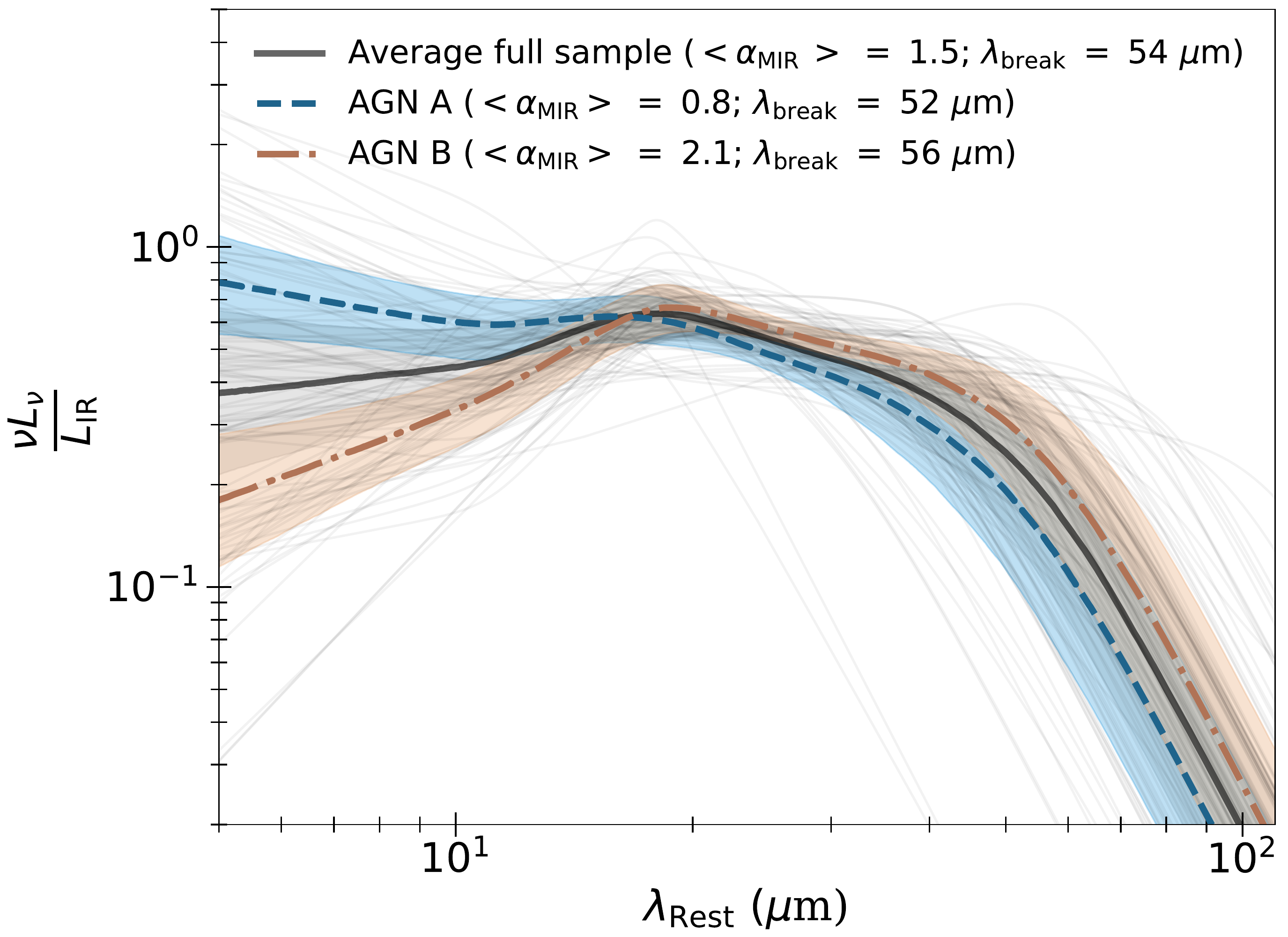}
 \caption{Our two AGN continuum templates, free of silicate emission, were built by combining the best fitting AGN contributions of our SEDs. The divide between the two was based on the MIR slope, $\alpha_{\rm MIR}$ (see \S\,\ref{subsubsec:AGNtempl}). The dashed blue and the dot-dashed brown lines are our ``AGN A'' and ``AGN B'' templates, respectively. We also show with shaded blue and brown areas an estimation of the 1$\sigma$ uncertainties for our ``AGN A'' and ``AGN B'' templates, respectively. The black line shows the average AGN continuum SED of the full sample of 100 AGNs, and the grey area shows an estimation of the 1$\sigma$ uncertainties. The thin black lines show each individual AGN continuum used to construct our templates. The main parameters of each of the templates are indicated in the keys. \label{fig:AGNtempl}}
\end{figure}

In \S\,\ref{subsubsec:model} we defined a flexible model for the IR emission of AGNs which, once combined with our galaxy templates, was able to fit the IR SEDs (consisting of a combination of {\it Spitzer}--IRS spectra and {\it Herschel} photometry) of our AGN sample. We now explain how we used the results of these fits to build our three AGN templates. The individual AGN contributions were defined by the best fitting parameters, which were calculated using a linear sum of all of the possible model attempts, weighted by their respective Akaike weights, as described in \S\,\ref{subsubsec:AGNfit}, and considering the AGN continuum and the silicate emission separately, as there no strong evidence for a relationship between the two \citep[e.g.][]{Deo2009, Hatziminaoglou2015}.

The extinction-corrected models for the continuum of our sample of 100 AGNs (free of silicate emission) are shown as thin lines in Fig.\,\ref{fig:AGNtempl}. In each case we have normalised the SEDs to \Lir\ (i.e. integrated from 8~\mum\ to 1000~\mum). This shows the large diversity found in the continuum emission of AGNs at IR wavelengths. In particular, we find that the position of the break, $\lambda_{\rm break}$, ranges from $\sim$11~\mum\footnote{We note that the position of the break was limited to $\lambda_{\rm break}~>~30$~\mum. However, some of our fits showed a steeply decreasing $\alpha_2$ and $\alpha_3$, which is equivalent to a break at $\lambda~=~$11~\mum.} to $\sim$100~\mum\ with a mean position of $\sim$54~\mum. We recall, however, that we placed a constraint on $\lambda_{\rm break}$ for which values at $\lambda~>~$40~\mum\ were less likely (see \S\,\ref{subsubsec:AGNfit}). We further find that the MIR slope ($\alpha_{\rm MIR}$), defined by the average value of $\alpha_1$ and $\alpha_2$ in our full AGN model (see \S\,\ref{subsubsec:model} for the AGN model), can take a large range of values, effectively producing a diverse range of SEDs, from flatter ones with $\alpha_{\rm MIR}~\sim$~0 to steeply rising ones with $\alpha_{\rm MIR}~\sim$~3. The mean $\alpha_{\rm MIR}$ is found to be $\sim$~1.5, and is slightly higher than the average value of 1.2 reported in \citeauthor{Mullaney2011} (\citeyear{Mullaney2011}; averaging their $\Gamma_1$ and $\Gamma_2$).

To build a set of templates for the AGN continuum, we separated our best fits into two bins using the slope $\alpha_{\rm MIR}$ {(see \S\,\ref{subsec:intDiff} for the motivation)}. By splitting our sample at the average value of $\alpha_{\rm MIR}~=~1.5$, we defined two template SEDs for the AGN continuum emission: ``AGN A'' and ``AGN B''. We note that the mean $\alpha_{\rm MIR}$ for the ``AGN A'' and the ``AGN B'' templates are 0.8 and 2.1, respectively, while the mean $\lambda_{\rm break}$ for the ``AGN A'' and ``AGN B'' templates are 52~\mum\ and 56~\mum, respectively. In Fig.\,\ref{fig:AGNtempl} we plot the ``AGN A'' template as a dashed blue line, and the ``AGN B'' template as a dot-dashed brown line. The uncertainties on the AGN templates were estimated using the first and the third quartiles of 5000 realisations, randomly drawing the best parameters defining our AGN templates to within 1$\sigma$ uncertainties. The average AGN SED for the whole sample is shown with a black line in Fig.\,\ref{fig:AGNtempl}.

By design, the silicate emission at 11~\mum\ and 18~\mum\ are similar (see \S\,\ref{subsubsec:model} and \S\,\ref{subsubsec:AGNfit}). We find that the average shift is $\lambda_{\rm shift}^{\rm Si}~\sim~0.1$~\mum, such that, the silicate features peak at $\sim$~11~\mum\ and $\sim$~18, consistent with previous work \citep[e.g.][]{Hatziminaoglou2015}. To build a template for the silicate emission, we simply averaged over the full sample, after normalising to \Lir, to extract the typical silicate emission observed in AGNs. The uncertainties on the average template for the silicate emission were built as for those on our AGN continuum templates.

We have, therefore, a total of three AGN templates, two of which correspond to AGN continuum (``AGN A'' and ``AGN B'') and one of which corresponds to silicate emission. We note that to build these templates we adopted a simple flexible model for the AGN continuum and silicate emissions. In fact, detailed investigations of the AGN contribution to the near-to-mid-IR reveals a rather complex dust geometry and composition (e.g. hot graphite dust, dusty narrow line regions, hollow cone; e.g. \citealt{Schweitzer2008, Mor2012, Honig2019}). However, we stress that for this work, we are only interested in accurately removing the AGN contribution to the IR, rather than performing a detailed analysis of the various dust components observed in AGNs.

\subsubsection{Comparing results from our AGN templates to our full AGN model}
\label{subsubsec:photoFit}

\begin{figure}
 \centering
 \includegraphics[width=0.5\textwidth]{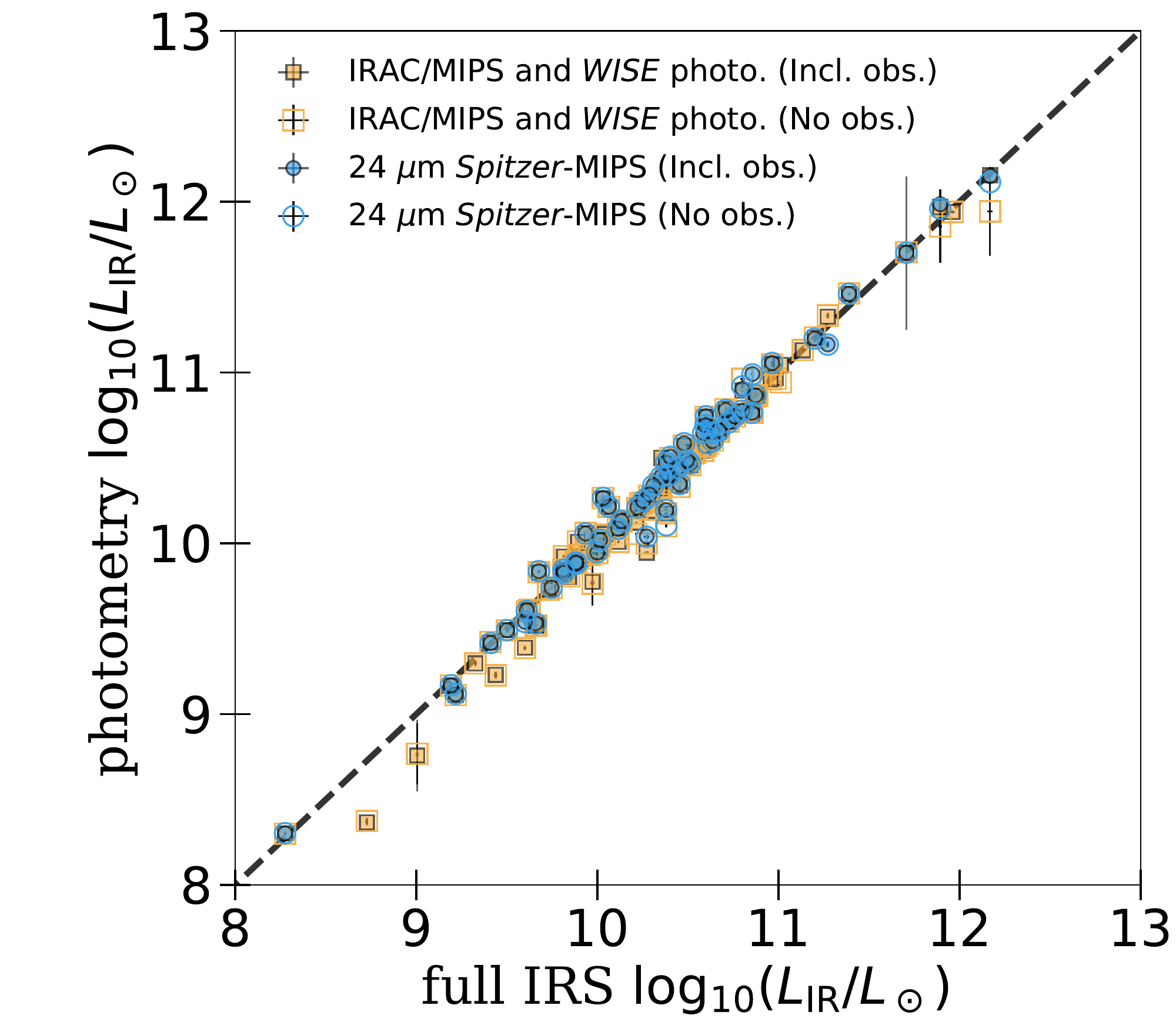}
 \caption{Comparison between the host galaxy (i.e after removing any AGN contamination) IR luminosities of our sample when our fit includes {\it Spitzer}--IRS spectra, against when we fit to photometric data alone. The blue circles show results obtained when we replaced {\it Spitzer}--IRS spectra with 24~\mum\ {\it Spitzer}--MIPS photometry alone, while the orange squares show those obtained when we replaced {\it Spitzer}--IRS spectra with 5~\mum\ and 8~\mum\ {\it Spitzer}--IRAC, 12~\mum\ and 22~\mum\ {\it WISE}, and 24~\mum\ {\it Spitzer}--MIPS photometry. For both, filled and open symbols correspond to when obscuration was accounted for, to when ignored, respectively (the open and filled symbols generally overlap). Irrespective of what photometry we replaced the {\it Spitzer}-IRS data with, and whether obscuration was accounted for, we find very good agreements between luminosities derived from when we included the {\it Spitzer}-IRS spectra versus when we did not. The dashed black line shows the 1:1 relationship. \label{fig:modelVStempl}}
\end{figure}

We now have a set of IR templates for AGNs, built using the best SED fits that includes {\it Spitzer}--IRS spectra and {\it Herschel} photometry. This set of templates can be used to fit AGN SEDs for which less rich datasets are available (e.g. photometry alone instead of IR spectra). In this section, we test that our set of templates (i.e. galaxy and AGNs) could, indeed, recover the host IR properties of our AGN sample when the data are reduced to only include photometry. To do this, we replaced the {\it Spitzer}--IRS spectra of our AGN sample with photometric data at MIR wavelengths. We used the {\it WISE} photometric fluxes at 12~\mum\ and 22~\mum, the {\it Spitzer}--IRAC photometric fluxes at 5.8~\mum\ and 8~\mum, and the {\it Spitzer}--MIPS photometric flux at 24~\mum, all collected from the IRSA database. We then re-fit the AGN sample (after replacing the {\it Spitzer}-IRS spectra with photometry) with a combination of each of our galaxy and AGN templates. Our fitting routine was of a similar design to that used when fitting our full AGN model (see \S\,\ref{subsubsec:AGNfit}), but replacing the six free parameters of the latter with a single normalisation for the AGN continuum and a single normalisation for the silicate emission. As we have measured the level of obscuration using the {\it Spitzer}-IRS spectra for our sample of AGNs, we subjected the templates to obscuration using the prescriptions of \S\,\ref{subsubsec:obsCorr} (see also \S\,\ref{subsubsec:AGNfit}).

In Fig.\,\ref{fig:modelVStempl} we show that the host IR luminosities (\Lir) inferred from photometric fits only, using our set of AGN and galaxy templates, are in very good agreements with those inferred from detailed fits (see \S\,\ref{subsubsec:model} for the full AGN model) in which {\it Spitzer}-IRS spectra were included (filled orange squares in Fig.\,\ref{fig:modelVStempl}). On average, the ratio between the two \Lir\ is one, and the standard deviation is 0.4. We note that the average standard deviation is dominated by a handful of sources which significantly deviates from the one-to-one relationship between the two inferred \Lir. These latter cases lack {\it Spitzer}-MIPS photometry, potentially reducing the accuracy of the fits. The standard deviation between the two inferred \Lir\ is reduced to 0.2 when only including sources that have measured {\it Spitzer}-MIPS photometry in addition to {\it Spitzer}-IRAC and {\it WISE} photometry.

To test our templates further, we also considered fits where the minimal amount of data is available by replacing the full {\it Spitzer}-IRS spectra by a single observed {\it Spitzer}-MIPS 24\mum\ flux. We find that, even in these minimal conditions, the AGN-free \Lir\ inferred from the photometric fits are in very good agreements with those measured in the detailed fits that included the {\it Spitzer}-IRS spectra, and the \Lir\ are also recovered to within 20~per~cent (see Fig.\,\ref{fig:modelVStempl}).

To test the importance of obscuration, we re-performed the two aforementioned photometric fits, this time ignoring obscuration. As shown in Fig.\,\ref{fig:modelVStempl}, this has no significant effect on the recovered \Lir\ for the vast majority of sources. In fact, the obscuration has a significant effect on the inferred \Lir\ only for the rare sources in our sample with large levels of obscuration, such as NGC3079. However, we stress that the difference is at most a factor of two. The lack of importance of the obscuration is likely related to the minimal effect of obscuration at most IR wavelengths, when compared to, for instance, optical wavelengths. Therefore, although accounting for obscuration is preferred when aiming to model the detailed features observed in the {\it Spitzer}-IRS spectra, it has no significant effects on the integrated properties of the hosts.

Finally, when performing fits with a {\it single photometric measurement} in the MIR, we have removed the silicate emission template as there were not enough data to constrain the galaxy (continuum + PAHs), the AGN continuum, and the silicate emission features. Nevertheless, this has no impact on the inferred \Lir, as shown in Fig.\,\ref{fig:modelVStempl}.

\subsection{{\sc iragnsep}: a fitting code to decompose AGN IR SEDs}
\label{subsec:pydecompir}

For the purpose of this work, we have developed {\sc iragnsep} which performs IR SED fits to separate AGN and galaxy contributions. Using {\sc iragnsep}, one can extract the host galaxy properties, such as SFRs, free of AGN contamination. The advantage of {\sc iragnsep} is that, in addition to fitting observed photometric fluxes, it also allows spectra to be incorporated in the fits which, if available, can greatly improve the robustness of the AGN--galaxy separation.

For the galaxy component {\sc iragnsep} uses the library of seven templates built from the sample of local star-forming galaxies described and presented in \S\,\ref{subsec:hostTempl} (see also Fig.\,\ref{fig:B18templ}). In terms of the AGN contribution, if the input dataset is a mixture of spectral and photometric data, then {\sc iragnsep} uses our full AGN model presented in \S\,\ref{subsubsec:model}. If, instead, the dataset contains photometric data alone, the AGN contribution is accounted-for using our library of three AGN templates presented in \S\,\ref{subsubsec:AGNtempl}. The models can be corrected for obscuration, either calculated by {\sc iragnsep} when spectral information in the MIR is provided, or set by the user (see \S\,\ref{subsubsec:obsCorr} for the method). The user can also easily modify the parameter space explored by {\sc iragnsep} by controlling the priors on each of the parameters defining the galaxy and the AGN emissions.

The advanced fitting techniques used by {\sc iragnsep} (i.e. MLE optimised with MCMC, see \S\,\ref{subsubsec:SFfit} and \S\,\ref{subsubsec:AGNfit}) combined with the powerful model comparison tests (i.e. AIC, see \S\,\ref{subsubsec:AGNfit}) allow {\sc iragnsep} to provide a statistically robust interpretation of IR SEDs in terms of AGN--galaxy contributions. {\sc iragnsep} also uses NUMBA \citep{numba} wrappers which translate Python functions to the LLVM compiler library, offering speeds that can approach those of C or FORTRAN, therefore allowing {\sc iragnsep} to be applied to a large number of sources. We stress, however, that {\sc iragnsep} has been designed on a sample of local (\z$<0.3$), low-to-high X-ray luminosity AGNs (i.e. \Lbol$~\sim~10^{42\--46}$~\ergps; see \S\,\ref{sec:sample}), with bright FIR emission. As such, care must be taken when applying these templates to higher \z\ and/or fainter FIR sources. Future work aims at testing {\sc iragnsep} on such sources.

{\sc iragnsep} is freely available at \url{https://pypi.org/project/iragnsep/}. We used the version 7.2.0 for this work. The templates associated with this work are available in the supplementary material, or directly at \,\url{https://tinyurl.com/yawp96qc}.

\section{Results and discussion}
\label{sec:results}

In this work, we have carried out a post-{\it Herschel} investigation of the intrinsic AGN emission at IR wavelengths using a sample of local (\z$<0.3$), low-to-high X-ray luminosity (i.e. \Lbol$~\sim~10^{42\--46}$~\ergps) AGNs. First, from a sample of local star-forming galaxy SEDs, we built a set of seven templates representing galaxy emission (see \S\,\ref{subsec:hostTempl}). Then, the IR contribution to the SEDs of AGN was extracted by successfully fitting the total SEDs with a combination of one of our galaxy templates and our full AGN model (see \S\,\ref{subsubsec:AGNfit}). Finally, using these best fits, we constructed a set of three IR templates to represent the AGN emission. Two of these three correspond to AGN IR continuum (coined ``AGN A'' and ``AGN B''), while the third corresponds to silicate emission (see \S\,\ref{subsubsec:AGNtempl}).

In this section, we compare our templates to previous work, and discuss the implications of our new library of templates. In \S\,\ref{subsec:compPrevWork}, we compare our AGN continuum templates to previous work. In \S\,\ref{subsec:intDiff}, we discuss the intrinsic differences of our two AGN continuum templates, which appear to be related to the level of nuclear obscuration along the line of sight. Finally, in \S\,\ref{subsec:AGNcool} we investigate the relationship between AGN luminosity and FIR emission.

\subsection{Comparison with previous IR templates for AGNs}
\label{subsec:compPrevWork}

\begin{figure*}
 \centering
 \includegraphics[width=\textwidth]{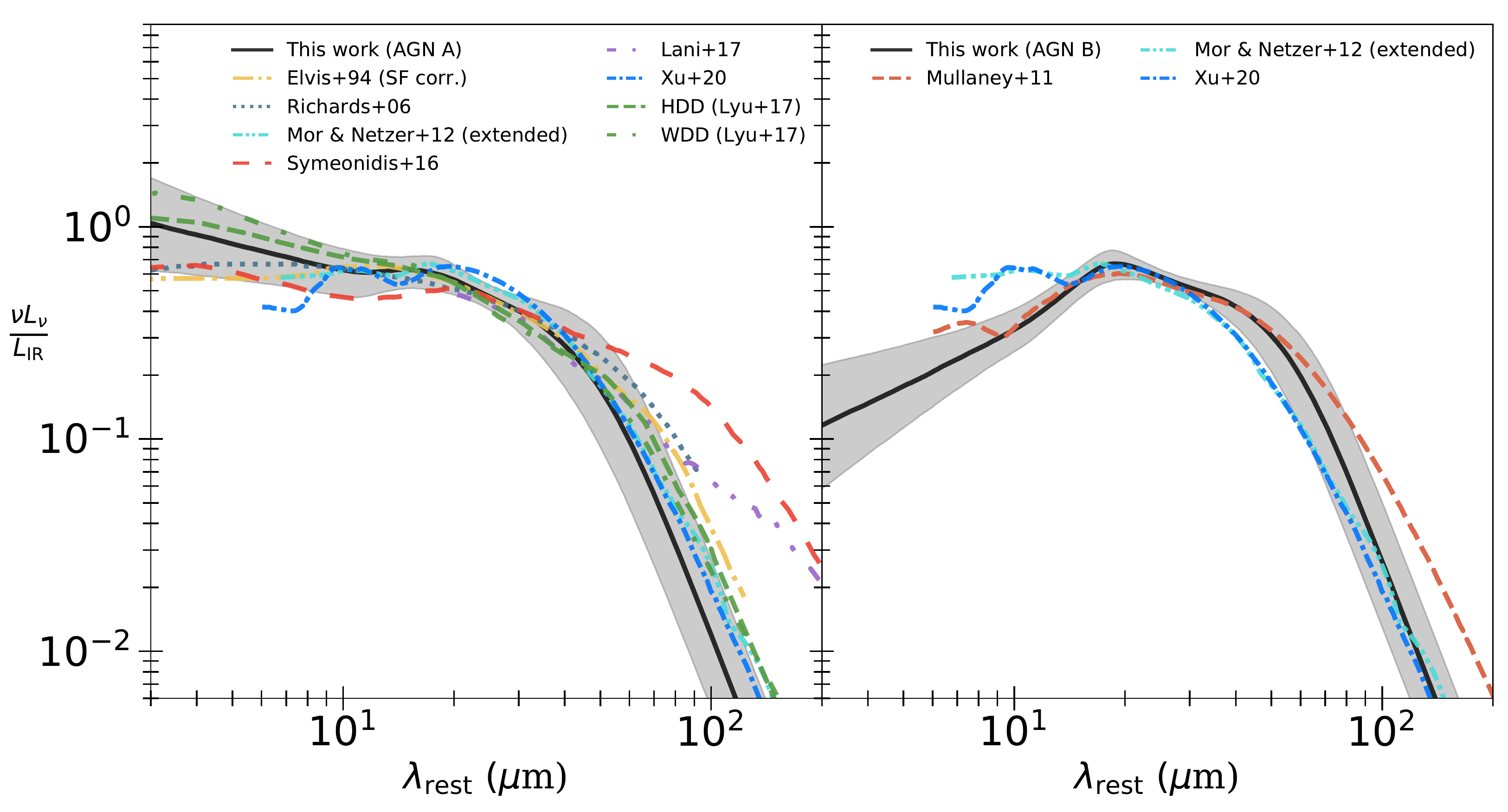}
 \caption{Our two AGN continuum templates, defined in \S\,\ref{subsubsec:AGNtempl}, compared to templates taken from previous studies. {\it Left panel:} Our ``AGN A'' template compared against the AGN templates for QSOs of \protect\citeauthor{Elvis1994} (\protect\citeyear{Elvis1994}; corrected for host contribution by \protect\citealt{Xu2015}), \protect\cite{Richards2006}, \protect\cite{Symeonidis2016}, \protect\cite{Lani2017}, and \protect\cite{Xu2020}, the template for optically selected Type-I AGNs of \protect\citeauthor{Mor2012} (\protect\citeyear{Mor2012}; extended at $\lambda~>~35$~\mum\ by \protect\citealt{Netzer2016}), and the templates for the hot dust deficient (HDD) and the warm dust deficient (WDD) QSOs of \protect\cite{Lyu2017b}. The grey area shows the 1$\sigma$ uncertainties of our ``AGN A'' template. {\it Right panel}: Our ``AGN B'' template compared against AGN templates for the more typical, less luminous AGNs of \protect\citealt{Mullaney2011}, the template for optically selected Type-I AGNs of \protect\cite{Mor2012}, and the template for Type-I QSOs of \protect\cite{Xu2020}. The grey area shows the 1$\sigma$ uncertainties of our ``AGN B'' template. Each of the AGN templates presented in this figure has been renormalised to \Lir\ (i.e. integrated from 8~\mum~to~1000~\mum) to avoid comparison biases due to the choice of the normalisation. \label{fig:B18vsM11AGN}}
\end{figure*}

In Fig.\,\ref{fig:B18vsM11AGN} we compare our two AGN continuum templates against those obtained in previous studies. We recall that our AGN templates have been corrected for obscuration, following prescriptions presented in \S\,\ref{subsubsec:obsCorr} (see also \S\,\ref{subsubsec:AGNfit}; although see \S\,\ref{subsec:intDiff} for a discussion on the obscuration), and that the silicate emission has been considered separately, so Fig.\,\ref{fig:B18vsM11AGN} shows the ``intrinsic'' AGN continua found in this work, free of silicate emission. To ease comparison, we chose to renormalise each of the templates to \Lir\ (i.e. integrated from 8~\mum\ to 1000~\mum), after extrapolating at shorter and/or longer wavelengths when necessary. We stress, however, that the extrapolations were only used to calculate \Lir, and to renormalise the templates, and Fig.\,\ref{fig:B18vsM11AGN} shows the templates within the range of wavelengths defined by the corresponding authors. This choice of normalisation allows us to directly compare the relative mid-to-far-IR output power amongst AGN templates. In Fig.\,\ref{fig:B18vsM11AGN}, we further separate our ``AGN A'' and ``AGN B'' templates into two panels for clarity.

We find that our ``AGN A'' template, and its uncertainties, is fully consistent with some recent average SEDs derived for Type-I AGNs \citep{Mor2012}, Type-I QSOs \citep{Xu2020}, and the hot dust deficient (HDD) and warm dust deficient (WDD) QSOs of \citeauthor{Lyu2017b} (\citeyear{Lyu2017b}; see left-hand side panel in Fig.\,\ref{fig:B18vsM11AGN}). The latter are populations of QSOs which appear to have lower emission in the near-to-far-IR, and the mid-to-far-IR, respectively, when compared to the typical QSO template of \cite{Elvis1994}. However, we note that, once normalised to \Lir, the HDD and WDD appear very similar, though with MIR whose is consistent with the upper band of the 1$\sigma$ uncertainties of our ``AGN A'' template (see Fig.\,\ref{fig:B18vsM11AGN}). 

More importantly, our ``AGN A'' template agrees well with studies finding hotter AGN SEDs (less contribution to the FIR), and do not show evidence supporting the recent claims that the IR emission of QSOs is cooler than previously thought \citep[i.e.][]{Symeonidis2016, Lani2017}. In fact, Fig.\,\ref{fig:B18vsM11AGN} shows that our average ``AGN A'' template offers even less FIR contribution than the hottest QSO SEDs of \cite{Mor2012} and \cite{Xu2020}. Nevertheless, we note that the differences in the FIR (i.e. integrated at $\lambda~>~$70~\mum) is typically below a few per~cent. This is consistent with the largest discrepancies found in the average QSO SEDs of \cite{Xu2020} while testing various libraries for galaxy emission.

To construct a library for AGN IR emission, \cite{Xu2020} used a grid of AGN templates built on the phenomenological torus model of \cite{Siebenmorgen2015}, which represents a completely different approach to that presented here. In particular, the grid of AGN models constructed in \cite{Xu2020} contains AGN templates which show a very large AGN contribution to the FIR, including that reported in \cite{Symeonidis2016}, which is the template that, like-for-like, has the largest relative contribution to the FIR. Nevertheless, the average quasar template from \cite{Xu2020} also shows a hotter SED, consistent with our ``AGN A'' template.

By contrast, our ``AGN B'' template shows a different distribution of power, where most of the IR output is coming from cooler dust when compared to our ``AGN A'' template (i.e. 20~\mum~$<~\lambda~<$~50~\mum). The MIR emission of our ``AGN B'' template is consistent with that previously reported for more typical, less luminous AGNs of \cite{Mullaney2011}. In particular, we note the presence of a strong IR bump at $\lambda~\sim~18$~\mum, consistent with that found in \cite{Mullaney2011} at $\sim$~19~\mum. However, we find that our ``AGN B'' template is hotter, and produces less relative FIR output than found in \cite{Mullaney2011}. We note that \cite{Mullaney2011} did not have readily available access to {\it Herschel} data, and as such their FIR SED is based on an extrapolation from shorter wavelengths.

One of the major challenges in extracting the AGN contribution to the FIR, {and which could potentially explain the differences between AGN templates}, is to accurately estimate, and preferably remove, that arising from the host galaxy. A popular method used in previous studies consisted of selecting optically bright QSOs, and assuming that the full IR SEDs is dominated by emission from the AGN, as in \cite{Elvis1994}, or to rely on generic host-to-SMBH relationships to estimate and remove the {\it minimum} host contamination, as in \cite{Richards2006}. {Both these methods tend to significantly over-estimate the AGN contribution to the FIR, as shown by \cite{Xu2015}, who corrected the \cite{Elvis1994} AGN SED for contribution from the host galaxy.}

Another popular method of removing the host contribution is to rely on the observed PAH features, since they are believed to be associated with star formation, and use star-forming galaxy templates to estimate the host contribution to the FIR \citep[e.g.][]{Xu2015, Symeonidis2016, Lani2017}. However, the accuracy of this method strongly depends on the quality of the MIR spectra, as well as the library of galaxy templates used to extrapolate to the FIR \citep{Lyu2017a, Xu2020}. In fact, \cite{Xu2020} showed that by using a library of galaxy templates which does not account for the silicate trough at 9.7~\mum\ \citep[e.g.][]{Chary2001, Dale2002, Dale2014}, the host contribution to the FIR is systematically under-estimated by a factor of $\sim$1.6-to-1.8, leading to larger estimates of the AGN contribution to the FIR. Our templates (galaxy and AGN) are less likely to be affected in such a way since we have factored-in obscuration. However, we note that the treatment of obscuration alone cannot fully explain the differences in the shapes of AGN SEDs in the FIR, as \cite{Lyu2017b} and \cite{Xu2020} both considered silicate absorption, yet found slightly cooler SEDs.

Finally, we note that the observed differences between ours and others’ FIR AGN template contribution to the FIR cannot also solely be attributed to the differences in galaxy templates. In fact, both \cite{Netzer2016}, which is a FIR extension of the \cite{Mor2012} template, and \cite{Lani2017} used the galaxy library of \cite{Chary2001}, and found significantly different AGN contributions to the FIR (see Fig.\,\ref{fig:B18vsM11AGN}). However, while \cite{Lani2017} used a sample of local bright QSOs with \Lbol$~\sim~4~\times~10^{43\--46}$~\ergps, \cite{Netzer2016} focused on a sample of very luminous optically selected AGNs with \Lbol$~\gtrsim~2~\times~10^{47}$~\ergps\ at 2~$<$~\z~$<$~3.5, which could explain the differences between the two. In addition, the modelling conducted by \cite{Netzer2016} required that the galaxy dominates over the AGN at wavelengths around 60~\mum. This is, in essence, very similar to our prior on the position of the break, implemented in our modelling, and which slowly drops at $\lambda_{\rm break}~>~40$~\mum\ (see \S\,\ref{subsubsec:AGNfit}). This could explain the similarities in the FIR between our AGN templates and that of \cite{Mor2012}, but also bias the typical AGN SED toward a smaller contribution to the FIR (i.e. hotter SED).

\subsection{Intrinsic differences between our ``AGN A'' and ``AGN B'' templates}
\label{subsec:intDiff}

Once normalised to \Lir, our ``AGN B'' template is cooler (i.e. shows a stronger relative FIR emission) than our ``AGN A'' template (see Fig.\,\ref{fig:B18vsM11AGN}). In this sub-section we investigate the potential reasons behind such an anisotropy. This could arise either from an excess of relative extinction, or by an ``intrinsic dust emission anisotropy'' resulting from the torus shape of the dusty structure around the AGN, {and unified via the AGN unification scheme} \citep{Antonucci1993, Urry1995}. We stress that for the former, although we have removed to some extent the extinction along the line of sight, it is possible that some degrees of extinction remain due to the complexity of the dusty structures in AGNs.

Several physical mechanisms are at work, each affecting the shape of the intrinsic, and/or the observed IR AGN SEDs. To better understand the physical mechanisms leading to such an anisotropy in the average IR emission of AGNs, we combined together several observations tracing different emission/absorption mechanisms for our sample of AGNs fit with {\sc iragnsep}. First, we considered the column densities of N$_{\rm H}$ (taken from \citealt{Ricci2017a}), mainly responsible for the absorption of X-rays. Then, we looked at the optical type of each of our AGNs (using the classifications reported in \citealt{Oh2018}), providing information on the optical ``visibility'' of the central engine. We further traced the amount of obscuring material present along the line of sight by using the measured total absorption at 9.7~\mum, $\Theta(\tau_{9.7})$, converted to S$_{\rm 9.7}$, using,

\begin{equation}
\label{eq:s9p7}
{\rm S}_{9.7}~\equiv~-~\ln\left(\frac{\rm F^{\rm Obs}_{\rm 9.7\mu m}}{\rm F^{\rm Int}_{\rm 9.7\mu m}}\right),
\end{equation}

\noindent where F$^{\rm Obs}_{\rm 9.7\mu m}$ and F$^{\rm Int}_{\rm 9.7\mu m}$ are the observed and the intrinsic (i.e. de-absorbed) fluxes at 9.7~\mum, respectively. Finally, we have the MIR slopes, $\alpha_{\rm MIR}$, used to separate our two AGN templates (see \S\,\ref{subsubsec:AGNtempl}) to probe the ``temperature'' of the AGN continuum, where flatter slopes correspond to hotter SEDs, and steeper slopes to cooler SEDs. For $\alpha_{\rm MIR}$, we also compared the intrinsic slopes, as corrected by our prescription for dust obscuration, to the observed ones.

\begin{figure}
 \centering
 \includegraphics[width=0.5\textwidth]{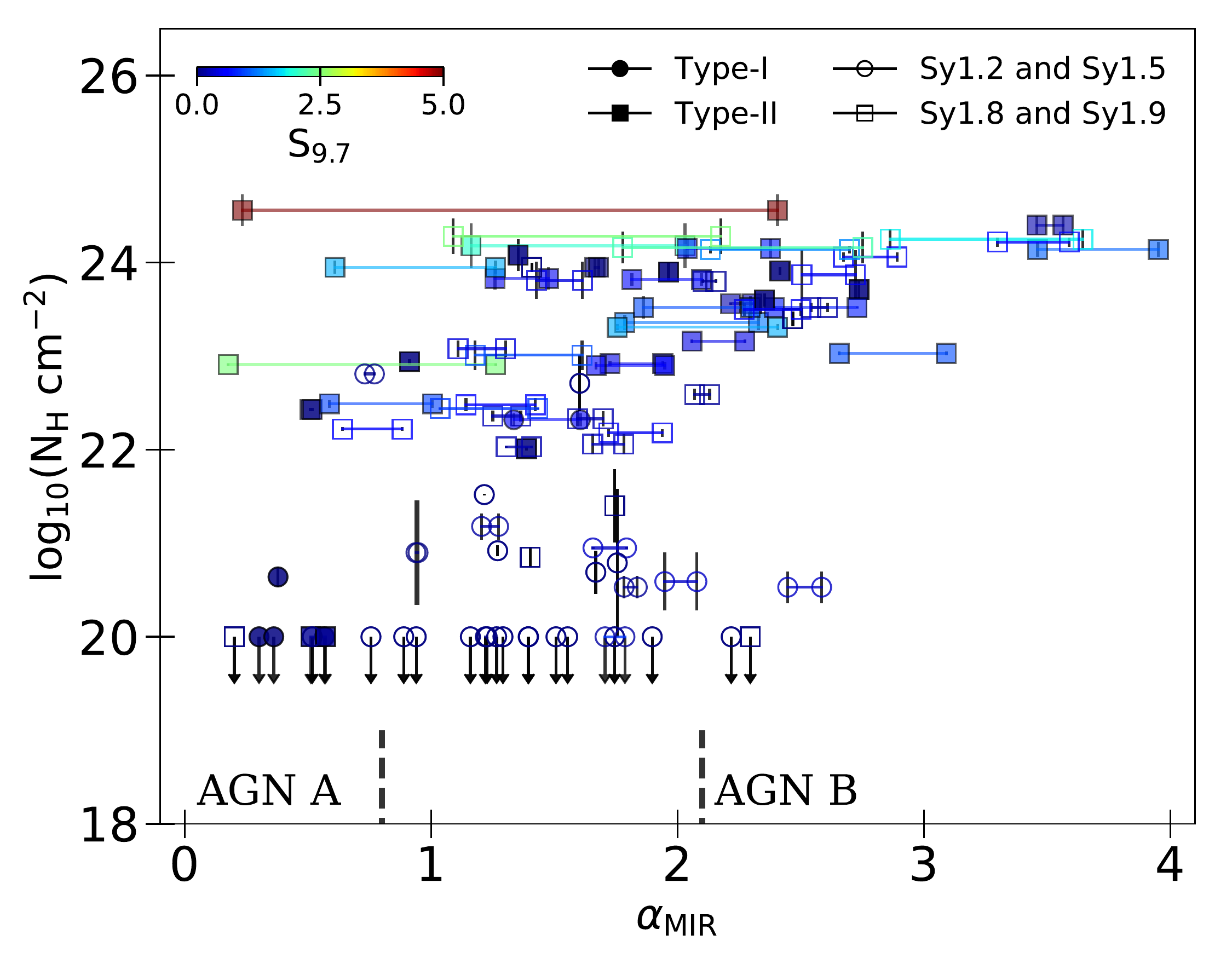}
 \caption{Plot showing the relationship between the column density of gas N$_{\rm H}$ and the average MIR slope $\alpha_{\rm MIR}$ (i.e. average between $\alpha_1$ and $\alpha_2$; see \S\,\ref{subsubsec:model}) found in our sample of AGNs, split in terms of AGN types (see keys), and colour coded by the strength of the observed silicate absorption at 9.7~\mum\ (see the colour bar in the top left-hand side corner). The lines connecting two symbols show the effect of dust obscuration on the slopes $\alpha_{\rm MIR}$, where the connected right-hand symbol and left-hand symbol correspond to the absorbed and the de-absorbed values, respectively. The arrows on the symbols indicate upper limits for N$_{\rm H}$. The dashed line and dot-dashed line show the average MIR slope of our ``AGN A'' and ``AGN B'' templates. \label{fig:logNhvsSlope}}
\end{figure}

In Fig\,\ref{fig:logNhvsSlope} we show all these parameters combined, where the abscissa is the slope $\alpha_{\rm MIR}$, and the ordinate is the column density of gas N$_{\rm H}$. The symbols show the optical types, where circles and squares are for Type-I and Type-II AGNs, respectively (intermediate types are shown with open symbols), and the colour code indicates the strength of the silicate absorption at 9.7~\mum. In Fig\,\ref{fig:logNhvsSlope}, each of the AGNs has two $\alpha_{\rm MIR}$ slopes connected with a line and indicating the absorbed (right-hand point) and de-absorbed (left-hand point) $\alpha_{\rm MIR}$ slope. The larger S$_{9.7}$, the larger the distance between the two.

Within the context of the simplest version of the AGN unification scheme, a single universal intrinsic AGN IR SED should be valid for all of our sources, and the differences observed should be attributed to dust obscuration along the line of sight. However, we note from Fig.\,\ref{fig:logNhvsSlope} that this is not the case, and our sources show a wide range of $\alpha_{\rm MIR}$, even after correcting for dust obscuration based on the strength of the silicate absorption at 9.7~\mum. This justifies the need for our ``AGN B'' template which represents sources with somewhat steeper ``intrinsic'' $\alpha_{\rm MIR}$.

Although obscuration does not appear to be the obvious reason behind the differences between our two AGN templates, we still find a mild trend for more dust-obscured sources (lighter blue in Fig.\,\ref{fig:logNhvsSlope}) to have larger column densities of gas (i.e. N$_{\rm H}>10^{22}\,{\rm cm}^{-2}$), and to be classified as optical Type-II AGNs (square symbols in Fig.\,\ref{fig:logNhvsSlope}). {In fact, it is believed that dust in the host galaxy is responsible for the silicate absorption trough at 9.7~\mum\ \cite[e.g.][]{Deo2009, Goulding2012}.} By using this feature to correct for absorption, it is likely that we are not sensitive to the level of nuclear absorption.

Furthermore, it is now clear that dusty structures in AGNs (i.e. those responsible for the nuclear obscuration) are more complex than the classical equatorial dusty torus. In fact, IR interferometry has recently shown that the parsec-scale MIR emission of some AGNs has a stronger polar-extended component, when compared to the classical equatorial dust \cite[e.g.][]{Honig2013, Leftley2018}. In addition, it was shown that this component can extend up to scales of tens-to-hundreds of parsecs, which is similar to scales observed for the narrow-line-emitting region of the AGN \citep[e.g.][]{Radomski2003, Asmus2016}. Therefore, the unified AGN orientation scenario has been re-visited to account for the discovery of the extended polar dust (see \citealt{Ogawa2021}, and references therein). It was further proposed that extended AGN polar dust creates a cooler component in the IR SEDs of AGNs, peaking at $\lambda~\sim~$30~\mum\ \citep[e.g.][]{Honig2011, Lyu2018}. In light of these, we suggest that our ``AGN A'' template represents the typical unobscured Type-I sources, for which both, the AGN and the galaxy are face-on. In addition, our ``AGN A'' template also represents the absorption-corrected Type-II sources for which the main dust obscuring material is located in the host galaxy (i.e. face-on AGN at the centre of an edge-on galaxy disk, or with coincident galactic dust lanes).

By contrast, we suggest that our ``AGN B'' template represents sources that, in addition to galactic intervening dust material, are suffering from significant nuclear obscuration. In such cases, the hotter dust contribution is fully or partly absorbed (i.e. depending on the amount of intervening dust in the line of sight), and only the extended AGN-heated dust components, such as the extended polar dust, are recovered. The source UGC3601 in Fig.\,\ref{fig:AGNfitEx} is a good example as it shows no evidence for silicate absorption, has a low column density of gas (i.e. N$_{\rm H}$~=~$2.5~\times~10^{21}~\rm cm^{-2}$) with significant silicate emission, though shows a rising $\alpha_{\rm MIR}$, and is classified as a Seyfert 1.9 galaxy, meaning that only broad H$\alpha$ lines are detected.

Perhaps more interestingly, we find that intermediate Seyfert-type galaxies follow a better relationship between the column density of gas, N$_{\rm H}$, and $\alpha_{\rm MIR}$ slope, yet show minimum silicate absorption (see Fig.\,\ref{fig:logNhvsSlope}). In fact, by inspecting their IR SEDs, we found that Seyfert 1.2 and Seyfert 1.5 AGNs mostly show a rising MIR continuum and mild silicate emission with the absence of strong silicate absorption, and Seyfert 1.8 and Seyfert 1.9 AGNs also show a rising MIR continuum, yet no silicate emission or absorption (see also \citealt{Deo2009}). It is possible that intermediate Seyfert galaxies are free of galactic dust extinction, and that only nuclear obscuration (i.e. via the torus and/or the extended polar dust) affects the light. We reserve further analysis regarding the connection between the shape of the IR SED of AGNs to the unified scheme to future study, and assume for the rest of this discussion that nuclear obscuration is responsible for the differences between our ``AGN A'' and ``AGN B'' templates, and that the latter is dominated by AGN-heated dust at larger scales.

We note that some studies included an extra parameter representing obscuration in their fits to account for the full diversity of MIR slopes \citep[e.g.][]{DelMoro2013, DelMoro2016}. In fact, we find that both our AGN templates are necessary to fit the photometric SEDs of our sample of AGNs, where the ``AGN A'' and ``AGN B'' templates {are best fitting 43~per~cent and 57~per~cent of the SEDs}, respectively, when sufficient photometry measurements are available to constrain the MIR slope. It is useful to bear in mind that using templates is less demanding, and less susceptible to degeneracies than adding a free obscuration parameter in the fit, as in \cite{DelMoro2016}.

\subsection{The relationship between AGN luminosity and the AGN far-IR emission}
\label{subsec:AGNcool}

It was suggested in \cite{Mullaney2011} that less powerful AGNs (i.e. where less powerful means \LxChandra$~<~10^{42.9}$~\ergps, i.e., below the median \LxChandra\ of their sample) tend to show cooler emission (i.e. contribute more to the FIR) than their higher X-ray luminosity counterparts. To assess whether our templates also show this trend, we split our sample into two sub-samples according to bolometric luminosity (\Lbol). We calculated \Lbol\ by first converting the 14--195~keV to 2--10~keV luminosities, using the photon indices found in \cite{Oh2018}, then convert these 2--10~keV luminosities to \Lbol, using the average bolometric correction of 22.4 reported in \cite{Vasudevan2007}. We split our sample at \Lbol$~=~10^{44.2}$~\ergps, which is equivalent to the value used in \cite{Mullaney2011}, after we converted their 2--10~keV luminosities to \Lbol.

\begin{figure}
 \centering
 \includegraphics[width=0.5\textwidth]{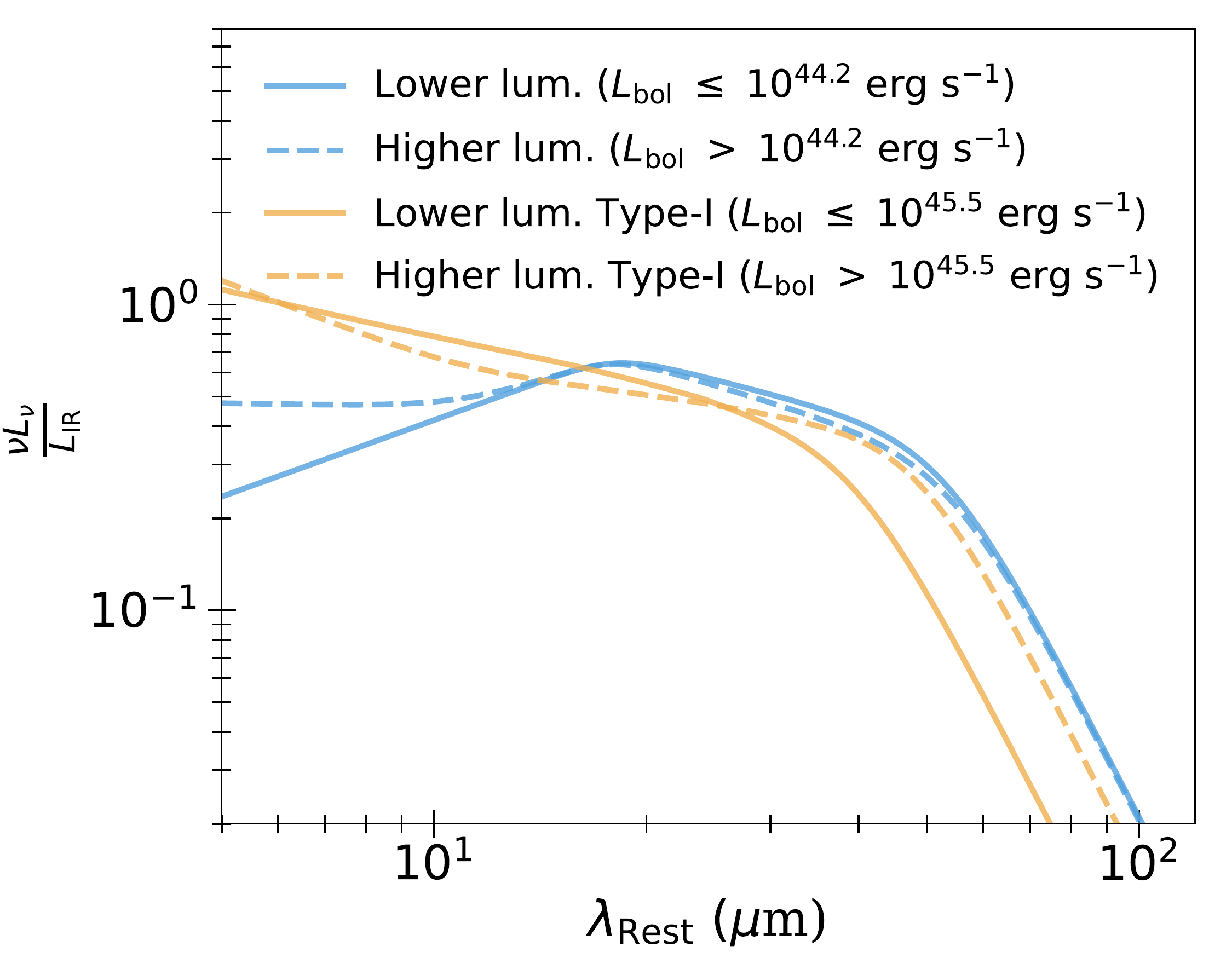}
 \caption{Plot showing the average AGN SEDs in bins of bolometric luminosities. The continuous and the dashed blue lines show the average SEDs of the full sample split at \Lbol$~=~10^{44.2}$~\ergps, which corresponds to the same split as in \protect\cite{Mullaney2011}, while the continuous and the dashed orange lines show the average SEDs for Type-I AGNs in our sample, split at \Lbol$~=~10^{45.5}$~\ergps, which corresponds to the same split as in \protect\cite{Xu2020}. \label{fig:AGNcompObs}}
\end{figure}

We show in Fig.\,\ref{fig:AGNcompObs} the average AGN continuum SEDs split into bins of \Lbol. Although we find that lower luminosity AGNs have a higher FIR-to-MIR relative emission when compared to their higher luminosity counterparts, we do not find any strong differences between the positions of the break, as suggested in \cite{Mullaney2011}.

The lack of a relationship between \Lbol\ and the contribution to the FIR, when considering our whole sample, is also in apparent contradiction with \cite{Xu2020}, which reported that more optically luminous quasars (i.e. using the 5100\AA\ luminosity; $L_{5100}$) show cooler SEDs than their optically less luminous counterparts. To test this in our AGN sample, we derived average SEDs for our Type-I AGNs alone, which are similar sources to those in \cite{Xu2020}, after splitting the sample at \Lbol$~=~10^{45.5}$~\ergps. The latter corresponds to the divide adopted in \cite{Xu2020}, after converting their $L_{5100}$ to \Lbol\ using a factor of ten \citep{Richards2006}. In doing this, we find that more powerful Type-I AGNs show, on average, a cooler SED when compared to their less powerful counterparts, in agreement with \citeauthor{Xu2020} (\citeyear{Xu2020}; see Fig.\,\ref{fig:AGNcompObs}). In fact, on average, the position of the break shifts from $\lambda_{\rm break}~\sim~$35~\mum\ to $\lambda_{\rm break}~\sim~$50~\mum, from lower-to-higher luminosity Type-I AGNs, while the MIR-to-FIR relative emission remains similar. We stress, however, that our sub-sample of Type-I AGNs is small, containing only four and three sources at higher and lower bolometric luminosities, respectively.

The differences in the relationship between \Lbol\ and the FIR AGN contribution observed for our full sample and the Type-I sources could potentially be related to the Eddington ratio. In fact, \cite{Ricci2017b} found that only higher Eddington ratio AGNs have the ability to clear-out the obscuring material, and, for those, a relationship is expected between AGN power and the FIR luminosity of the AGN (see also \citealt{Ogawa2021}). A full analysis of the shapes of IR AGN SEDs and how it changes with \Ledd\ is beyond the scope of this paper, and left for future work.

\section{Conclusion}
\label{sec:conclusion}

In this work we have investigated the intrinsic AGN IR emission of AGNs selected from the 105-month {\it Swift}--BAT all-sky hard X-ray survey. Our sample is typical of local (\z~$<$~0.3) low-to-high X-ray luminosity AGNs (i.e. \Lbol$~\sim~\times10^{42\--46}$~\ergps; see Fig.\,\ref{fig:sample}). For the IR counterparts, we used archival {\it Spitzer}--IRS spectra (5~\mum~$<~\lambda~<$~40~\mum), as well as {\it Herschel}--PACS and {\it Herschel}--SPIRE photometry (70~\mum--500~\mum) for which we have corrected the flux for spatially extended emission (see \S\,\ref{subsubsec:dataFIRphoto}). Our final sample consists of 100 AGN SEDs covering most of the IR portion of the electromagnetic spectrum (see Fig.\,\ref{fig:allSEDs}).

Prior to extracting the intrinsic AGN IR emission we defined a set of templates for galaxy emission using a sample of 110 star-forming galaxies pre-selected based on the \eqwspd\ and the WISE colours W1 and W2 (see \S\,\ref{subsec:hostTempl}). After fitting their SEDs, we discarded an extra 50~per~cent of this sample as they show potential mild AGN contamination (see \S\,\ref{subsubsec:SFfit}), and kept 55 galaxies that we used to build eight templates for star-forming galaxies (see \S\,\ref{subsubsec:Templ}). Specifically, we have one template that represents the PAH emission, and six templates that represent the dust continuum emission. Using the best fits, we defined Eq.\,\ref{eq:PAHcorrDust} to evaluate the normalisation of the dust continuum using the PAH features observed in the MIR to within a scatter of 0.3~dex (see \S\,\ref{subsubsec:Templ}).

To account for the AGN emission we defined a flexible model that represented continuum emission and silicate features (see \S\,\ref{subsubsec:model}). By fitting our AGN SEDs with a combination of each of our galaxy templates plus our full AGN model (see \S\,\ref{subsubsec:AGNfit}), we extracted a new set of three IR templates for AGN (see \S\,\ref{subsubsec:AGNtempl}). Two of the templates correspond to continuum emission (``AGN A'' and ``AGN B''), and one corresponds to silicate emission (see Fig.\,\ref{fig:AGNtempl}). The most important features of our two AGN continuum templates is the MIR slope, with ``AGN B'' showing a rising MIR emission compared to ``AGN A''.

We compared our two AGN continuum templates to previous work and found that our ``AGN A'' template was consistent with some of the most recent AGN templates derived for (local) Type-I AGNs and QSOs, for which the host contributions have been correctly accounted-for. By contrast, our ``AGN B'' template was in better agreement with AGN SEDs measured in low-to-moderate X-ray luminosity AGNs (see \S\,\ref{subsec:compPrevWork} and Fig.\,\ref{fig:B18vsM11AGN}). We further found that, overall, our AGN templates show a hotter SED when compared to previous work, yet {the differences in the FIR are minimal, and can be explained by the use of different libraries of galaxy templates.}

We also found that, in general, the differences between our ``AGN A'' and ``AGN B'' templates can be explained by nuclear obscuration, and the presence of extended AGN polar dust. We suggested that, while our ``AGN A'' template is typical of sources for which there is no, or minimal, amount of nuclear obscuration, our ``AGN B'' template represents sources which suffer from nuclear obscuration which cannot be easily removed. For the latter, the SED is dominated by the emission of the extended AGN-heated polar dust, and the hotter dust contributions are, to some extent, suppressed.

After splitting our AGN sample into two bins of X-ray luminosity, we found that less powerful AGNs have a stronger relative FIR-to-MIR energy balance, when compared to more powerful AGNs, but the former do not show any evidence for an excess of FIR emission, in contrast with results reported in \cite{Mullaney2011}. However, we found that the average SED for more powerful Type-I sources is cooler (i.e. contributes more to the FIR), when compared to their less powerful counterparts. This is in agreement with the findings reported in \cite{Xu2020}, where more optically luminous AGNs show cooler SEDs (i.e. contribute more to the FIR) when compared to less optically luminous AGNs (see \S\,\ref{subsec:AGNcool}).

Lastly, we have developed {\sc iragnsep} -- a python-based package that can be used to decompose the IR SEDs of AGNs (see \S\,\ref{subsec:pydecompir}). The default templates included in {\sc iragnsep} are those that we have presented in this work, and can be downloaded via \url{https://tinyurl.com/yawp96qc}. {\sc iragnsep} is freely available at \url{https://pypi.org/project/iragnsep/}.

\section*{Acknowledgements}
We thank the anonymous referee for the useful comments that help to significantly improve the quality of the paper. EB, JRM, CT acknowledge STFC grant R/151397. DJR and DMA acknowledge the Science and Technology Facilities Council (grant codes ST/P000541/1 and ST/T000244/1). EB thanks Mouyuan Sun for the valuable discussion on the relationship between AGN luminosity and AGN contribution to the FIR. This research has made use of the NASA/IPAC Infrared Science Archive which is operated by the Jet Propulsion Laboratory, California Institute of Technology, under contract with the National Aeronautics and Space Administration. This research has made use of the SIMBAD database, operated at CDS, Strasbourg, France. The following packages were used for the data reduction and analysis: MATPLOTLIB \citep{Hunter2007}, ASTROPY \citep{Astropy2018}, NUMPY, SCIPY \citep{scipy}, PANDAS \citep{pandas}, and NUMBA \citep{numba}. This research made use of Photutils, an Astropy package for detection and photometry of astronomical sources \citep{Bradley2020}.

\section*{Data availability statement}
The data underlying this article were accessed from the IRSA database (\url{https://irsa.ipac.caltech.edu/frontpage/}), the {\it Swift} observatory archive (\url{https://swift.gsfc.nasa.gov/results/bs105mon/}), and the CASSIS database (\url{https://cassis.sirtf.com}). Our analysis led to new data that are shared in the supplementary material, or directly available at \url{https://github.com/epbernhard/iragnsep_paper}.




\bibliographystyle{mnras}
\bibliography{./biblio}

\begin{thebibliography}{}
\makeatletter
\relax
\def\mn@urlcharsother{\let\do\@makeother \do\$\do\&\do\#\do\^\do\_\do\%\do\~}
\def\mn@doi{\begingroup\mn@urlcharsother \@ifnextchar [ {\mn@doi@}
  {\mn@doi@[]}}
\def\mn@doi@[#1]#2{\def\@tempa{#1}\ifx\@tempa\@empty \href
  {http://dx.doi.org/#2} {doi:#2}\else \href {http://dx.doi.org/#2} {#1}\fi
  \endgroup}
\def\mn@eprint#1#2{\mn@eprint@#1:#2::\@nil}
\def\mn@eprint@arXiv#1{\href {http://arxiv.org/abs/#1} {{\tt arXiv:#1}}}
\def\mn@eprint@dblp#1{\href {http://dblp.uni-trier.de/rec/bibtex/#1.xml}
  {dblp:#1}}
\def\mn@eprint@#1:#2:#3:#4\@nil{\def\@tempa {#1}\def\@tempb {#2}\def\@tempc
  {#3}\ifx \@tempc \@empty \let \@tempc \@tempb \let \@tempb \@tempa \fi \ifx
  \@tempb \@empty \def\@tempb {arXiv}\fi \@ifundefined
  {mn@eprint@\@tempb}{\@tempb:\@tempc}{\expandafter \expandafter \csname
  mn@eprint@\@tempb\endcsname \expandafter{\@tempc}}}

\bibitem[\protect\citeauthoryear{{Aird}, {Coil}, {Georgakakis}, {Nandra},
  {Barro}  \& {P{\'e}rez-Gonz{\'a}lez}}{{Aird} et~al.}{2015}]{Aird2015}
{Aird} J.,  {Coil} A.~L.,  {Georgakakis} A.,  {Nandra} K.,  {Barro} G.,
  {P{\'e}rez-Gonz{\'a}lez} P.~G.,  2015, \mn@doi [\mnras]
  {10.1093/mnras/stv1062}, \href
  {http://adsabs.harvard.edu/abs/2015MNRAS.451.1892A} {451, 1892}

\bibitem[\protect\citeauthoryear{{Akaike}}{{Akaike}}{1973}]{Akaike1973}
{Akaike} H.,  1973, \mn@doi [Biometrika] {10.1093/biomet/60.2.255}, 60, 255

\bibitem[\protect\citeauthoryear{{Akaike}}{{Akaike}}{1994}]{Akaike1994}
{Akaike} H.,  1994, in Bozdogan H.,  Sclove S.~L.,  Gupta A.~K.,  Haughton D.,
  Kitagawa G.,  Ozaki T.,   Tanabe K.,  eds, , Proceedings of the First
  US/Japan Conference on the Frontiers of Statistical Modeling: An
  Informational Approach: Volume 3 Engineering and Scientific Applications.
Springer Netherlands, Dordrecht, pp 27--38,
  \mn@doi{10.1007/978-94-011-0854-6_2}, \url
  {http://dx.doi.org/10.1007/978-94-011-0854-6_2}

\bibitem[\protect\citeauthoryear{{Antonucci}}{{Antonucci}}{1993}]{Antonucci1993}
{Antonucci} R.,  1993, \mn@doi [\araa] {10.1146/annurev.aa.31.090193.002353},
  \href {http://adsabs.harvard.edu/abs/1993ARA%26A..31..473A} {31, 473}

\bibitem[\protect\citeauthoryear{{Armus} et~al.,}{{Armus}
  et~al.}{2007}]{Armus2007}
{Armus} L.,  et~al., 2007, \mn@doi [\apj] {10.1086/510107}, \href
  {http://adsabs.harvard.edu/abs/2007ApJ...656..148A} {656, 148}

\bibitem[\protect\citeauthoryear{{Asmus}, {H{\"o}nig}  \& {Gandhi}}{{Asmus}
  et~al.}{2016}]{Asmus2016}
{Asmus} D.,  {H{\"o}nig} S.~F.,   {Gandhi} P.,  2016, \mn@doi [\apj]
  {10.3847/0004-637X/822/2/109}, \href
  {https://ui.adsabs.harvard.edu/abs/2016ApJ...822..109A} {822, 109}

\bibitem[\protect\citeauthoryear{{Assef} et~al.,}{{Assef}
  et~al.}{2013}]{Assef2013}
{Assef} R.~J.,  et~al., 2013, \mn@doi [\apj] {10.1088/0004-637X/772/1/26},
  \href {https://ui.adsabs.harvard.edu/abs/2013ApJ...772...26A} {772, 26}

\bibitem[\protect\citeauthoryear{{Astropy Collaboration} et~al.,}{{Astropy
  Collaboration} et~al.}{2018}]{Astropy2018}
{Astropy Collaboration} et~al., 2018, \mn@doi [\aj] {10.3847/1538-3881/aabc4f},
  \href {https://ui.adsabs.harvard.edu/abs/2018AJ....156..123A} {156, 123}

\bibitem[\protect\citeauthoryear{{Barthelmy} et~al.,}{{Barthelmy}
  et~al.}{2005}]{Barthelmy2005}
{Barthelmy} S.~D.,  et~al., 2005, \mn@doi [\ssr] {10.1007/s11214-005-5096-3},
  \href {http://adsabs.harvard.edu/abs/2005SSRv..120..143B} {120, 143}

\bibitem[\protect\citeauthoryear{Bradley et~al.,}{Bradley
  et~al.}{2020}]{Bradley2020}
Bradley L.,  et~al., 2020, \mn@doi [Zenodo] {10.5281/zenodo.4044744}

\bibitem[\protect\citeauthoryear{{Brandl} et~al.,}{{Brandl}
  et~al.}{2006}]{Brandl2006}
{Brandl} B.~R.,  et~al., 2006, \mn@doi [\apj] {10.1086/508849}, \href
  {http://adsabs.harvard.edu/abs/2006ApJ...653.1129B} {653, 1129}

\bibitem[\protect\citeauthoryear{{Burgarella}, {Buat}  \&
  {Iglesias-P{\'a}ramo}}{{Burgarella} et~al.}{2005}]{Burgarella2005}
{Burgarella} D.,  {Buat} V.,   {Iglesias-P{\'a}ramo} J.,  2005, \mn@doi
  [\mnras] {10.1111/j.1365-2966.2005.09131.x}, \href
  {https://ui.adsabs.harvard.edu/abs/2005MNRAS.360.1413B} {360, 1413}

\bibitem[\protect\citeauthoryear{{Calistro Rivera}, {Lusso}, {Hennawi}  \&
  {Hogg}}{{Calistro Rivera} et~al.}{2016}]{CalistroRivera2016}
{Calistro Rivera} G.,  {Lusso} E.,  {Hennawi} J.~F.,   {Hogg} D.~W.,  2016,
  \mn@doi [\apj] {10.3847/1538-4357/833/1/98}, \href
  {https://ui.adsabs.harvard.edu/abs/2016ApJ...833...98C} {833, 98}

\bibitem[\protect\citeauthoryear{{Calzetti}}{{Calzetti}}{2001}]{Calzetti2001}
{Calzetti} D.,  2001, \mn@doi [\pasp] {10.1086/324269}, \href
  {http://adsabs.harvard.edu/abs/2001PASP..113.1449C} {113, 1449}

\bibitem[\protect\citeauthoryear{{Chabrier}}{{Chabrier}}{2003}]{Chabrier2003}
{Chabrier} G.,  2003, \mn@doi [\pasp] {10.1086/376392}, \href
  {http://cdsads.u-strasbg.fr/abs/2003PASP..115..763C} {115, 763}

\bibitem[\protect\citeauthoryear{{Chary} \& {Elbaz}}{{Chary} \&
  {Elbaz}}{2001}]{Chary2001}
{Chary} R.,  {Elbaz} D.,  2001, \mn@doi [\apj] {10.1086/321609}, \href
  {http://cdsads.u-strasbg.fr/abs/2001ApJ...556..562C} {556, 562}

\bibitem[\protect\citeauthoryear{{Chen}, {Yang}, {Liu}  \& {Shan}}{{Chen}
  et~al.}{2018}]{Chen2018}
{Chen} P.~S.,  {Yang} X.~H.,  {Liu} J.~Y.,   {Shan} H.~G.,  2018, \mn@doi [\aj]
  {10.3847/1538-3881/aa988c}, \href
  {https://ui.adsabs.harvard.edu/abs/2018AJ....155...17C} {155, 17}

\bibitem[\protect\citeauthoryear{{Chiar} \& {Tielens}}{{Chiar} \&
  {Tielens}}{2006}]{Chiar2006}
{Chiar} J.~E.,  {Tielens} A.~G.~G.~M.,  2006, \mn@doi [\apj] {10.1086/498406},
  \href {https://ui.adsabs.harvard.edu/abs/2006ApJ...637..774C} {637, 774}

\bibitem[\protect\citeauthoryear{{Ciesla} et~al.,}{{Ciesla}
  et~al.}{2015}]{Ciesla2015}
{Ciesla} L.,  et~al., 2015, \mn@doi [\aap] {10.1051/0004-6361/201425252}, \href
  {http://adsabs.harvard.edu/abs/2015A%26A...576A..10C} {576, A10}

\bibitem[\protect\citeauthoryear{{Cresci}, {Vanzi}, {Telles}, {Lanzuisi},
  {Brusa}, {Mingozzi}, {Sauvage}  \& {Johnson}}{{Cresci}
  et~al.}{2017}]{Cresci2017}
{Cresci} G.,  {Vanzi} L.,  {Telles} E.,  {Lanzuisi} G.,  {Brusa} M.,
  {Mingozzi} M.,  {Sauvage} M.,   {Johnson} K.,  2017, \mn@doi [\aap]
  {10.1051/0004-6361/201730876}, \href
  {http://adsabs.harvard.edu/abs/2017A%26A...604A.101C} {604, A101}

\bibitem[\protect\citeauthoryear{{Dale} \& {Helou}}{{Dale} \&
  {Helou}}{2002}]{Dale2002}
{Dale} D.~A.,  {Helou} G.,  2002, \mn@doi [\apj] {10.1086/341632}, \href
  {http://cdsads.u-strasbg.fr/abs/2002ApJ...576..159D} {576, 159}

\bibitem[\protect\citeauthoryear{{Dale}, {Helou}, {Contursi}, {Silbermann}  \&
  {Kolhatkar}}{{Dale} et~al.}{2001}]{Dale2001}
{Dale} D.~A.,  {Helou} G.,  {Contursi} A.,  {Silbermann} N.~A.,   {Kolhatkar}
  S.,  2001, \mn@doi [\apj] {10.1086/319077}, \href
  {https://ui.adsabs.harvard.edu/abs/2001ApJ...549..215D} {549, 215}

\bibitem[\protect\citeauthoryear{Dale, Helou, Magdis, Armus, D{\'{\i}}az-Santos
   \& Shi}{Dale et~al.}{2014}]{Dale2014}
Dale D.~A.,  Helou G.,  Magdis G.~E.,  Armus L.,  D{\'{\i}}az-Santos T.,   Shi
  Y.,  2014, \mn@doi [The Astrophysical Journal] {10.1088/0004-637x/784/1/83},
  784, 83

\bibitem[\protect\citeauthoryear{{Del Moro} et~al.,}{{Del Moro}
  et~al.}{2013}]{DelMoro2013}
{Del Moro} A.,  et~al., 2013, \mn@doi [\aap] {10.1051/0004-6361/201219880},
  \href {https://ui.adsabs.harvard.edu/abs/2013A&A...549A..59D} {549, A59}

\bibitem[\protect\citeauthoryear{{Del Moro} et~al.,}{{Del Moro}
  et~al.}{2016}]{DelMoro2016}
{Del Moro} A.,  et~al., 2016, \mn@doi [\mnras] {10.1093/mnras/stv2748}, \href
  {https://ui.adsabs.harvard.edu/abs/2016MNRAS.456.2105D} {456, 2105}

\bibitem[\protect\citeauthoryear{{Deo}, {Richards}, {Crenshaw}  \&
  {Kraemer}}{{Deo} et~al.}{2009}]{Deo2009}
{Deo} R.~P.,  {Richards} G.~T.,  {Crenshaw} D.~M.,   {Kraemer} S.~B.,  2009,
  \mn@doi [\apj] {10.1088/0004-637X/705/1/14}, \href
  {https://ui.adsabs.harvard.edu/abs/2009ApJ...705...14D} {705, 14}

\bibitem[\protect\citeauthoryear{{D{\'\i}az-Santos} et~al.,}{{D{\'\i}az-Santos}
  et~al.}{2017}]{Diaz2017}
{D{\'\i}az-Santos} T.,  et~al., 2017, \mn@doi [\apj]
  {10.3847/1538-4357/aa81d7}, \href
  {https://ui.adsabs.harvard.edu/abs/2017ApJ...846...32D} {846, 32}

\bibitem[\protect\citeauthoryear{{Dicken}, {Tadhunter}, {Axon}, {Morganti},
  {Inskip}, {Holt}, {Gonz{\'a}lez Delgado}  \& {Groves}}{{Dicken}
  et~al.}{2009}]{Dicken2009}
{Dicken} D.,  {Tadhunter} C.,  {Axon} D.,  {Morganti} R.,  {Inskip} K.~J.,
  {Holt} J.,  {Gonz{\'a}lez Delgado} R.,   {Groves} B.,  2009, \mn@doi [\apj]
  {10.1088/0004-637X/694/1/268}, \href
  {http://adsabs.harvard.edu/abs/2009ApJ...694..268D} {694, 268}

\bibitem[\protect\citeauthoryear{{Dudik}, {Satyapal}  \& {Marcu}}{{Dudik}
  et~al.}{2009}]{Dudik2009}
{Dudik} R.~P.,  {Satyapal} S.,   {Marcu} D.,  2009, \mn@doi [\apj]
  {10.1088/0004-637X/691/2/1501}, \href
  {http://adsabs.harvard.edu/abs/2009ApJ...691.1501D} {691, 1501}

\bibitem[\protect\citeauthoryear{{Elvis} et~al.,}{{Elvis}
  et~al.}{1994}]{Elvis1994}
{Elvis} M.,  et~al., 1994, \mn@doi [\apjs] {10.1086/192093}, \href
  {http://adsabs.harvard.edu/abs/1994ApJS...95....1E} {95, 1}

\bibitem[\protect\citeauthoryear{{Foreman-Mackey}, {Hogg}, {Lang}  \&
  {Goodman}}{{Foreman-Mackey} et~al.}{2013}]{Foreman-Mackey2013}
{Foreman-Mackey} D.,  {Hogg} D.~W.,  {Lang} D.,   {Goodman} J.,  2013, \mn@doi
  [\pasp] {10.1086/670067}, \href
  {http://adsabs.harvard.edu/abs/2013PASP..125..306F} {125, 306}

\bibitem[\protect\citeauthoryear{{Galliano} et~al.,}{{Galliano}
  et~al.}{2011}]{Galliano2011}
{Galliano} F.,  et~al., 2011, \mn@doi [\aap] {10.1051/0004-6361/201117952},
  \href {http://adsabs.harvard.edu/abs/2011A%26A...536A..88G} {536, A88}

\bibitem[\protect\citeauthoryear{{Gehrels} et~al.,}{{Gehrels}
  et~al.}{2004}]{Gehrels2004}
{Gehrels} N.,  et~al., 2004, \mn@doi [\apj] {10.1086/422091}, \href
  {http://adsabs.harvard.edu/abs/2004ApJ...611.1005G} {611, 1005}

\bibitem[\protect\citeauthoryear{Goodman \& Weare}{Goodman \&
  Weare}{2010}]{Goodman2010}
Goodman J.,  Weare J.,  2010, Communications in applied mathematics and
  computational science, 5, 65

\bibitem[\protect\citeauthoryear{{Goulding} \& {Alexander}}{{Goulding} \&
  {Alexander}}{2009}]{Goulding2009}
{Goulding} A.~D.,  {Alexander} D.~M.,  2009, \mn@doi [\mnras]
  {10.1111/j.1365-2966.2009.15194.x}, \href
  {http://adsabs.harvard.edu/abs/2009MNRAS.398.1165G} {398, 1165}

\bibitem[\protect\citeauthoryear{{Goulding}, {Alexander}, {Bauer}, {Forman},
  {Hickox}, {Jones}, {Mullaney}  \& {Trichas}}{{Goulding}
  et~al.}{2012}]{Goulding2012}
{Goulding} A.~D.,  {Alexander} D.~M.,  {Bauer} F.~E.,  {Forman} W.~R.,
  {Hickox} R.~C.,  {Jones} C.,  {Mullaney} J.~R.,   {Trichas} M.,  2012,
  \mn@doi [\apj] {10.1088/0004-637X/755/1/5}, \href
  {https://ui.adsabs.harvard.edu/abs/2012ApJ...755....5G} {755, 5}

\bibitem[\protect\citeauthoryear{{Griffin} et~al.,}{{Griffin}
  et~al.}{2010}]{Griffin2010}
{Griffin} M.~J.,  et~al., 2010, \mn@doi [\aap] {10.1051/0004-6361/201014519},
  \href {http://cdsads.u-strasbg.fr/abs/2010A%26A...518L...3G} {518, L3+}

\bibitem[\protect\citeauthoryear{{Hao} et~al.,}{{Hao} et~al.}{2005}]{Hao2005}
{Hao} L.,  et~al., 2005, \mn@doi [\apjl] {10.1086/431227}, \href
  {https://ui.adsabs.harvard.edu/abs/2005ApJ...625L..75H} {625, L75}

\bibitem[\protect\citeauthoryear{{Hao}, {Weedman}, {Spoon}, {Marshall},
  {Levenson}, {Elitzur}  \& {Houck}}{{Hao} et~al.}{2007}]{Hao2007}
{Hao} L.,  {Weedman} D.~W.,  {Spoon} H.~W.~W.,  {Marshall} J.~A.,  {Levenson}
  N.~A.,  {Elitzur} M.,   {Houck} J.~R.,  2007, \mn@doi [\apjl]
  {10.1086/511973}, \href {http://adsabs.harvard.edu/abs/2007ApJ...655L..77H}
  {655, L77}

\bibitem[\protect\citeauthoryear{{Harrison}}{{Harrison}}{2017}]{Harrison2017}
{Harrison} C.~M.,  2017, \mn@doi [Nature Astronomy] {10.1038/s41550-017-0165},
  \href {http://adsabs.harvard.edu/abs/2017NatAs...1E.165H} {1, 0165}

\bibitem[\protect\citeauthoryear{{Hatziminaoglou}, {Hern{\'a}n-Caballero},
  {Feltre}  \& {Pi{\~n}ol Ferrer}}{{Hatziminaoglou}
  et~al.}{2015}]{Hatziminaoglou2015}
{Hatziminaoglou} E.,  {Hern{\'a}n-Caballero} A.,  {Feltre} A.,   {Pi{\~n}ol
  Ferrer} N.,  2015, \mn@doi [\apj] {10.1088/0004-637X/803/2/110}, \href
  {https://ui.adsabs.harvard.edu/abs/2015ApJ...803..110H} {803, 110}

\bibitem[\protect\citeauthoryear{{Hebbar}, {Heinke}, {Sivakoff}  \&
  {Shaw}}{{Hebbar} et~al.}{2019}]{Hebbar2019}
{Hebbar} P.~R.,  {Heinke} C.~O.,  {Sivakoff} G.~R.,   {Shaw} A.~W.,  2019,
  \mn@doi [\mnras] {10.1093/mnras/stz553}, \href
  {http://adsabs.harvard.edu/abs/2019MNRAS.tmp..547H} {}

\bibitem[\protect\citeauthoryear{{Hinshaw} et~al.,}{{Hinshaw}
  et~al.}{2013}]{Hinshaw2013}
{Hinshaw} G.,  et~al., 2013, \mn@doi [\apjs] {10.1088/0067-0049/208/2/19},
  \href {https://ui.adsabs.harvard.edu/abs/2013ApJS..208...19H} {208, 19}

\bibitem[\protect\citeauthoryear{{Hirashita}, {Deng}  \& {Murga}}{{Hirashita}
  et~al.}{2020}]{Hirashita2020}
{Hirashita} H.,  {Deng} W.,   {Murga} M.~S.,  2020, \mn@doi [\mnras]
  {10.1093/mnras/staa3101}, \href
  {https://ui.adsabs.harvard.edu/abs/2020MNRAS.tmp.2909H} {}

\bibitem[\protect\citeauthoryear{{H{\"o}nig}}{{H{\"o}nig}}{2019}]{Honig2019}
{H{\"o}nig} S.~F.,  2019, \mn@doi [\apj] {10.3847/1538-4357/ab4591}, \href
  {https://ui.adsabs.harvard.edu/abs/2019ApJ...884..171H} {884, 171}

\bibitem[\protect\citeauthoryear{{H{\"o}nig} \& {Kishimoto}}{{H{\"o}nig} \&
  {Kishimoto}}{2010}]{Honig2010}
{H{\"o}nig} S.~F.,  {Kishimoto} M.,  2010, \mn@doi [\aap]
  {10.1051/0004-6361/200912676}, \href
  {https://ui.adsabs.harvard.edu/abs/2010A&A...523A..27H} {523, A27}

\bibitem[\protect\citeauthoryear{{H{\"o}nig}, {Leipski}, {Antonucci}  \&
  {Haas}}{{H{\"o}nig} et~al.}{2011}]{Honig2011}
{H{\"o}nig} S.~F.,  {Leipski} C.,  {Antonucci} R.,   {Haas} M.,  2011, \mn@doi
  [\apj] {10.1088/0004-637X/736/1/26}, \href
  {https://ui.adsabs.harvard.edu/abs/2011ApJ...736...26H} {736, 26}

\bibitem[\protect\citeauthoryear{{H{\"o}nig} et~al.,}{{H{\"o}nig}
  et~al.}{2013}]{Honig2013}
{H{\"o}nig} S.~F.,  et~al., 2013, \mn@doi [\apj] {10.1088/0004-637X/771/2/87},
  \href {http://adsabs.harvard.edu/abs/2013ApJ...771...87H} {771, 87}

\bibitem[\protect\citeauthoryear{{Houck} et~al.,}{{Houck}
  et~al.}{2004}]{Houck2004}
{Houck} J.~R.,  et~al., 2004, \mn@doi [\apjs] {10.1086/423134}, \href
  {http://cdsads.u-strasbg.fr/abs/2004ApJS..154...18H} {154, 18}

\bibitem[\protect\citeauthoryear{{Hunter}}{{Hunter}}{2007}]{Hunter2007}
{Hunter} J.~D.,  2007, \mn@doi [Computing in Science and Engineering]
  {10.1109/MCSE.2007.55}, \href
  {https://ui.adsabs.harvard.edu/abs/2007CSE.....9...90H} {9, 90}

\bibitem[\protect\citeauthoryear{{Johnson}, {Leitherer}, {Vacca}  \&
  {Conti}}{{Johnson} et~al.}{2000}]{Johnson2000}
{Johnson} K.~E.,  {Leitherer} C.,  {Vacca} W.~D.,   {Conti} P.~S.,  2000,
  \mn@doi [\aj] {10.1086/301541}, \href
  {http://adsabs.harvard.edu/abs/2000AJ....120.1273J} {120, 1273}

\bibitem[\protect\citeauthoryear{{Kemper}, {de Koter}, {Waters}, {Bouwman}  \&
  {Tielens}}{{Kemper} et~al.}{2002}]{Kemper2002}
{Kemper} F.,  {de Koter} A.,  {Waters} L.~B.~F.~M.,  {Bouwman} J.,   {Tielens}
  A.~G.~G.~M.,  2002, \mn@doi [\aap] {10.1051/0004-6361:20020036}, \href
  {https://ui.adsabs.harvard.edu/abs/2002A&A...384..585K} {384, 585}

\bibitem[\protect\citeauthoryear{{Kennicutt}}{{Kennicutt}}{1998}]{Kennicutt1998}
{Kennicutt} Jr. R.~C.,  1998, \mn@doi [\apj] {10.1086/305588}, \href
  {http://cdsads.u-strasbg.fr/abs/1998ApJ...498..541K} {498, 541}

\bibitem[\protect\citeauthoryear{{Kr{\"u}gel}}{{Kr{\"u}gel}}{2009}]{Krugel2009}
{Kr{\"u}gel} E.,  2009, \mn@doi [\aap] {10.1051/0004-6361:200809976}, \href
  {https://ui.adsabs.harvard.edu/abs/2009A&A...493..385K} {493, 385}

\bibitem[\protect\citeauthoryear{Lam, Pitrou  \& Seibert}{Lam
  et~al.}{2015}]{numba}
Lam S.~K.,  Pitrou A.,   Seibert S.,  2015, in Proceedings of the Second
  Workshop on the LLVM Compiler Infrastructure in HPC. LLVM ’15.
Association for Computing Machinery, New York, NY, USA,
  \mn@doi{10.1145/2833157.2833162}, \url
  {https://doi.org/10.1145/2833157.2833162}

\bibitem[\protect\citeauthoryear{{Lambrides}, {Petric}, {Tchernyshyov},
  {Zakamska}  \& {Watts}}{{Lambrides} et~al.}{2019}]{Lambrides2019}
{Lambrides} E.~L.,  {Petric} A.~O.,  {Tchernyshyov} K.,  {Zakamska} N.~L.,
  {Watts} D.~J.,  2019, \mn@doi [\mnras] {10.1093/mnras/stz1316}, \href
  {https://ui.adsabs.harvard.edu/abs/2019MNRAS.487.1823L} {487, 1823}

\bibitem[\protect\citeauthoryear{{Lani}, {Netzer}  \& {Lutz}}{{Lani}
  et~al.}{2017}]{Lani2017}
{Lani} C.,  {Netzer} H.,   {Lutz} D.,  2017, \mn@doi [\mnras]
  {10.1093/mnras/stx1374}, \href
  {https://ui.adsabs.harvard.edu/abs/2017MNRAS.471...59L} {471, 59}

\bibitem[\protect\citeauthoryear{{Laurent}, {Mirabel}, {Charmandaris},
  {Gallais}, {Madden}, {Sauvage}, {Vigroux}  \& {Cesarsky}}{{Laurent}
  et~al.}{2000}]{Laurent2000}
{Laurent} O.,  {Mirabel} I.~F.,  {Charmandaris} V.,  {Gallais} P.,  {Madden}
  S.~C.,  {Sauvage} M.,  {Vigroux} L.,   {Cesarsky} C.,  2000, \aap, \href
  {http://adsabs.harvard.edu/abs/2000A%26A...359..887L} {359, 887}

\bibitem[\protect\citeauthoryear{{Lebouteiller}, {Bernard-Salas}, {Sloan}  \&
  {Barry}}{{Lebouteiller} et~al.}{2010}]{Lebouteiller2010}
{Lebouteiller} V.,  {Bernard-Salas} J.,  {Sloan} G.~C.,   {Barry} D.~J.,  2010,
  \mn@doi [\pasp] {10.1086/650426}, \href
  {http://adsabs.harvard.edu/abs/2010PASP..122..231L} {122, 231}

\bibitem[\protect\citeauthoryear{{Lebouteiller}, {Barry}, {Spoon},
  {Bernard-Salas}, {Sloan}, {Houck}  \& {Weedman}}{{Lebouteiller}
  et~al.}{2011}]{Lebouteiller2011}
{Lebouteiller} V.,  {Barry} D.~J.,  {Spoon} H.~W.~W.,  {Bernard-Salas} J.,
  {Sloan} G.~C.,  {Houck} J.~R.,   {Weedman} D.~W.,  2011, \mn@doi [\apjs]
  {10.1088/0067-0049/196/1/8}, \href
  {http://adsabs.harvard.edu/abs/2011ApJS..196....8L} {196, 8}

\bibitem[\protect\citeauthoryear{{Leftley}, {Tristram}, {H{\"o}nig},
  {Kishimoto}, {Asmus}  \& {Gandhi}}{{Leftley} et~al.}{2018}]{Leftley2018}
{Leftley} J.~H.,  {Tristram} K. R.~W.,  {H{\"o}nig} S.~F.,  {Kishimoto} M.,
  {Asmus} D.,   {Gandhi} P.,  2018, \mn@doi [\apj] {10.3847/1538-4357/aac8e5},
  \href {https://ui.adsabs.harvard.edu/abs/2018ApJ...862...17L} {862, 17}

\bibitem[\protect\citeauthoryear{{Li} \& {Draine}}{{Li} \&
  {Draine}}{2001}]{Li2001}
{Li} A.,  {Draine} B.~T.,  2001, \mn@doi [\apj] {10.1086/323147}, \href
  {https://ui.adsabs.harvard.edu/abs/2001ApJ...554..778L} {554, 778}

\bibitem[\protect\citeauthoryear{{Lyu} \& {Rieke}}{{Lyu} \&
  {Rieke}}{2017}]{Lyu2017a}
{Lyu} J.,  {Rieke} G.~H.,  2017, \mn@doi [\apj] {10.3847/1538-4357/aa7051},
  \href {https://ui.adsabs.harvard.edu/abs/2017ApJ...841...76L} {841, 76}

\bibitem[\protect\citeauthoryear{{Lyu} \& {Rieke}}{{Lyu} \&
  {Rieke}}{2018}]{Lyu2018}
{Lyu} J.,  {Rieke} G.~H.,  2018, \mn@doi [\apj] {10.3847/1538-4357/aae075},
  \href {https://ui.adsabs.harvard.edu/abs/2018ApJ...866...92L} {866, 92}

\bibitem[\protect\citeauthoryear{{Lyu}, {Rieke}  \& {Shi}}{{Lyu}
  et~al.}{2017}]{Lyu2017b}
{Lyu} J.,  {Rieke} G.~H.,   {Shi} Y.,  2017, \mn@doi [\apj]
  {10.3847/1538-4357/835/2/257}, \href
  {https://ui.adsabs.harvard.edu/abs/2017ApJ...835..257L} {835, 257}

\bibitem[\protect\citeauthoryear{{Magdis} et~al.,}{{Magdis}
  et~al.}{2013}]{Magdis2013}
{Magdis} G.~E.,  et~al., 2013, \mn@doi [\aap] {10.1051/0004-6361/201322226},
  \href {https://ui.adsabs.harvard.edu/abs/2013A&A...558A.136M} {558, A136}

\bibitem[\protect\citeauthoryear{{Mart{\'\i}nez-Paredes}
  et~al.,}{{Mart{\'\i}nez-Paredes} et~al.}{2020}]{MartinezParedes2020}
{Mart{\'\i}nez-Paredes} M.,  et~al., 2020, \mn@doi [\apj]
  {10.3847/1538-4357/ab6732}, \href
  {https://ui.adsabs.harvard.edu/abs/2020ApJ...890..152M} {890, 152}

\bibitem[\protect\citeauthoryear{{Marton} et~al.,}{{Marton}
  et~al.}{2017}]{Marton2017}
{Marton} G.,  et~al., 2017, arXiv e-prints, \href
  {http://adsabs.harvard.edu/abs/2017arXiv170505693M} {}

\bibitem[\protect\citeauthoryear{McKinney}{McKinney}{2010}]{pandas}
McKinney W.,  2010, in van~der Walt S.,  Millman J.,  eds, Proceedings of the
  9th Python in Science Conference. pp 51 -- 56

\bibitem[\protect\citeauthoryear{{Mel{\'e}ndez}, {Mushotzky}, {Shimizu},
  {Barger}  \& {Cowie}}{{Mel{\'e}ndez} et~al.}{2014}]{Melendez2014}
{Mel{\'e}ndez} M.,  {Mushotzky} R.~F.,  {Shimizu} T.~T.,  {Barger} A.~J.,
  {Cowie} L.~L.,  2014, \mn@doi [\apj] {10.1088/0004-637X/794/2/152}, \href
  {https://ui.adsabs.harvard.edu/abs/2014ApJ...794..152M} {794, 152}

\bibitem[\protect\citeauthoryear{{Messias}, {Afonso}, {Salvato}, {Mobasher}  \&
  {Hopkins}}{{Messias} et~al.}{2012}]{Messias2012}
{Messias} H.,  {Afonso} J.,  {Salvato} M.,  {Mobasher} B.,   {Hopkins} A.~M.,
  2012, \mn@doi [\apj] {10.1088/0004-637X/754/2/120}, \href
  {http://adsabs.harvard.edu/abs/2012ApJ...754..120M} {754, 120}

\bibitem[\protect\citeauthoryear{{Min}, {Waters}, {de Koter}, {Hovenier},
  {Keller}  \& {Markwick-Kemper}}{{Min} et~al.}{2007}]{Min2007}
{Min} M.,  {Waters} L.~B.~F.~M.,  {de Koter} A.,  {Hovenier} J.~W.,  {Keller}
  L.~P.,   {Markwick-Kemper} F.,  2007, \mn@doi [\aap]
  {10.1051/0004-6361:20065436}, \href
  {https://ui.adsabs.harvard.edu/abs/2007A&A...462..667M} {462, 667}

\bibitem[\protect\citeauthoryear{{Mor} \& {Netzer}}{{Mor} \&
  {Netzer}}{2012}]{Mor2012}
{Mor} R.,  {Netzer} H.,  2012, \mn@doi [\mnras]
  {10.1111/j.1365-2966.2011.20060.x}, \href
  {https://ui.adsabs.harvard.edu/abs/2012MNRAS.420..526M} {420, 526}

\bibitem[\protect\citeauthoryear{{Mor}, {Netzer}  \& {Elitzur}}{{Mor}
  et~al.}{2009}]{Mor2009}
{Mor} R.,  {Netzer} H.,   {Elitzur} M.,  2009, \mn@doi [\apj]
  {10.1088/0004-637X/705/1/298}, \href
  {https://ui.adsabs.harvard.edu/abs/2009ApJ...705..298M} {705, 298}

\bibitem[\protect\citeauthoryear{{Mullaney}, {Alexander}, {Goulding}  \&
  {Hickox}}{{Mullaney} et~al.}{2011}]{Mullaney2011}
{Mullaney} J.~R.,  {Alexander} D.~M.,  {Goulding} A.~D.,   {Hickox} R.~C.,
  2011, \mn@doi [\mnras] {10.1111/j.1365-2966.2011.18448.x}, \href
  {http://adsabs.harvard.edu/abs/2011MNRAS.414.1082M} {414, 1082}

\bibitem[\protect\citeauthoryear{{Nenkova}, {Sirocky}, {Nikutta}, {Ivezi{\'c}}
  \& {Elitzur}}{{Nenkova} et~al.}{2008}]{Nenkova2008}
{Nenkova} M.,  {Sirocky} M.~M.,  {Nikutta} R.,  {Ivezi{\'c}} {\v Z}.,
  {Elitzur} M.,  2008, \mn@doi [\apj] {10.1086/590483}, \href
  {http://adsabs.harvard.edu/abs/2008ApJ...685..160N} {685, 160}

\bibitem[\protect\citeauthoryear{{Netzer} et~al.,}{{Netzer}
  et~al.}{2007}]{Netzer2007}
{Netzer} H.,  et~al., 2007, \mn@doi [\apj] {10.1086/520716}, \href
  {http://adsabs.harvard.edu/abs/2007ApJ...666..806N} {666, 806}

\bibitem[\protect\citeauthoryear{{Netzer}, {Lani}, {Nordon}, {Trakhtenbrot},
  {Lira}  \& {Shemmer}}{{Netzer} et~al.}{2016}]{Netzer2016}
{Netzer} H.,  {Lani} C.,  {Nordon} R.,  {Trakhtenbrot} B.,  {Lira} P.,
  {Shemmer} O.,  2016, \mn@doi [\apj] {10.3847/0004-637X/819/2/123}, \href
  {https://ui.adsabs.harvard.edu/abs/2016ApJ...819..123N} {819, 123}

\bibitem[\protect\citeauthoryear{{Noll}, {Burgarella}, {Giovannoli}, {Buat},
  {Marcillac}  \& {Mu{\~n}oz-Mateos}}{{Noll} et~al.}{2009}]{Noll2009}
{Noll} S.,  {Burgarella} D.,  {Giovannoli} E.,  {Buat} V.,  {Marcillac} D.,
  {Mu{\~n}oz-Mateos} J.~C.,  2009, \mn@doi [\aap]
  {10.1051/0004-6361/200912497}, \href
  {http://cdsads.u-strasbg.fr/abs/2009A%26A...507.1793N} {507, 1793}

\bibitem[\protect\citeauthoryear{{Ogawa}, {Ueda}, {Tanimoto}  \&
  {Yamada}}{{Ogawa} et~al.}{2021}]{Ogawa2021}
{Ogawa} S.,  {Ueda} Y.,  {Tanimoto} A.,   {Yamada} S.,  2021, \mn@doi [\apj]
  {10.3847/1538-4357/abccce}, \href
  {https://ui.adsabs.harvard.edu/abs/2021ApJ...906...84O} {906, 84}

\bibitem[\protect\citeauthoryear{{Oh} et~al.,}{{Oh} et~al.}{2018}]{Oh2018}
{Oh} K.,  et~al., 2018, \mn@doi [\apjs] {10.3847/1538-4365/aaa7fd}, \href
  {http://adsabs.harvard.edu/abs/2018ApJS..235....4O} {235, 4}

\bibitem[\protect\citeauthoryear{{Orellana} et~al.,}{{Orellana}
  et~al.}{2017}]{Orellana2017}
{Orellana} G.,  et~al., 2017, \mn@doi [\aap] {10.1051/0004-6361/201629009},
  \href {https://ui.adsabs.harvard.edu/abs/2017A&A...602A..68O} {602, A68}

\bibitem[\protect\citeauthoryear{{Peeters}, {Mattioda}, {Hudgins}  \&
  {Allamandola}}{{Peeters} et~al.}{2004}]{Peeters2004}
{Peeters} E.,  {Mattioda} A.~L.,  {Hudgins} D.~M.,   {Allamandola} L.~J.,
  2004, \mn@doi [\apjl] {10.1086/427186}, \href
  {http://adsabs.harvard.edu/abs/2004ApJ...617L..65P} {617, L65}

\bibitem[\protect\citeauthoryear{{Pilbratt} et~al.,}{{Pilbratt}
  et~al.}{2010}]{Pilbratt2010}
{Pilbratt} G.~L.,  et~al., 2010, \mn@doi [\aap] {10.1051/0004-6361/201014759},
  \href {http://cdsads.u-strasbg.fr/abs/2010A%26A...518L...1P} {518, L1+}

\bibitem[\protect\citeauthoryear{{Poglitsch} et~al.,}{{Poglitsch}
  et~al.}{2010}]{Poglitsch2010}
{Poglitsch} A.,  et~al., 2010, \mn@doi [\aap] {10.1051/0004-6361/201014535},
  \href {http://adsabs.harvard.edu/abs/2010A%26A...518L...2P} {518, L2}

\bibitem[\protect\citeauthoryear{{Prieto}, {Reunanen}, {Tristram}, {Neumayer},
  {Fernandez-Ontiveros}, {Orienti}  \& {Meisenheimer}}{{Prieto}
  et~al.}{2010}]{Prieto2010}
{Prieto} M.~A.,  {Reunanen} J.,  {Tristram} K.~R.~W.,  {Neumayer} N.,
  {Fernandez-Ontiveros} J.~A.,  {Orienti} M.,   {Meisenheimer} K.,  2010,
  \mn@doi [\mnras] {10.1111/j.1365-2966.2009.15897.x}, \href
  {https://ui.adsabs.harvard.edu/abs/2010MNRAS.402..724P} {402, 724}

\bibitem[\protect\citeauthoryear{{Radomski}, {Pi{\~n}a}, {Packham}, {Telesco},
  {De Buizer}, {Fisher}  \& {Robinson}}{{Radomski} et~al.}{2003}]{Radomski2003}
{Radomski} J.~T.,  {Pi{\~n}a} R.~K.,  {Packham} C.,  {Telesco} C.~M.,  {De
  Buizer} J.~M.,  {Fisher} R.~S.,   {Robinson} A.,  2003, \mn@doi [\apj]
  {10.1086/367612}, \href
  {https://ui.adsabs.harvard.edu/abs/2003ApJ...587..117R} {587, 117}

\bibitem[\protect\citeauthoryear{{Reines}, {Sivakoff}, {Johnson}  \&
  {Brogan}}{{Reines} et~al.}{2011}]{Reines2011}
{Reines} A.~E.,  {Sivakoff} G.~R.,  {Johnson} K.~E.,   {Brogan} C.~L.,  2011,
  \mn@doi [\nat] {10.1038/nature09724}, \href
  {http://adsabs.harvard.edu/abs/2011Natur.470...66R} {470, 66}

\bibitem[\protect\citeauthoryear{{Reines}, {Reynolds}, {Miller}, {Sivakoff},
  {Greene}, {Hickox}  \& {Johnson}}{{Reines} et~al.}{2016}]{Reines2016}
{Reines} A.~E.,  {Reynolds} M.~T.,  {Miller} J.~M.,  {Sivakoff} G.~R.,
  {Greene} J.~E.,  {Hickox} R.~C.,   {Johnson} K.~E.,  2016, \mn@doi [\apjl]
  {10.3847/2041-8205/830/2/L35}, \href
  {http://adsabs.harvard.edu/abs/2016ApJ...830L..35R} {830, L35}

\bibitem[\protect\citeauthoryear{Ricci et~al.,}{Ricci
  et~al.}{2017a}]{Ricci2017a}
Ricci C.,  et~al., 2017a, \mn@doi [The Astrophysical Journal Supplement Series]
  {10.3847/1538-4365/aa96ad}, 233, 17

\bibitem[\protect\citeauthoryear{{Ricci} et~al.,}{{Ricci}
  et~al.}{2017b}]{Ricci2017b}
{Ricci} C.,  et~al., 2017b, \mn@doi [\nat] {10.1038/nature23906}, \href
  {https://ui.adsabs.harvard.edu/abs/2017Natur.549..488R} {549, 488}

\bibitem[\protect\citeauthoryear{{Richards} et~al.,}{{Richards}
  et~al.}{2006}]{Richards2006}
{Richards} G.~T.,  et~al., 2006, \mn@doi [\apjs] {10.1086/506525}, \href
  {http://adsabs.harvard.edu/abs/2006ApJS..166..470R} {166, 470}

\bibitem[\protect\citeauthoryear{{Rieke}, {Alonso-Herrero}, {Weiner},
  {P{\'e}rez-Gonz{\'a}lez}, {Blaylock}, {Donley}  \& {Marcillac}}{{Rieke}
  et~al.}{2009}]{Rieke2009}
{Rieke} G.~H.,  {Alonso-Herrero} A.,  {Weiner} B.~J.,  {P{\'e}rez-Gonz{\'a}lez}
  P.~G.,  {Blaylock} M.,  {Donley} J.~L.,   {Marcillac} D.,  2009, \mn@doi
  [\apj] {10.1088/0004-637X/692/1/556}, \href
  {https://ui.adsabs.harvard.edu/abs/2009ApJ...692..556R} {692, 556}

\bibitem[\protect\citeauthoryear{{Riffel} et~al.,}{{Riffel}
  et~al.}{2019}]{Riffel2019}
{Riffel} R.,  et~al., 2019, \mn@doi [\mnras] {10.1093/mnras/stz1077}, \href
  {https://ui.adsabs.harvard.edu/abs/2019MNRAS.486.3228R} {486, 3228}

\bibitem[\protect\citeauthoryear{{Robotham}, {Bellstedt}, {Lagos}, {Thorne},
  {Davies}, {Driver}  \& {Bravo}}{{Robotham} et~al.}{2020}]{Robotham2020}
{Robotham} A.~S.~G.,  {Bellstedt} S.,  {Lagos} C. d.~P.,  {Thorne} J.~E.,
  {Davies} L.~J.,  {Driver} S.~P.,   {Bravo} M.,  2020, \mn@doi [\mnras]
  {10.1093/mnras/staa1116}, \href
  {https://ui.adsabs.harvard.edu/abs/2020MNRAS.495..905R} {495, 905}

\bibitem[\protect\citeauthoryear{{Rosario} et~al.,}{{Rosario}
  et~al.}{2012}]{Rosario2012}
{Rosario} D.~J.,  et~al., 2012, \mn@doi [\aap] {10.1051/0004-6361/201219258},
  \href {http://adsabs.harvard.edu/abs/2012A%26A...545A..45R} {545, A45}

\bibitem[\protect\citeauthoryear{{Sales}, {Pastoriza}  \& {Riffel}}{{Sales}
  et~al.}{2010}]{Sales2010}
{Sales} D.~A.,  {Pastoriza} M.~G.,   {Riffel} R.,  2010, \mn@doi [\apj]
  {10.1088/0004-637X/725/1/605}, \href
  {https://ui.adsabs.harvard.edu/abs/2010ApJ...725..605S} {725, 605}

\bibitem[\protect\citeauthoryear{{Samsonyan}, {Weedman}, {Lebouteiller},
  {Barry}  \& {Sargsyan}}{{Samsonyan} et~al.}{2016}]{Samsonyan2016}
{Samsonyan} A.,  {Weedman} D.,  {Lebouteiller} V.,  {Barry} D.,   {Sargsyan}
  L.,  2016, \mn@doi [\apjs] {10.3847/0067-0049/226/1/11}, \href
  {http://adsabs.harvard.edu/abs/2016ApJS..226...11S} {226, 11}

\bibitem[\protect\citeauthoryear{{Sanders}, {Mazzarella}, {Kim}, {Surace}  \&
  {Soifer}}{{Sanders} et~al.}{2003}]{Sanders2003b}
{Sanders} D.~B.,  {Mazzarella} J.~M.,  {Kim} D.-C.,  {Surace} J.~A.,   {Soifer}
  B.~T.,  2003, \mn@doi [\aj] {10.1086/376841}, \href
  {http://adsabs.harvard.edu/abs/2003AJ....126.1607S} {126, 1607}

\bibitem[\protect\citeauthoryear{{Sargsyan}, {Weedman}, {Lebouteiller},
  {Houck}, {Barry}, {Hovhannisyan}  \& {Mickaelian}}{{Sargsyan}
  et~al.}{2011}]{Sargsyan2011}
{Sargsyan} L.,  {Weedman} D.,  {Lebouteiller} V.,  {Houck} J.,  {Barry} D.,
  {Hovhannisyan} A.,   {Mickaelian} A.,  2011, \mn@doi [\apj]
  {10.1088/0004-637X/730/1/19}, \href
  {http://adsabs.harvard.edu/abs/2011ApJ...730...19S} {730, 19}

\bibitem[\protect\citeauthoryear{{Schreiber} et~al.,}{{Schreiber}
  et~al.}{2015}]{Schreiber2015}
{Schreiber} C.,  et~al., 2015, \mn@doi [\aap] {10.1051/0004-6361/201425017},
  \href {http://adsabs.harvard.edu/abs/2015A%26A...575A..74S} {575, A74}

\bibitem[\protect\citeauthoryear{{Schreiber}, {Elbaz}, {Pannella}, {Ciesla},
  {Wang}  \& {Franco}}{{Schreiber} et~al.}{2018}]{Schreiber2018}
{Schreiber} C.,  {Elbaz} D.,  {Pannella} M.,  {Ciesla} L.,  {Wang} T.,
  {Franco} M.,  2018, \mn@doi [\aap] {10.1051/0004-6361/201731506}, \href
  {http://adsabs.harvard.edu/abs/2018A%26A...609A..30S} {609, A30}

\bibitem[\protect\citeauthoryear{{Schulz} et~al.,}{{Schulz}
  et~al.}{2017}]{Schulz2017}
{Schulz} B.,  et~al., 2017, arXiv e-prints, \href
  {http://adsabs.harvard.edu/abs/2017arXiv170600448S} {}

\bibitem[\protect\citeauthoryear{{Schulze} et~al.,}{{Schulze}
  et~al.}{2019}]{Schulze2019}
{Schulze} A.,  et~al., 2019, \mn@doi [\mnras] {10.1093/mnras/stz1746}, \href
  {https://ui.adsabs.harvard.edu/abs/2019MNRAS.488.1180S} {488, 1180}

\bibitem[\protect\citeauthoryear{{Schweitzer} et~al.,}{{Schweitzer}
  et~al.}{2008}]{Schweitzer2008}
{Schweitzer} M.,  et~al., 2008, \mn@doi [\apj] {10.1086/587097}, \href
  {https://ui.adsabs.harvard.edu/abs/2008ApJ...679..101S} {679, 101}

\bibitem[\protect\citeauthoryear{{Shimizu}, {Mel{\'e}ndez}, {Mushotzky},
  {Koss}, {Barger}  \& {Cowie}}{{Shimizu} et~al.}{2016}]{Shimizu2016}
{Shimizu} T.~T.,  {Mel{\'e}ndez} M.,  {Mushotzky} R.~F.,  {Koss} M.~J.,
  {Barger} A.~J.,   {Cowie} L.~L.,  2016, \mn@doi [\mnras]
  {10.1093/mnras/stv2828}, \href
  {https://ui.adsabs.harvard.edu/abs/2016MNRAS.456.3335S} {456, 3335}

\bibitem[\protect\citeauthoryear{{Siebenmorgen}, {Haas}, {Kruegel}  \&
  {Schulz}}{{Siebenmorgen} et~al.}{2005}]{Siebenmorgen2005}
{Siebenmorgen} R.,  {Haas} M.,  {Kruegel} E.,   {Schulz} B.,  2005,
  Astronomische Nachrichten, \href
  {https://ui.adsabs.harvard.edu/abs/2005AN....326R.556S} {326, 556}

\bibitem[\protect\citeauthoryear{{Siebenmorgen}, {Heymann}  \&
  {Efstathiou}}{{Siebenmorgen} et~al.}{2015}]{Siebenmorgen2015}
{Siebenmorgen} R.,  {Heymann} F.,   {Efstathiou} A.,  2015, \mn@doi [\aap]
  {10.1051/0004-6361/201526034}, \href
  {https://ui.adsabs.harvard.edu/abs/2015A&A...583A.120S} {583, A120}

\bibitem[\protect\citeauthoryear{{Smail}, {Swinbank}, {Ivison}  \&
  {Ibar}}{{Smail} et~al.}{2011}]{Smail2011}
{Smail} I.,  {Swinbank} A.~M.,  {Ivison} R.~J.,   {Ibar} E.,  2011, \mn@doi
  [\mnras] {10.1111/j.1745-3933.2011.01064.x}, \href
  {http://adsabs.harvard.edu/abs/2011MNRAS.414L..95S} {414, L95}

\bibitem[\protect\citeauthoryear{{Smith} et~al.,}{{Smith}
  et~al.}{2007}]{Smith2007}
{Smith} J.~D.~T.,  et~al., 2007, \mn@doi [\apj] {10.1086/510549}, \href
  {http://adsabs.harvard.edu/abs/2007ApJ...656..770S} {656, 770}

\bibitem[\protect\citeauthoryear{{Spoon}, {Marshall}, {Houck}, {Elitzur},
  {Hao}, {Armus}, {Brandl}  \& {Charmandaris}}{{Spoon}
  et~al.}{2007}]{Spoon2007}
{Spoon} H.~W.~W.,  {Marshall} J.~A.,  {Houck} J.~R.,  {Elitzur} M.,  {Hao} L.,
  {Armus} L.,  {Brandl} B.~R.,   {Charmandaris} V.,  2007, \mn@doi [\apjl]
  {10.1086/511268}, \href {http://adsabs.harvard.edu/abs/2007ApJ...654L..49S}
  {654, L49}

\bibitem[\protect\citeauthoryear{{Stanley}, {Harrison}, {Alexander}, {Simpson},
  {Knudsen}, {Mullaney}, {Rosario}  \& {Scholtz}}{{Stanley}
  et~al.}{2018}]{Stanley2018}
{Stanley} F.,  {Harrison} C.~M.,  {Alexander} D.~M.,  {Simpson} J.,  {Knudsen}
  K.~K.,  {Mullaney} J.~R.,  {Rosario} D.~J.,   {Scholtz} J.,  2018, \mn@doi
  [\mnras] {10.1093/mnras/sty1044}, \href
  {https://ui.adsabs.harvard.edu/abs/2018MNRAS.478.3721S} {478, 3721}

\bibitem[\protect\citeauthoryear{{Stierwalt} et~al.,}{{Stierwalt}
  et~al.}{2014}]{Stierwalt2014}
{Stierwalt} S.,  et~al., 2014, \mn@doi [\apj] {10.1088/0004-637X/790/2/124},
  \href {http://adsabs.harvard.edu/abs/2014ApJ...790..124S} {790, 124}

\bibitem[\protect\citeauthoryear{{Sturm} et~al.,}{{Sturm}
  et~al.}{2005}]{Sturm2005}
{Sturm} E.,  et~al., 2005, \mn@doi [\apjl] {10.1086/444359}, \href
  {http://adsabs.harvard.edu/abs/2005ApJ...629L..21S} {629, L21}

\bibitem[\protect\citeauthoryear{{Su}, {Irwin}, {White}  \& {Cooper}}{{Su}
  et~al.}{2015}]{Su2015}
{Su} Y.,  {Irwin} J.~A.,  {White} Raymond~E. I.,   {Cooper} M.~C.,  2015,
  \mn@doi [\apj] {10.1088/0004-637X/806/2/156}, \href
  {https://ui.adsabs.harvard.edu/abs/2015ApJ...806..156S} {806, 156}

\bibitem[\protect\citeauthoryear{{Symeonidis}, {Giblin}, {Page}, {Pearson},
  {Bendo}, {Seymour}  \& {Oliver}}{{Symeonidis} et~al.}{2016}]{Symeonidis2016}
{Symeonidis} M.,  {Giblin} B.~M.,  {Page} M.~J.,  {Pearson} C.,  {Bendo} G.,
  {Seymour} N.,   {Oliver} S.~J.,  2016, \mn@doi [\mnras]
  {10.1093/mnras/stw667}, \href
  {http://adsabs.harvard.edu/abs/2016MNRAS.459..257S} {459, 257}

\bibitem[\protect\citeauthoryear{{Tadhunter} et~al.,}{{Tadhunter}
  et~al.}{2007}]{Tadhunter2007}
{Tadhunter} C.,  et~al., 2007, \mn@doi [\apjl] {10.1086/518421}, \href
  {http://adsabs.harvard.edu/abs/2007ApJ...661L..13T} {661, L13}

\bibitem[\protect\citeauthoryear{{Urry} \& {Padovani}}{{Urry} \&
  {Padovani}}{1995}]{Urry1995}
{Urry} C.~M.,  {Padovani} P.,  1995, \mn@doi [\pasp] {10.1086/133630}, \href
  {http://adsabs.harvard.edu/abs/1995PASP..107..803U} {107, 803}

\bibitem[\protect\citeauthoryear{{Vasudevan} \& {Fabian}}{{Vasudevan} \&
  {Fabian}}{2007}]{Vasudevan2007}
{Vasudevan} R.~V.,  {Fabian} A.~C.,  2007, \mn@doi [\mnras]
  {10.1111/j.1365-2966.2007.12328.x}, \href
  {http://adsabs.harvard.edu/abs/2007MNRAS.381.1235V} {381, 1235}

\bibitem[\protect\citeauthoryear{{Vika}, {Ciesla}, {Charmandaris}, {Xilouris}
  \& {Lebouteiller}}{{Vika} et~al.}{2017}]{Vika2017}
{Vika} M.,  {Ciesla} L.,  {Charmandaris} V.,  {Xilouris} E.~M.,
  {Lebouteiller} V.,  2017, \mn@doi [\aap] {10.1051/0004-6361/201629031}, \href
  {https://ui.adsabs.harvard.edu/abs/2017A&A...597A..51V} {597, A51}

\bibitem[\protect\citeauthoryear{{Virtanen} et~al.,}{{Virtanen}
  et~al.}{2020}]{scipy}
{Virtanen} P.,  et~al., 2020, \mn@doi [Nature Methods]
  {https://doi.org/10.1038/s41592-019-0686-2}, \href {https://rdcu.be/b08Wh} {}

\bibitem[\protect\citeauthoryear{{Wenger} et~al.,}{{Wenger}
  et~al.}{2000}]{Wenger2000}
{Wenger} M.,  et~al., 2000, \mn@doi [\aaps] {10.1051/aas:2000332}, \href
  {https://ui.adsabs.harvard.edu/abs/2000A&AS..143....9W} {143, 9}

\bibitem[\protect\citeauthoryear{Xie, Ho, Li  \& Shangguan}{Xie
  et~al.}{2018}]{Xie2018}
Xie Y.,  Ho L.~C.,  Li A.,   Shangguan J.,  2018, \mn@doi [The Astrophysical
  Journal] {10.3847/1538-4357/aac3dc}, 860, 154

\bibitem[\protect\citeauthoryear{{Xu}, {Rieke}, {Egami}, {Pereira}, {Haines}
  \& {Smith}}{{Xu} et~al.}{2015}]{Xu2015}
{Xu} L.,  {Rieke} G.~H.,  {Egami} E.,  {Pereira} M.~J.,  {Haines} C.~P.,
  {Smith} G.~P.,  2015, \mn@doi [\apjs] {10.1088/0067-0049/219/2/18}, \href
  {https://ui.adsabs.harvard.edu/abs/2015ApJS..219...18X} {219, 18}

\bibitem[\protect\citeauthoryear{{Xu}, {Sun}  \& {Xue}}{{Xu}
  et~al.}{2020}]{Xu2020}
{Xu} J.,  {Sun} M.,   {Xue} Y.,  2020, arXiv e-prints, \href
  {https://ui.adsabs.harvard.edu/abs/2020arXiv200310078X} {p. arXiv:2003.10078}

\bibitem[\protect\citeauthoryear{{Yang} et~al.,}{{Yang}
  et~al.}{2020}]{Yang2020}
{Yang} G.,  et~al., 2020, \mn@doi [\mnras] {10.1093/mnras/stz3001}, \href
  {https://ui.adsabs.harvard.edu/abs/2020MNRAS.491..740Y} {491, 740}

\makeatother
\end{thebibliography}




\appendix
\section{Calculating IR fluxes for spatially extended sources, and comparison with literature}
\label{app:IRflux}

To calculate IR fluxes we downloaded observation plates from the ``{\it Herschel} High Level Images'' products found in the IRSA database which contain a large collection of {\it Herschel} observations. We only used the data product levels 2.0 or above (maximum of 3.0) which correspond to levels on which scientific analysis can be performed. We used pointed observations when available, and survey data otherwise. While the maps for PACS (i.e. 70~\mum, 100~\mum, and 160~\mum) are provided in Janskys (Jy) per pixel, those for SPIRE (i.e. 250~\mum, 350~\mum, and 500~\mum) are given in Jy per beam. Therefore, we first converted the SPIRE maps from Jy per beam to Jy per pixel, using the size of the beams (469.35~arsec$^2$, 831.27~arsec$^2$, and 1804.31~arsec$^2$), and the size of the pixels (6~\arcsec$\times$6~\arcsec, 10~\arcsec$\times$10~\arcsec, and 14~\arcsec$\times$14~\arcsec) at each of the SPIRE wavelengths (250~\mum, 350~\mum, and 500~\mum), as reported in the {\it Herschel} observer manual.

Once the images were converted to Jy per pixel, we used the Python package Photutils \citep{Bradley2020} to analyse the images and to extract fluxes and their uncertainties. First, a median background map was created using sigma-clip statistics, after degrading the image with a five-by-five Gaussian kernel. Using this background map, we calculated the detection threshold of the image, using a one sigma limit. We then created a segmentation map (i.e. map flagging pixels above a certain value), using the detection threshold, and after degrading the image using a three-by-three Gaussian kernel. To identify the detections in survey images, we used the known position of the source. The extent of the source was calculated on the segmentation map by Photutils, and characterised by a semi-major and a semi-minor axis (1$\sigma$).

To extract the fluxes, we used an elliptical aperture on each cut-out, the size of which corresponded to three times the size of the corresponding semi-axes. We then summed the pixels within the elliptical aperture, and removed the median background to obtain the flux of the source in Jy. The uncertainties were estimated in an elliptical annulus where the inner and the outer edges were 1.5 and 1.7 times the elliptical aperture used to extract the flux, respectively. We added in quadrature an extra 3~per~cent of the flux to the uncertainties, as we found that the true edges of extended sources were challenging to define, and a change of aperture to include more extended flux corresponded to a change in flux of roughly 3~per~cent.

\begin{figure}
\begin{subfigure}[t]{0.5\textwidth}
\centering
\includegraphics[width=\textwidth]{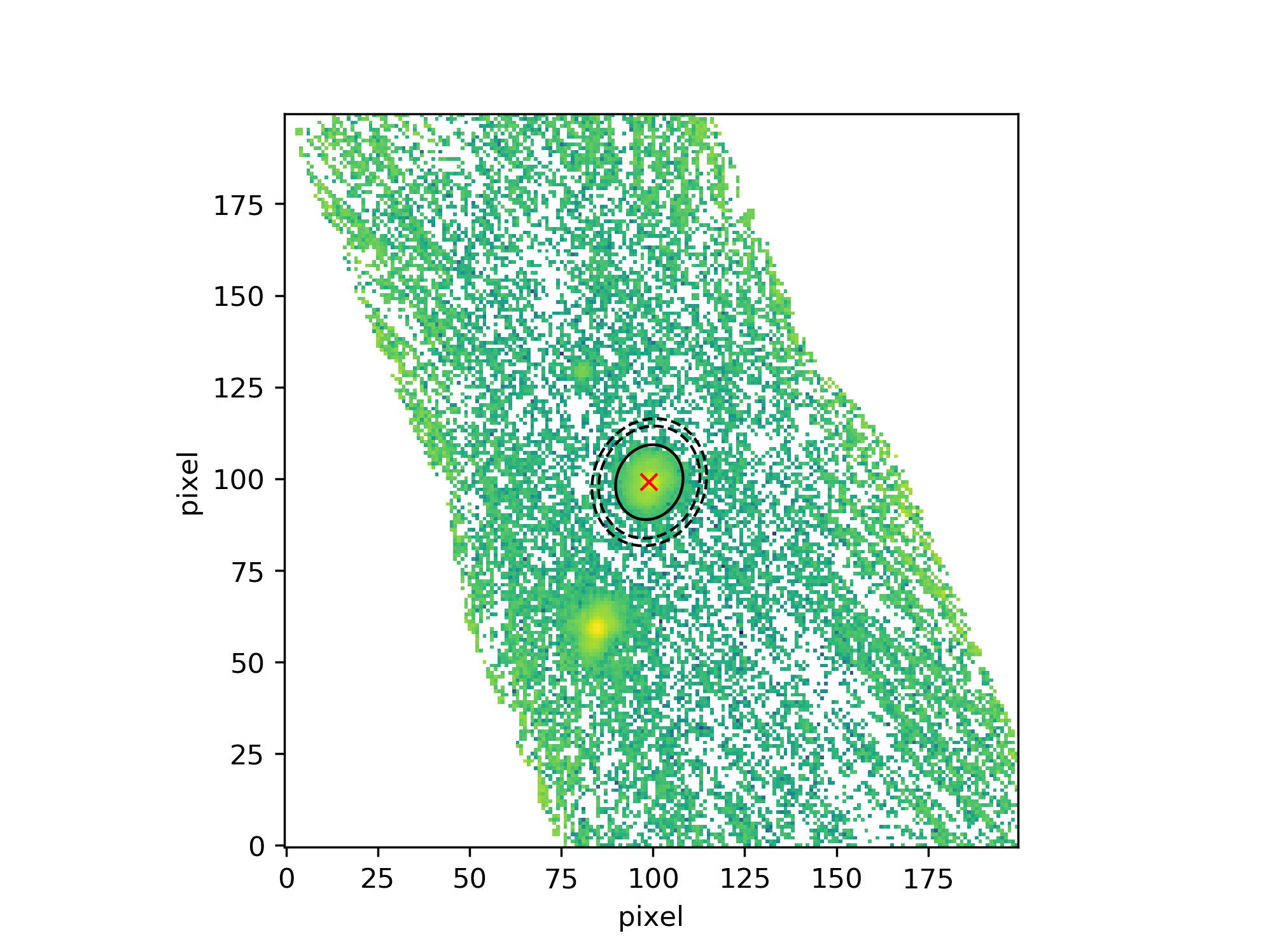}
\caption{UGC 7064 at 70\mum.}
\end{subfigure}

\begin{subfigure}[t]{0.5\textwidth}
\centering
\includegraphics[width=\textwidth]{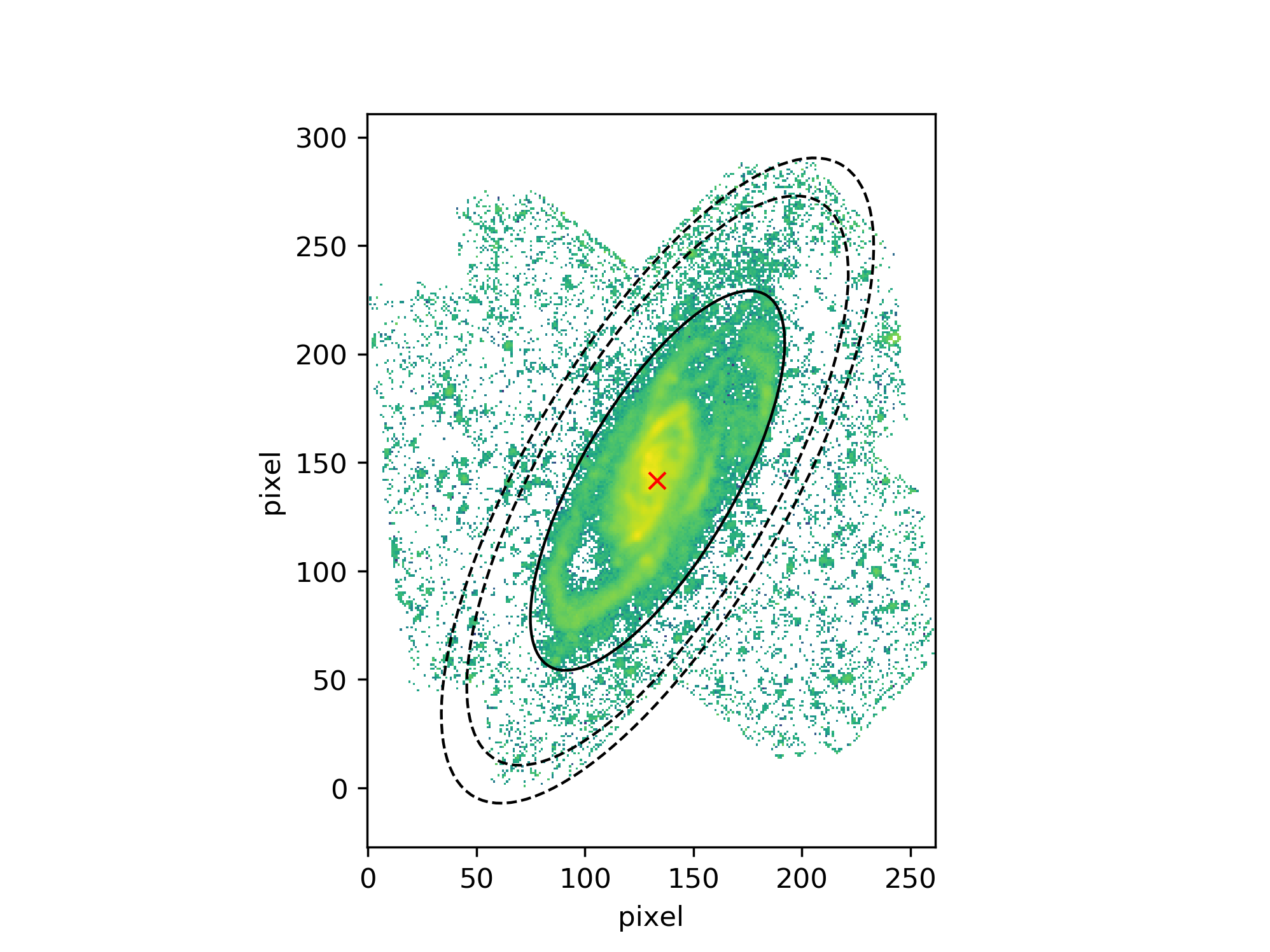} 
\caption{NGC 4258 at 250\mum.}
\end{subfigure}
\caption{Two examples of extended sources found in our AGN sample. The panel (a) shows an example of a barely extended source, and the panel (b) shows a fully extended source where features like spiral arms are visible. Both these sources were classified as point sources in the HPDPs. The inner solid ellipse represents the aperture that were used to extract the flux, and the dashed elliptical annulus shows that of the uncertainties. The red cross corresponds to the centre of the ellipses.} \label{Fig1appAperture}
\end{figure}

We show in Fig.\,\ref{Fig1appAperture} two examples of extended sources found in our AGN sample which were classified as point sources in the HPDPs. We also show the apertures used to extract the fluxes and their uncertainties. The top panel in Fig.\,\ref{Fig1appAperture} shows an example of a source that is barely extended, and for which our automated aperture flux extraction seemed to capture the full extent of the source. Instead, the bottom panel in Fig.\,\ref{Fig1appAperture} shows a source that is fully extended and for which we had to increase the size of the apertures to match the physical extent of the source. The fraction of the elliptical annulus that fell outside of the frame was ignored in the calculation of the uncertainties. In total, we had to modify the size of the aperture for roughly 30~per~cent of our sources to ensure that all our apertures encompassed the totality of the source, and that no extended flux was missed. The full sets of images for our extended sources are available in the supplementary material, or directly at \url{https://tinyurl.com/yazhbx7p} and \url{https://tinyurl.com/yde5eptk} for the AGN and the star-forming galaxy samples, respectively. 

\begin{figure*}
 \centering
 \includegraphics[width=\textwidth]{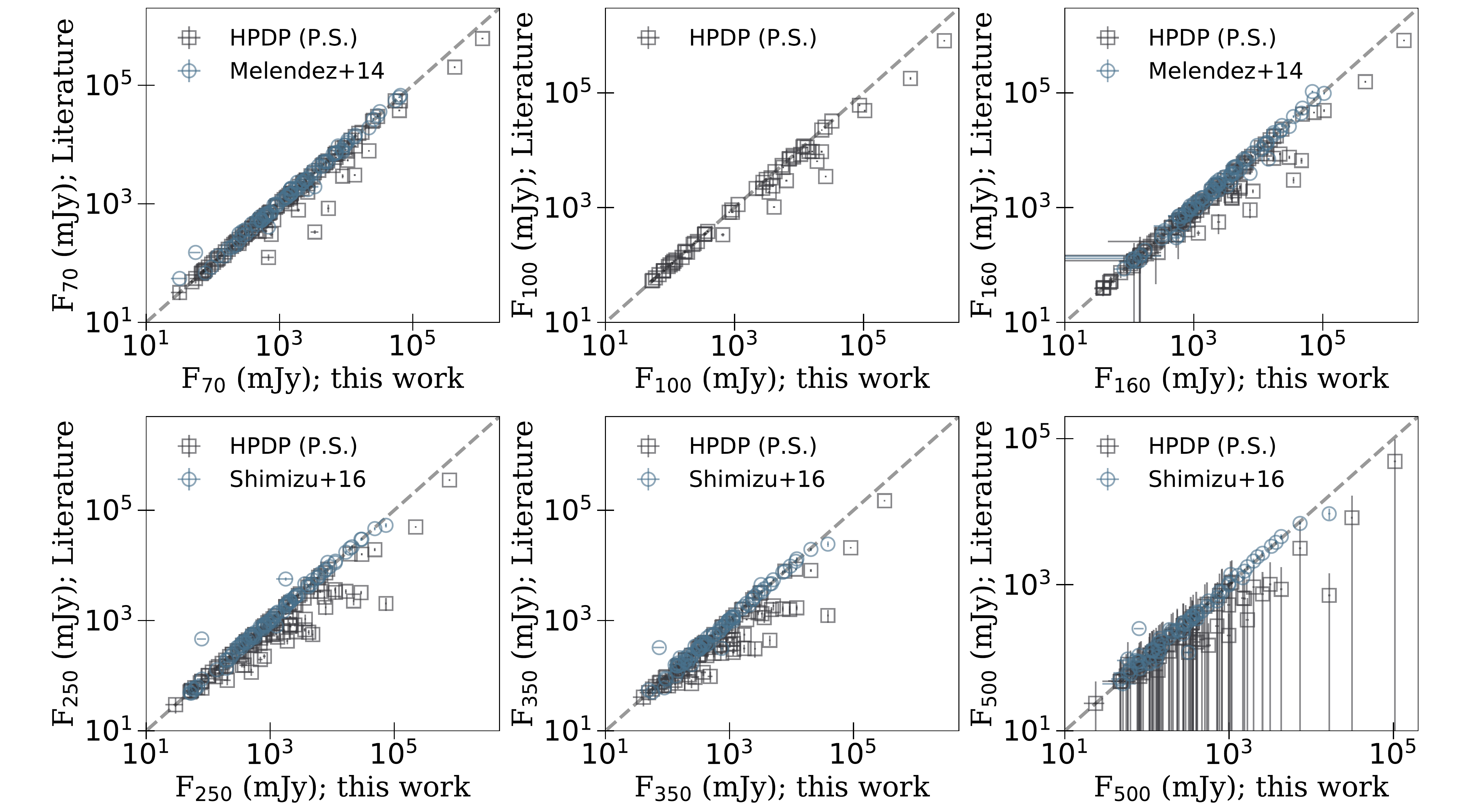}
 \caption{The fluxes used in this work, corresponding to the HPDP fluxes for point sources and to our own fluxes for spatially extended sources, against the HPDP fluxes (open grey squares), and the fluxes from \protect\cite{Melendez2014} for PACS at 70~\mum\ and 160~\mum, and from \protect\cite{Shimizu2016} for SPIRE (open blue circles). Each panel corresponds to an {\it Herschel} wavelength, from 70~\mum\ to 500~\mum, and increases in reading order. The dashed grey line in each of the panels shows the 1:1 relationship.
\label{fig:fluxCompAGN}}
\end{figure*}

Overall, we find that 30~per~cent, 20~per~cent, 55~per~cent, 62~per~cent, 58~per~cent, and 45~per~cent of our sources have been misclassified as point sources in the HPDPs at 70~\mum, 100~\mum, 160~\mum, 250~\mum, 350~\mum, and 500~\mum, respectively, as shown in Fig.\,\ref{fig:fluxCompAGN} with open grey squares. We note that this mostly affects bright sources (likely to be at lower redshifts) at SPIRE wavelengths. In addition, a vast majority of sources that were undetected at 500~\mum\ in the HPDP are now detected due to the larger aperture used to calculate the fluxes. In addition, we find that all of our {\it Herschel} fluxes are fully consistent with those overlapping with \cite{Melendez2014} for PACS at 70~\mum\ and 160~\mum, and from \cite{Shimizu2016} for SPIRE, where extended sources were correctly accounted for (see Fig.\,\ref{fig:fluxCompAGN}).

\section{Detailed analysis of the failed fits}
\label{app:AGNinter}

\begin{figure*}
 \centering
 \includegraphics[width=\textwidth]{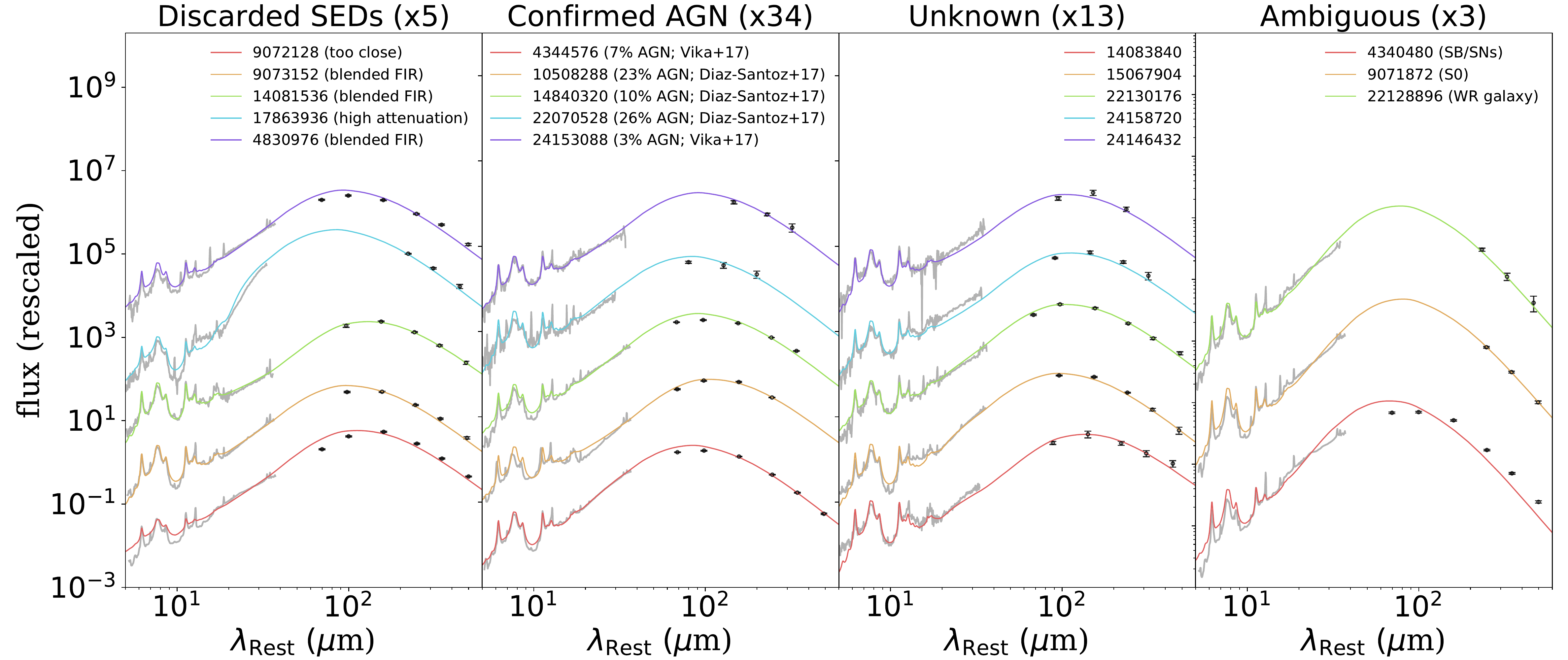}
 \caption{The SEDs that could not be fit by using the galaxy model of S18. We separated the full set of failed SEDs into four bins (see panels), as indicated at the top of each panel. The ``discarded SEDs'' were SEDs for which we found intrinsic issues (i.e. blended photometry). The ``confirmed AGNs'' were found to have IR contamination in some previous studies. The ``unknown'' are SEDs for which we did not find any information regarding the AGN/galaxy contribution. Finally, the ``ambiguous'' are SEDs for which we found no evidence for AGN contribution in the IR based on previous work. The number next to the name of the bin indicates the total number of sources in the bin.
\label{fig:AGNinterlop}}
\end{figure*}

While attempting to fit the IR SEDs of our pre-selected star-forming galaxies (based on \eqwspd and WISE colours), we found a large fraction (i.e. 50~per~cent) for which the galaxy model of S18, split in terms of PAHs and dust continuum, failed. For each of these sources, we carefully inspected their SEDs and searched the NED database to find evidence of misclassification.

First, we directly discarded IC 342 (aorkey~=~9072128) which is a well known star-forming galaxy, yet its redshift of \z~=~0.0001 and spatial extent are such that, at the resolution of the {\it Spitzer}-IRS spectra, only the central starburst was observed, and a simple re-scaling was not sufficient. We then also directly discarded NGC4676 (aorkey~=~9073152), NGC1141 (aorkey~=~4830976), and MCG+06-32-058 (aorkey~=~14081536) as their {\it Herschel} fluxes were enhanced by blended companions which could not be easily de-blended (the images can be found in the supplementary material and at \url{https://tinyurl.com/yde5eptk}). Finally, we directly discarded WISEAJ000024.24-525031.0 (aorkey~=~17863936) as its large attenuation is such that our prescriptions were not able to recover the observed fluxes. These five SEDs and their fits are shown in the leftmost panel of Fig.\,\ref{fig:AGNinterlop}.

Second, we found 34 galaxies with potential AGN contamination in the IR. In particular, we considered the work of \cite{Vika2017}, where the SED fitting code CIGALE \citep{Burgarella2005, Noll2009} was used on photometry spanning the UV to IR (up to 22\mum), including an AGN contribution. In addition, we used the work from \cite{Diaz2017} who combined up to five IR diagnostics, including emission line diagnostics, \eqwspd, MIR flux ratios, and the diagram from \cite{Laurent2000}, to calculate an average IR AGN contamination. Finally, we also used AGN fractions reported in \cite{Magdis2013} where DECOMPIR \citep{Mullaney2011} was used to decompose the IR SEDs of galaxies, including an AGN contribution. Although we found a large number of galaxies that could not be fit with S18 to have AGN IR contribution in the aforementioned studies, the fractions were generally low (i.e. typically less than 10--20~per~cent) and ambiguous. For instance, WISEAJ103450.50+584418.2 (aorkey~=~24153088) was found to have 2.1$\pm$3.5~per~cent or 64$\pm$10~per~cent AGN IR contributions in \cite{Vika2017} or \cite{Magdis2013}, respectively. This shows the difficulties of accurately measuring the AGN fraction at IR wavelengths, as well as the importance of distinguishing the AGN component from that of the galaxy, even when the AGN contributes at less than $\sim$10~per~cent of the IR.

We note that these are mainly unveiled via SED fits which, in turn, depend upon the accuracy of the star-forming galaxy templates used. However, by inspecting the SEDs, we noted the presence of multiple faint high-ionisation emission lines, such as [NeV] at 24~\mum\ and [OIV] at 26~\mum, in half of these galaxies, with a confirmed [NeV]-detected galaxy (i.e. ESO593-IG008 or aorkey~=~20311296) in \cite{Dudik2009}. Although this strongly suggests that these are genuine IR AGNs, we cannot exclude the possibility that some SEDs are less well fit due the spatial extent of the source and the resolution of the {\it Spitzer}-IRS instrument. The second leftmost panel of Fig.\,\ref{fig:AGNinterlop} shows five random examples amongst our 34 galaxies with potential IR AGN contribution. The full set of 34 individual SEDs can be found in the supplementary material, or directly at \url{https://tinyurl.com/ybu9jegl}.

Third, we found 13 SEDs for which we could not find any references regarding an AGN (or lack of) at MIR wavelengths. For these we matched the SEDs with those from our confirmed AGNs, after normalising both at 15~\mum. We found that, for each of these SEDs, we could find an identical matching MIR SED with confirmed AGN contamination, and which was differing in the FIR. Therefore, these are likely to also be contaminated by AGNs, and were discarded. The second to last panel of Fig.\,\ref{fig:AGNinterlop} show an example of five of these sources. The entire set of 13 SEDs can be found in the supplementary material, or directly at \url{https://tinyurl.com/ybu9jegl}.

Finally, we were left with three sources for which we did not find any evidence of any AGN contribution, yet the model of S18 could not fit their SEDs. Firstly, Henize 2-10 (aorkey~=~4340480; He2-10) is a remarkably compact starburst galaxy containing an abundant number of young super star clusters \citep[e.g.][]{Johnson2000}. While \cite{Reines2011} speculated the presence of an AGN at the centre of He2-10 by finding an X-ray point source that coincides with a steep-spectrum radio source, further X-ray observations by \cite{Reines2016} revealed the presence of two sources, one coincident with the radio emission (with an X-ray luminosity of \Lx$~\sim~10^{38}$~\ergps), and believed to be an AGN, and the other one showing no radio emission, and believed to be an X-ray binary. However, \cite{Cresci2017} found that the optical emission line diagnostics at the centre of He2-10 was fully consistent with star formation, and showed that the X-ray luminosity found previously was consistent with emission from supernovae. Finally, \cite{Hebbar2019} have recently found that SN models of X-ray emission explain much better the observed X-ray spectrum of He2-10 than an AGN power-law. 

We then found that NGC1222 (aorkey~=~9071872) has a matching SED to He2-10, suggesting the presence of a starburst. In fact, NGC1222 is reported to be a lenticular S0 galaxy \citep{Riffel2019}, and was undetected in X-rays \citep{Su2015}. In addition, we found that WISEAJ155616.05+395137.8 (aorkey~=~22128896) also has an SED which is identical to He2-10 and NGC1222, yet is reported as a Wolf-Rayet galaxy \citep{Chen2018}, and is therefore consistent with IR powered by star formation. Although these three galaxies were found to be consistent with starburst galaxies, we discarded them from our star-forming galaxy sample as (1) their SEDs were very similar to that of galaxies for which the MIR is contaminated by AGNs, and (2) because of their peculiar nature. The rightmost panel of Fig.\,\ref{fig:AGNinterlop} show the SEDs of these three galaxies.

\section{Tables}
\onecolumn

\begin{ThreePartTable}
\begin{longtable}{ccccccccc}
\caption {The properties of our sample of 112 AGNs selected from the 105-month {\it Swift}–BAT all-sky hard X-ray survey of \protect\cite{Oh2018} (see \S\,\ref{sec:sample}). All of the columns are from \protect\cite{Oh2018}, aside for the column densities of N$_{\rm H}$, which are from \protect\cite{Ricci2017a}. The positions are those of the counterparts. A machine-readable version of this table is available in the supplementary material, or directly at \url{https://tinyurl.com/yah2os8v}.}\label{table:AGNprop}\\\toprule
\endfirsthead
\caption*{Table \ref{table:AGNprop}: $continued.$}\\\toprule
\endhead
\endfoot
\bottomrule
\endlastfoot
ID  & {\it Swift}-BAT Name  & Counterpart Name    & R.A. (h:m:s) & Dec. (d:m:s)  & \z      & log(\Lx~\ergps) & Type  & $\log$(N$_{\rm H}$~cm$^{-2}$)  \\
\hline
1   & J0900.9+2095  & 2MASXJ09003684+2053402  & 09:00:36.864 & +20:53:40.560 & 0.235 & 45.1 & Sy2   &  --   \\ 
2   & J0508.1+1727  & 2MASXJ05081967+1721483  & 05:08:19.680 & +17:21:46.800 & 0.017 & 43.2 & Sy1.9 & 22.33 \\ 
3   & J0516.2-0009  & Ark120                  & 05:16:11.424 & -00:08:59.280 & 0.032 & 44.2 & Sy1   & 20.0  \\ 
4   & J0138.6-4001  & ESO297-18               & 01:38:37.152 & -40:00:41.040 & 0.025 & 43.9 & Sy2   & 23.83 \\ 
5   & J1306.4-4025A & ESO323-77               & 13:06:26.136 & -40:24:52.560 & 0.015 & 43.1 & Sy1.5 & 22.81 \\ 
6   & J0623.8-3215  & ESO426-2                & 06:23:46.416 & -32:12:59.760 & 0.022 & 43.4 & Sy2   & 23.95 \\ 
7   & J0601.9-8636  & ESO5-4                  & 06:05:41.640 & -86:37:54.840 & 0.006 & 42.4 & Sy2   & 24.18 \\
\end{longtable}
\end{ThreePartTable}

\begin{landscape}
\begin{ThreePartTable}
\begin{longtable}{cccccccccccccccccc}
\caption {The {\it Herschel} fluxes of our sample of 112 AGNs. A star is used to flag the fluxes of the two sources with potential IR blending. The symbol \dag\ is used to flag sources with potential issues in the FIR fluxes, as found to give inconsistent SEDs. A machine-readable version of this table is available in the supplementary material, or directly at \url{https://tinyurl.com/yah2os8v}.}\label{table:AGNFlux}\\\toprule
\endfirsthead
\caption* {Table \ref{table:AGNFlux}: $continued.$}\\\toprule
\endhead
\endfoot
\bottomrule
\endlastfoot
ID$^{\rm a}$  & F70 (mJy) & Flag$_{70}$~$^{\rm b}$ & F100 (mJy) & Flag$_{100}$ & F160 (mJy) & Flag$_{160}$ & F250 (mJy) & Flag$_{250}$ & F350 (mJy) & Flag$_{350}$ & F500 (mJy) & Flag$_{500}$ \\
\hline
1   & 71.62     $^{\pm 3.79           }$ & P.S.     & 53.98      $^{\pm 4.19            }$ & P.S.      & 39.61      $^{\pm 10.07          }$ & P.S.      & 0.0                                 & N/A       & 0.0                                 & N/A       & 0.0                                 & N/A       \\
2   & 2265.97   $^{\pm 157.88         }$ & P.S.     & 3105.46    $^{\pm 221.24          }$ & P.S.      & 3313.94    $^{\pm 127.05         }$ & P.S.      & 0.0                                 & N/A       & 0.0                                 & N/A       & 0.0                                 & N/A       \\
3   & 627.83    $^{\pm 16.89          }$ & P.S.     & 0.0                                  & N/A       & 1012.0     $^{\pm 194.93         }$ & P.S.      & 565.99     $^{\pm 28.3           }$ & Ext.      & 315.76     $^{\pm 30.98          }$ & Ext.      & 108.7      $^{\pm 13.9           }$ & P.S.      \\
4   & 668.01    $^{\pm 37.27          }$ & Ext.     & 0.0                                  & N/A       & 1939.47    $^{\pm 79.6           }$ & Ext.      & 1280.13    $^{\pm 45.99          }$ & Ext.      & 680.99     $^{\pm 37.55          }$ & Ext.      & 232.99     $^{\pm 20.87          }$ & Ext.      \\
5   & 6814.31   $^{\pm 127.0          }$ & P.S.     & 0.0                                  & N/A       & 7087.19    $^{\pm 244.47         }$ & P.S.      & 2857.6     $^{\pm 45.0           }$ & P.S.      & 1135.6     $^{\pm 42.94          }$ & Ext.      & 398.1      $^{\pm 25.5           }$ & P.S.      \\
\end{longtable}
\begin{tablenotes}
\footnotesize
\item [a] ID of the source as in Table\,\ref{table:AGNprop}.
\item [b] Indicates whether it is a point source (P.S.) or a spatially extended source (Ext.) at a given wavelength. The fluxes for point sources are from the HPDPs, and the fluxes for spatially extended sources have been re-calculated to include the extended flux (see Appendix\,\ref{app:IRflux}).
\end{tablenotes}
\end{ThreePartTable}
\end{landscape}

\begin{ThreePartTable}
\begin{longtable}{ccccc}
\caption{The IR properties of our sample of 100 AGNs as derived using {\it iragnsep}. These are the sources selected to build our AGN templates (see \S\,\ref{subsubsec:AGNtempl}). A machine-readable version of this table is available in the supplementary material, or directly at \url{https://tinyurl.com/y86on7wk}.}\label{table:AGNfitprop}\\\toprule
\endfirsthead
\caption* {Table \ref{table:AGNfitprop}: $continued.$}\\\toprule
\endhead
\endfoot
\bottomrule
\endlastfoot
ID$\rm ^a$ & S$_{9.7}$~$^{\rm b}$ & $\log_{10}$(\Lir/\Lsun) & IR AGN$_{\rm frac}$ & SFR (\Msun$~yr^{-1}$)  \\
\hline
1   & 0.385  & 10.801  $^{\pm 0.003 }$ & 0.843   & 6.969   $^{\pm 0.043}$\\
2   & 0.24   & 10.399  $^{\pm 0.002 }$ & 0.28    & 2.733   $^{\pm 0.009}$\\
3   & 0.0    & 10.443  $^{\pm 0.001 }$ & 0.622   & 3.024   $^{\pm 0.01 }$\\
4   & 0.5*   & 10.421  $^{\pm 0.001 }$ & 0.21    & 2.871   $^{\pm 0.009}$\\
5   & 0.09   & 10.724  $^{\pm 0.002 }$ & 0.32    & 5.779   $^{\pm 0.021}$\\
6   & 0.035  & 9.932   $^{\pm 0.001 }$ & 0.54    & 0.931   $^{\pm 0.003}$\\
7   & 2.0*   & 10.131  $^{\pm 0.001 }$ & 0.13    & 1.474   $^{\pm 0.005}$\\
\end{longtable}
\begin{tablenotes}
\footnotesize
\item [a] ID of the source as in Table\,\ref{table:AGNprop}.
\item [b] The absorption coefficient at 9.7~\mum. A star indicates that it has been fixed prior to the fit.
\end{tablenotes}
\end{ThreePartTable}

\begin{ThreePartTable}
\begin{longtable}{cclcccccl}
\caption{The properties of our sample of 110 star-forming galaxies pre-selected based on \eqwspd\ and WISE colours (see \S\,\ref{subsubsec:galSample}). A machine-readable version of this table is available in the supplementary material, or directly at \url{https://tinyurl.com/ybe3ujdu}.}\label{table:preSF}\\\toprule
\endfirsthead
\caption* {Table \ref{table:preSF}: $continued.$}\\\toprule
\endhead
\endfoot
\bottomrule
\endlastfoot
ID  & aorkey$^{\rm a}$   & Counterpart Name                   & R.A. (h:m:s) & Dec. (d:m:s)  & \z        & \eqwspd$^{\rm b}$  & W1~-~W2$^{\rm c}$ & Notes$^{\rm d}$ \\
\hline
1   & 3756800  & NGC7714                     & 23:36:14.092 & +02:09:18.518 & 0.00933  & 1.024    & 0.38   & AGN?     \\
2   & 3758080  & NGC1097                     & 02:46:19.098 & -30:16:27.664 & 0.00424  & 0.585    & 0.33   & SF       \\
3   & 3856128  & NGC1808                     & 05:07:42.182 & -37:30:44.962 & 0.00332  & 1.015    & 0.404  & SF       \\
4   & 4340480  & He2-10                      & 08:36:15.220 & -26:24:33.699 & 0.00291  & 1.411    & 0.422  & Ambiguous\\
5   & 4344576  & Mkn331                      & 23:51:26.704 & +20:35:10.467 & 0.01848  & 0.965    & 0.467  & AGN?     \\
6   & 4830976  & NGC1140                     & 02:54:33.520 & -10:01:40.319 & 0.00501  & 0.977    & 0.178  & Discarded\\
7   & 4934400  & NGC2993                     & 09:45:48.311 & -14:22:04.883 & 0.00811  & 1.466    & 0.435  & SF       \\
\end{longtable}
\begin{tablenotes}
\footnotesize
\item [a] aorkey of the source as referenced in the {\sc CASSIS} database.
\item [b] The \eqwspd\ are from L19, and \eqwspd~$>$~0.54 was used to pre-select star-forming galaxies.
\item [c] The WISE colours are from L19 and W1~-~W2~$\leq$~0.5 was also used to pre-select star-forming galaxies.
\item [d] Notes on the source. Only sources flagged ``SF'' were kept in our final sample of pure star-forming galaxies, and used to build our galaxy templates. For the rest, ``Discarded'' means that an issue with the SED was found, ``Ambiguous'' are those sources for which we did not find any evidence of AGN contamination, yet could not be fit with the model of S18, and ``AGN?'' are sources that could not be fit with the model of S18 and for which we found potential AGN contamination (see Appendix\,\ref{app:AGNinter}).
\end{tablenotes}
\end{ThreePartTable}

\begin{landscape}
\begin{ThreePartTable}
\begin{longtable}{cccccccccccccccccc}
\caption {The {\it Herschel} fluxes of our sample of 110 pre-selected star-forming galaxies. A star is used to flag the fluxes of sources with potential IR blending. A machine-readable version of this table is available in the supplementary material, or directly at \url{https://tinyurl.com/ybe3ujdu}.}\label{table:galFlux}\\\toprule
\endfirsthead
\caption* {Table \ref{table:galFlux}: $continued.$}\\\toprule
\endhead
\endfoot
\bottomrule
\endlastfoot
ID$^{\rm a}$  & F70 (mJy) & Flag$_{70}$~$^{\rm b}$ & F100 (mJy) & Flag$_{100}$ & F160 (mJy) & Flag$_{160}$ & F250 (mJy) & Flag$_{250}$ & F350 (mJy) & Flag$_{350}$ & F500 (mJy) & Flag$_{500}$ \\
\hline
1   & 9741.47  $^{\pm293.51    }$     & Ext.     & 10859.77  $^{\pm351.16      }$    & Ext.      & 7846.16    $^{\pm 247.35      }$    & Ext.      & 0.0                                 & N/A       & 0.0                                 & N/A       & 0.0                                 & N/A       \\
2   & 76828.31 $^{\pm2322.1    }$     & Ext.     & 122960.04 $^{\pm3706.27     }$    & Ext.      & 132490.04  $^{\pm 3981.52     }$    & Ext.      & 66667.01   $^{\pm 2004.41     }$    & Ext.      & 28547.82   $^{\pm 861.78       }$   & Ext.      & 9746.59    $^{\pm 299.05      }$    & Ext.      \\
3   & 132596.84$^{\pm3943.52   }$     & Ext.     & 165345.79 $^{\pm4863.89     }$    & Ext.      & 144007.39  $^{\pm 4322.35     }$    & Ext.      & 58112.6    $^{\pm 1746.66     }$    & Ext.      & 22945.77   $^{\pm 693.47       }$   & Ext.      & 7141.51    $^{\pm 220.95      }$    & Ext.      \\
4   & 24455.75 $^{\pm734.81    }$     & Ext.     & 25054.55  $^{\pm755.45      }$    & Ext.      & 18442.76   $^{\pm 556.46      }$    & Ext.      & 6072.96    $^{\pm 185.17      }$    & Ext.      & 2538.51    $^{\pm 80.86        }$   & Ext.      & 872.16     $^{\pm 36.01       }$    & Ext.      \\
5   & 20726.44 $^{\pm372.34    }$     & P.S.     & 22639.5   $^{\pm509.78      }$    & P.S.      & 16445.94   $^{\pm 268.48      }$    & P.S.      & 6168.0     $^{\pm 66.2        }$    & P.S.      & 2340.4     $^{\pm 47.7         }$   & P.S.      & 746.1      $^{\pm 26.7        }$    & P.S.      \\
\end{longtable}
\begin{tablenotes}
\footnotesize
\item [a] ID of the source as in Table\,\ref{table:preSF}.
\item [b] Indicates whether it is a point source (P.S.) or an extended source (Ext.) at a given wavelength. The fluxes for point sources are from the HPDP, and the fluxes for extended sources have been re-calculated to include the extended flux (see Appendix\,\ref{app:IRflux}).
\end{tablenotes}
\end{ThreePartTable}
\end{landscape}

\begin{ThreePartTable}
\begin{table}
\caption {The best fitting parameters of the model of S18 for our sample of 55 pure star-forming galaxies. A machine-readable version of this table is available in the supplementary material, or directly at \url{https://tinyurl.com/ybkypz94}.}\label{table:fitgalParam}
\begin{tabular}{cccccc}
ID$\rm ^a$  & $\log_{10}(M_{\rm dust}$/\Msun) & \Tdust~(K)  & $\log_{10}(M_{\rm PAH}$/\Msun) & $T_{\rm PAH}$~(K)   & $\tau_{9.7}$~$^{\rm b}$ \\
\hline
2   & 9.613    & 26.37  & 9.641  & 63.825 & 1.416  \\
3   & 9.665    & 31.493 & 9.812  & 14.54  & 2.47   \\
7   & 9.471    & 33.992 & 9.577  & 18.732 & 1.68   \\
13  & 10.103   & 29.542 & 10.381 & 32.63  & 7.0*   \\
14  & 9.183    & 32.153 & 9.313  & 23.295 & 1.949  \\
15  & 9.945    & 30.171 & 10.483 & 50.462 & 1.299  \\
16  & 9.112    & 24.147 & 9.658  & 20.067 & 1.245  \\
18  & 9.389    & 31.287 & 9.762  & 26.777 & 1.83   \\
19  & 9.692    & 25.488 & 10.335 & 57.678 & 15.0*  \\
23  & 10.565   & 23.737 & 11.289 & 29.541 & 0.392  \\
25  & 10.626   & 33.491 & 10.869 & 14.837 & 2.653  \\
26  & 10.52    & 31.496 & 10.933 & 42.753 & 1.758  \\
27  & 9.98     & 29.866 & 10.097 & 15.971 & 2.534  \\
28  & 9.971    & 28.347 & 10.184 & 30.026 & 5.0*   \\
29  & 10.363   & 31.508 & 10.252 & 35.126 & 2.124  \\
33  & 10.021   & 26.393 & 10.477 & 22.377 & 30.0*  \\
37  & 11.94    & 35.427 & 11.997 & 19.931 & 20.0*  \\
39  & 11.643   & 34.955 & 11.436 & 58.968 & 1.339  \\
40  & 10.104   & 35.148 & 10.269 & 15.632 & 2.343  \\
41  & 10.045   & 29.106 & 10.335 & 15.103 & 2.283  \\
42  & 9.64     & 25.247 & 10.274 & 16.793 & 1.592  \\
43  & 9.798    & 29.059 & 10.023 & 24.476 & 2.257  \\
44  & 9.679    & 33.539 & 10.098 & 21.426 & 1.943  \\
46  & 9.037    & 21.069 & 9.406  & 25.324 & 0.749  \\
51  & 9.746    & 26.772 & 10.278 & 61.16  & 1.509  \\
54  & 9.694    & 27.477 & 10.183 & 63.484 & 1.812  \\
55  & 10.223   & 31.781 & 10.444 & 15.139 & 2.169  \\
58  & 9.803    & 30.919 & 10.215 & 52.785 & 2.038  \\
59  & 9.726    & 25.231 & 9.904  & 63.709 & 1.584  \\
61  & 10.444   & 32.638 & 10.662 & 32.084 & 2.218  \\
62  & 9.814    & 29.856 & 9.97   & 35.195 & 2.142  \\
63  & 10.104   & 26.775 & 10.661 & 61.115 & 2.427  \\
65  & 9.836    & 30.524 & 10.218 & 50.89  & 1.855  \\
66  & 9.755    & 32.127 & 10.189 & 34.13  & 2.035  \\
68  & 10.099   & 29.869 & 10.442 & 24.211 & 1.613  \\
70  & 8.067    & 24.118 & 8.34   & 36.41  & 1.59   \\
72  & 7.656    & 22.141 & 8.189  & 21.43  & 0.549  \\
76  & 9.604    & 28.653 & 9.996  & 37.796 & 1.989  \\
77  & 9.796    & 43.835 & 9.499  & 47.666 & 0.011  \\
78  & 10.224   & 42.41  & 10.001 & 41.998 & 1.033  \\
79  & 9.835    & 38.42  & 9.793  & 36.339 & 0.979  \\
81  & 9.406    & 30.926 & 9.712  & 26.414 & 2.268  \\
87  & 10.036   & 31.265 & 10.299 & 20.76  & 2.105  \\
88  & 8.964    & 27.401 & 9.328  & 19.478 & 6.0*   \\
90  & 9.455    & 31.279 & 9.687  & 36.823 & 2.505  \\
94  & 8.62     & 26.393 & 9.117  & 22.912 & 1.369  \\
95  & 9.989    & 29.873 & 10.545 & 19.164 & 2.357  \\
96  & 9.995    & 27.08  & 9.843  & 23.777 & 2.257  \\
99  & 9.857    & 25.242 & 10.525 & 16.653 & 1.498  \\
102 & 9.735    & 30.159 & 10.156 & 28.824 & 1.731  \\
104 & 9.632    & 22.628 & 10.045 & 32.576 & 1.46   \\
105 & 9.489    & 27.113 & 9.781  & 24.969 & 1.392  \\
108 & 10.719   & 33.626 & 11.237 & 39.105 & 2.552  \\
109 & 10.761   & 25.788 & 11.338 & 32.098 & 2.25   \\
110 & 8.622    & 21.556 & 9.313  & 25.185 & 0.0 
\end{tabular}
\begin{tablenotes}
\footnotesize
\item [a] ID of the source as in Table\,\ref{table:preSF}.
\item [b] Total optical depth at 9.7~\mum. Values with a star have been fixed prior to the fit.
\end{tablenotes}
\end{table}
\end{ThreePartTable}

\bsp	
\label{lastpage}
\end{document}